\documentclass[aps,prx,reprint,showpacs,superscriptaddress, english]{revtex4-1}
\usepackage{color}
\usepackage{array}
\usepackage{verbatim}
\usepackage{longtable}
\usepackage{multirow}
\usepackage{amsmath}
\usepackage{amssymb}
\usepackage{graphicx}
\usepackage{dsfont}
\usepackage{esint}
\usepackage{enumitem}
\usepackage[unicode=true,bookmarks=true,bookmarksnumbered=true,bookmarksopen=false,breaklinks=true,pdfborder={0 0 0},backref=false,colorlinks=true]{hyperref}

\def \hlines  {\cline{2-10}}
\def \hliness {\cline{4-10}}
\def \figwid  {1.9cm}
\def \figwids {1.5cm}

\definecolor{red}{rgb}{1,0,0} 
\definecolor{blue}{rgb}{0,0,1}
\definecolor{black}{rgb}{0,0,0}

\newcommand{\PreserveBackslash}[1]{\let\temp=\\#1\let\\=\temp}
\newcolumntype{C}[1]{>{\PreserveBackslash\centering}p{#1}}
\newcolumntype{R}[1]{>{\PreserveBackslash\raggedleft}p{#1}}
\newcolumntype{L}[1]{>{\PreserveBackslash\raggedright}p{#1}}

\begin{document}
\title{Diagnosis for nonmagnetic topological semimetals in the absence of spin-orbital coupling}
\author{Zhida Song}
\affiliation{Beijing National Laboratory for Condensed Matter Physics,
and Institute of Physics, Chinese Academy of Sciences, Beijing 100190, China}
\affiliation{University of Chinese Academy of Sciences, Beijing 100049, China}
\author{Tiantian Zhang}
\affiliation{Beijing National Laboratory for Condensed Matter Physics,
and Institute of Physics, Chinese Academy of Sciences, Beijing 100190, China}
\affiliation{University of Chinese Academy of Sciences, Beijing 100049, China}
\author{Chen Fang}\email{cfang@iphy.ac.cn}
\affiliation{Beijing National Laboratory for Condensed Matter Physics, and Institute of Physics, Chinese Academy of Sciences, Beijing 100190, China}
\affiliation{CAS Center for Excellence in Topological Quantum Computation, Beijing, China}
\begin{abstract}
Topological semimetals are under intensive theoretical and experimental studies. 
The first step of these studies is always the theoretical (numerical) predication of one or several candidate materials, based on first principles numerics. 
In these calculations, it is crucial that all topological band crossings, including their types and positions in the Brillouin zone, are found.
While band crossings along high-symmetry lines, which are routinely scanned in numerics, are simple to locate, the ones at generic momenta are notoriously time-consuming to find, and may be easily missed.
In this paper, we establish a theoretical scheme of diagnosis for topological semimetals where all band crossings are at generic momenta in systems with time-reversal symmetry and negligible spin-orbital coupling.
The scheme only uses the symmetry (inversion and rotation) eigenvalues of the valence bands at high-symmetry points in the Brillouin zone as input, and provides the types (lines or points), topological charges, numbers and configurations of all robust topological band crossings, if any, at generic momenta.
The nature of new diagnosis scheme allows for full automation and parallelization, and paves way to high throughput numerical predictions of topological semimetals.
\end{abstract}
\maketitle
\section{Introduction}
\label{sec:intro}
Topologically protected Weyl points \cite{Murakami2007,Wan2011,Burkov2016} and nodal lines \cite{Burkov2011} are being actively searched for and studied in boson bands, such as photons \cite{Lu2013,Lu2015}, phonons \cite{Stenull2016,Zhang2017} and magnons \cite{Li2016,Fransson2016,Li2017,Yao2017,Bao2017}, as well as in electronic bands \cite{Lv2015,Xu2015a,Bian2016}.
On one hand, these topological band crossings bring about novel physical properties such as the existence of ``arcs'' for equal energy contours on the surface \cite{Wan2011,Fang2016}, and the quantum anomalies in the bulk \cite{Son2013}.
On the other hand, they are considered the ``parent states'' of many an interesting topological gapped and gapless state, if certain symmetries are broken either by natural or applied perturbation, such as spin-orbital coupling (SOC) or external strain. 
For example, a 2D honeycomb lattice hosts two Dirac points, but when SOC is present, it becomes a topological insulator characterized by the famous Kane-Mele model \cite{Kane2005,Kane2005a}; TaAs is a 3D nodal line semimetal without considering SOC, but when it is included (even only perturbatively), the nodal lines break into Weyl points \cite{Huang2015,Weng2015}.

Some topological band crossings appear at high-symmetry momenta.
These include point-type crossings (Weyl and Dirac points) along high-symmetry lines, or line-type crossings on high-symmetry planes.
The numerical diagnosis of these crossings, that is, predicting their existence, types and configurations from first principles, is relatively easy, as only a limited number of lines in the band structure should be scanned, a standard protocol integrated in most first principles implementations nowadays \cite{Kresse1996}.
On the other hand, the diagnosis for the topological band crossings at \textit{generic momenta} is difficult, because in principle the entire Brillouin zone needs to be scanned; and in order to confirm the type of a crossing point, advanced methods such as Wilson loops \cite{Yu2011} must be invoked to calculate its topological invariant.
This involved process severely slows down the numerical search for more topological semimetals in real materials.

Very recently, several seminal works have established the mathematical relations between the connectivity of bands in a band structure and the irreducible representations of space groups in the valence bands \cite{Po2017,Bradlyn2017,Kruthoff2016}.
For each space group, a set of equations called the ``compatibility relations'' are found, which are satisfied if and only if a band structure does not have any (non-accidental) crossing between the conduction and the valence bands along all high-symmetry lines in the Brillouin zone.
If any one of the compatibility relations is violated, the system \textit{must} be a topological semimetal with band crossing at high-symmetry momenta.
On the other hand, if all equations are met, there are two possibilities: the band structure is either fully gapped at all momenta (insulator), or has band crossings at generic momenta.
Ref. [\onlinecite{Po2017}] and Ref. [\onlinecite{Bradlyn2017}] further study the band structures of atomic insulators (or their superpositions), which form a linear space spanned by basis vectors called the ``elementary band representations''.
These authors point out that if a band structure satisfies all compatibility relations but cannot linear-decompose into elementary band representations with integer coefficients, it must be either a topological gapped state, or a topological semimetal.
Both the compatibility relations and the explicit expressions of elementary band representations are now fully available online at BILBAO thanks to the effort of Ref. \cite{Vergniory2017,Elcoro2017,Bradlyn2017}.

Po, Vishwanath and Watanabe (PVW) \cite{Po2017} showed that given that all compatibility relations satisfied, the numbers of the appearance of each irreducible representation at all high-symmetry momenta in the valence bands (defined as ``symmetry data'' for short) can be converted into a very small set containing at most four $\mathbb{Z}_n$ numbers with $n=2,4$ for the orthogonal Hamtiltonians (time-reversal without SOC) and $n=2,3,4,6,8,12$ for the symplectic ones (time-reversal with SOC).
These numbers are termed ``symmetry-based indicators'' (indicator for short).
Indicators are indicative of nontrivial topology in a band structure, such that any system with nonzero indicators \textit{cannot} be an atomic insulator, but can only be either a topologically gapped state or a topological semimetal.
Mark here that in the latter possibility, all band crossings must appear at generic momenta because the compatibility relations are given as satisfied.
More importantly, any system having zero indicators has the symmetry data that cannot be distinguished from that of an atomic insulator \cite{Po2017,Bradlyn2017}.
This property means that all information on band topology contained in symmetry data can be extracted from indicators alone.
However, the mathematical framework used by PVW does not tell us what topological states each nonzero set of indicators corresponds to: is the material an insulator or a semimetal? if insulator what are the topological invariants? if semimetal what are the types and configurations of the band crossings?
Also, Ref. [\onlinecite{Po2017}] only contains the groups formed by indicators (or the indicator groups) without giving the explicit isomorphism between the symmetry eigenvalues and the generators of that group.

The paper is tasked with answering these questions in systems with time-reversal and negligible SOC, filling the gaps between symmetry data and band topology by (i) giving each generator of each indicator group an explicit formula in terms of symmetry eigenvalues of valence bands at high-symmetry momenta and (ii) ``translating'' each nonzero set of indicators into a set of well-defined topological states.
The main results can be parsed into two statements:
\begin{enumerate}
\item All nonzero sets of indicators in all space groups necessarily correspond to topological semimetals, but not insulators.\\
\item The types, numbers and configurations of the band crossings can be partly predicted by the indicators.\\
\end{enumerate}
Both results are obtained by examining each indicator in every space group, where we first find the explicit formula for that indicator, prove that a band crossing must exist whenever it becomes nonzero, determine the type of the crossing, and finally give a possible configuration.
In centrosymmetric space groups, all these topological semimetals are nodal lines semimetals, where the nodal lines are away from high-symmetry points, lines or planes.
Particularly, we find that for many space groups, there is always one set of indicators that corresponds to nodal line topological semimetals where nodal loops have nontrivial $\mathbb{Z}_2$-monopole charge, a special type of nodal loops that have not been predicted in any realistic electronic materials.
In non-centrosymmetric space groups, all nonzero sets of indicators correspond to Weyl semimetals, where the Weyl points are away from high-symmetry lines.

In finding the expressions for all indicators, we happily discover, without having such expectations to begin with, that most of them are in fact Berry phases along certain loops formed by high-symmetry lines, and others the difference of topological charge between two high-symmetry planes.
They are all (except two) \textit{topological invariants} of some sub-manifolds of the Brillouin zone (BZ), and if nonzero, they give precise information on the number and positions of band crossings in the sub-manifold.

\section{Preliminaries: symmetry-based indicators}

In this section, we show how the generators of an indicator group are chosen and how the corresponding formulae are derived.
In doing this, we first introduce an abstract framework, and then give two examples showing the calculations step by step.

\subsection{General theory}\label{sub:SI}

As introduced in the previous section, the symmetry data of a band structure is given by the number of each irreducible representation in the occupied bands.
In the following we denote this number as  $n^{K_i}_{\xi_j}$, where the $K_i$ represents the high symmetry momentum and $\xi_j$ represents an irreducible representation at $K_i$.
Then the symmetry data of a band structure can be written as a ``vector''
\begin{equation}
\mathbf{n} = (n^{\mathbf{K}_1}_{\xi_1},n^{\mathbf{K}_1}_{\xi_2},\cdots,n^{\mathbf{K}_2}_{\xi_1},\cdots)^T.
\end{equation}
In principle $n^{\mathbf{K}_j}_{\xi_i}$'s should be nonnegative integers, however, here we generalize them to arbitrary integers \mbox{\cite{Po2017}}.
Since we are only interested in semimetals with crossing points at generic momenta, we require $\mathbf{n}$ to satisfy the compatibility relations \mbox{\cite{Bradlyn2017,Po2017}}, which are a set of constraints such as symmetry eigenvalue conservation along high symmetry lines, such that the corresponding band structure has no symmetry protected crossings at high symmetry momenta.
Compatibility relations for all space groups are available online at BILBAO derived in Ref. \mbox{%DIFAUXCMD
\cite{Bradlyn2017,Vergniory2017,Elcoro2017}}.
Mathematically, these constraints are described by a set of linear equations, and the compatibility-relation-allowed symmetry data form the solution subspace of these linear equations.
Because all entries of $\mathbf{n}$ should be integer, this solution space is an infinite abelian group, $\mathbb{Z}^{d_\mathrm{BS}}$ \cite{Kruthoff2016}.
In the following we denote this abelian group as $\{\mathrm{BS}\}$ and denote its generators as $\mathbf{b}_{i=1\cdots d_\mathrm{BS}}$.
Here $d_\mathrm{BS}$ is the rank of \{BS\}.

Another perspective to understand symmetry data is from atomic insulators.
Atomic insulators are defined as insulators consisting of uncoupled atoms with closed-shells, where, to meet the symmetry, atomic orbitals form the irreducible representations of the corresponding on-site symmetry group. 
By definition, the generated symmetry data satisfy the compatibility relations and thus form a subgroup, denoted as $\{\mathrm{AI}\}$, of $\{\mathrm{BS}\}$.
The authors of Ref. \cite{Po2017} prove that \{AI\} has same rank with $\{\mathrm{BS}\}$ and thus the quotient group $\{\mathrm{BS}\}/\{\mathrm{AI}\}$ is finite. 
Concretely, for each $\mathbf{b}_i$ there exists a minimal integer $\kappa_i$ giving $\kappa_i \mathbf{b}_i \in \{\mathrm{AI}\}$.
Therefore, any symmetry data in \{BS\} can be decomposed into a part belonging to \{AI\} and a part not belonging to \{AI\}
\begin{align}
\forall & \mathbf{n}=\sum_i c_i\mathbf{b}_i \in \{\mathrm{BS}\}, \quad \exists \mathbf{n}^\prime\in\{\mathrm{AI}\},\quad s.t., \nonumber\\
& \mathbf{n} = \mathbf{n}^\prime + \sum_i (c_i\;\mathrm{mod}\;\kappa_i)\mathbf{b}_i, \label{eq:SI}
\end{align}
where $\sum_i (c_i\;\mathrm{mod}\;\kappa_i)\mathbf{b}_i$ is the part not belonging to \{AI\}.
Such a decomposition implies that the quotient group $\{\mathrm{BS}\}/\{\mathrm{AI}\}$ is $\mathbb{Z}_{\kappa_1}\times\mathbb{Z}_{\kappa_2}\times\cdots$.
The integers, $(c_i\;\mathrm{mod}\;\kappa_i)$'s, which identifies $\mathbf{n}$ with one element of $\{\mathrm{BS}\}/\{\mathrm{AI}\}$, are defined as the indicators of $\mathbf{n}$, and hence the quotient group $\{\mathrm{BS}\}/\{\mathrm{AI}\}$ is also referred as the indicator group.
Band structures having different indicators must be topologically distinct from each other because their difference, the symmetry data of which does not belong to \{AI\} according to Eq. (\ref{eq:SI}), cannot be an atomic insulator.
Generally, indicators depend on the choice of \{BS\} generators.
Upon a change of the choice of \{BS\} generators, the values of indicators change, but the indicator group remains the same.
More importantly, the redefined indicators with the new \{BS\} generators are still valid because distinct indicators still correspond to distinct topologies.
In Sec. \ref{sub:SI-P-1} we take space group \#2 as an example to show such redefinition in more detail.
In the rest of this paper, we call the choice of \{BS\} generators as the convention of indicators.
For concreteness, in the following two subsections and Sec. \ref{sec:centro}-\ref{sec:noncentro}, we fix the conventions of indicators in all space groups.

The formulae to calculate indicators indeed form a linear mapping from \{BS\} to the indicator group.
Thus we express the formulae of the $i$-th indicator as $F_i (\mathbf{n}) \equiv \mathbf{f}^T_i \mathbf{n}\;\mathrm{mod}\;\kappa_i$, where $\mathbf{f}_i$ gives the explicit form of the formulae and satisfies $\mathbf{f}_i^T\mathbf{b}_j = \delta_{ij}\;\mathrm{mod}\;\kappa_i$.
We emphasize that, even for a fixed indicator convention, the explicit forms of the formulae, i.e., $\mathbf{f}_i$'s, are in general not unique due to compatibility relations.
Since all symmetry data in \{BS\} satisfy compatibility relations, which take the form $\mathbf{c}^T\mathbf{n}=0$, we can add $\mathbf{c}$ to $\mathbf{f}_i$'s without changing the results, i.e., $F^\prime_i(\mathbf{n})\equiv(\mathbf{f}_i+\mathbf{c})^T\mathbf{n}\;\mathrm{mod}\;\kappa_i=F_i(\mathbf{n})$. 
A general strategy to get one of the many equivalent explicit forms is to solve the left inverse of the nonsquare matrix $B=(\mathbf{b}_1,\mathbf{b}_2,\cdots)$,  $B^{-1}_\mathrm{left}$, such that $B^{-1}_\mathrm{left}B=\mathbb{I}$.
Then the $\mathbf{f}_i$ vector is given by the $i$-th column of $B^{-1 T}_\mathrm{left}$.

{In order to simplify finding explicit indicator formulae and get a consistent understanding of indicators in all space groups, here we represent a method to get indicators of a space group from indicators of its subgroup.
We denote the group and its subgroup as $G$ and $H$, respectively.
We first notice that the indicators of $H$, which are expressed as functions of the symmetry data in $H$, are also well defined for that in $G$, as each irreducible representation of $G$ necessarily reduces into one or multiple irreducible representations in $H$.
Then if two band structures in $G$ have distinct indicators of $H$, which imply topological distinction, they must have distinct indicators of $G$, because, by definition, topologies that can be distinguished from symmetry data must have distinct indicators.
In other words, any two distinct \textit{realizable} indicators of $H$ in $G$ correspond to two distinct indicators of $G$. 
Here by realizable indicators of $H$ in $G$ we mean that there is at least one symmetry data of $G$ that (i) satisfies all compatibility relations of $G$ and (ii) has such a set of indicators of $H$.
If the realizable indicators of $H$ happen to form the indicator group of $G$, we say that the indicator group of $G$ is \textit{completely} induced from $H$; otherwise, we say that the indicator group of $G$ is \textit{partly} induced from $H$.
In the latter case new formulae for the remaining indicators must be defined.
In Sec. \ref{sub:SI-P2/m} we give a concrete example for indicator induction.

\subsection{Space group \#2: the simplest example}\label{sub:SI-P-1}

Space group \#2 is generated by three lattice translations along three lattice vectors $\mathbf{a}_{1,2,3}$ and inversion symmetry \cite{Aroyo2009}.
The eight high symmetry momenta are the eight time-reversal invariant momenta (TRIMs), i.e. $(k_1,k_2,k_3)$ with $k_{1,2,3}=0,\pi$. 
(Hereafter we use $(k_1,k_2,k_3)$ to denote momentum in reciprocal lattices, i.e., $\mathbf{k}=k_1\mathbf{g}_1+k_2\mathbf{g}_2+k_3\mathbf{g}_3$, where $\mathbf{g}_{i=1,2,3}$ are the reciprocal bases, and use $(k_x,k_y,k_z)$ to denote momentum in cartesian coordinate. 
Definitions of the reciprocal lattices in all space groups can be found on the BILBAO website \cite{aroyo_BZ_2014}, and the part used in this paper are tabulated in table \ref{tab:1}.)  
On each of them there are two kinds of one-dimensional irreducible representations: the even under inversion and the odd under inversion.
Thus the symmetry data takes the form
$ \mathbf{n} = (n^{\mathbf{K}_1}_+, n^{\mathbf{K}_1}_-, \cdots, n^{\mathbf{K}_8}_+, n^{\mathbf{K}_8}_-)^T$,
where $n^{\mathbf{K}_i}_{\pm}$ represents the number of occupied states having the inversion eigenvalue $\pm 1$ at the $i$-th TRIM.
The only compatibility relation in space group \#2 is to require the same occupation numbers at different TRIMs, i.e., 
\begin{equation}
n^{\mathbf{K}_i}_+ + n^{\mathbf{K}_i}_- - n^{\mathbf{K}_j}_+ - n^{\mathbf{K}_j}_- = 0, \qquad (i,j=1,\cdots,8).    
\end{equation}
The solution space of these equations are easy to get, and we list the nine generators in table \ref{tab:P-1}.

\begin{table}
\begin{tabular}{|c|l|c|}
\hline
\multicolumn{3}{|c|}{Generators of \{BS\}} \\
\hline
Basis & Symmetry data & $\kappa_i$\\
\hline
$\mathbf{b}_1$ & $\mathbf{e}^\Gamma_- + \mathbf{e}^\mathrm{R}_- + \mathbf{e}^\mathrm{T}_+ + \mathbf{e}^\mathrm{U}_+ + \mathbf{e}^\mathrm{V}_- + \mathbf{e}^\mathrm{X}_+ + \mathbf{e}^\mathrm{Y}_+ + \mathbf{e}^\mathrm{Z}_-$ & 1\\
\hline
$\mathbf{b}_2$ & $\mathbf{e}^\Gamma_- + \mathbf{e}^\mathrm{R}_- + \mathbf{e}^\mathrm{T}_- + \mathbf{e}^\mathrm{U}_+ + \mathbf{e}^\mathrm{V}_+ + \mathbf{e}^\mathrm{X}_- + \mathbf{e}^\mathrm{Y}_+ + \mathbf{e}^\mathrm{Z}_+$ & 1\\
\hline
$\mathbf{b}_3$ & $\mathbf{e}^\mathrm{T}_+ - \mathbf{e}^\mathrm{T}_- - \mathbf{e}^\mathrm{U}_+ + \mathbf{e}^\mathrm{U}_- + \mathbf{e}^\mathrm{X}_+ - \mathbf{e}^\mathrm{X}_- - \mathbf{e}^\mathrm{Y}_+ + \mathbf{e}^\mathrm{Y}_-$ & 1\\
\hline
$\mathbf{b}_4$ & $-\mathbf{e}^\mathrm{R}_+ + \mathbf{e}^\mathrm{R}_- +  \mathbf{e}^\mathrm{U}_+ - \mathbf{e}^\mathrm{U}_- + \mathbf{e}^\mathrm{V}_+ - \mathbf{e}^\mathrm{V}_- - \mathbf{e}^\mathrm{X}_+ + \mathbf{e}^\mathrm{X}_-$ & 1\\
\hline
$\mathbf{b}_5$ & $-\mathbf{e}^\Gamma_+ + \mathbf{e}^\Gamma_- -\mathbf{e}^\mathrm{T}_+ + \mathbf{e}^\mathrm{T}_- -\mathbf{e}^\mathrm{U}_+ + \mathbf{e}^\mathrm{U}_- -\mathbf{e}^\mathrm{V}_+ + \mathbf{e}^\mathrm{V}_-  $ & 1\\
\hline
$\mathbf{b}_6$ & $-\mathbf{e}^\Gamma_+ + \mathbf{e}^\Gamma_- +\mathbf{e}^\mathrm{X}_+ - \mathbf{e}^\mathrm{X}_- $ & 2\\
\hline
$\mathbf{b}_7$ & $-\mathbf{e}^\Gamma_+ + \mathbf{e}^\Gamma_- +\mathbf{e}^\mathrm{Y}_+ - \mathbf{e}^\mathrm{Y}_- $ & 2\\
\hline
$\mathbf{b}_8$ & $-\mathbf{e}^\Gamma_+ + \mathbf{e}^\Gamma_- +\mathbf{e}^\mathrm{Z}_+ - \mathbf{e}^\mathrm{Z}_- $ & 2\\
\hline
$\mathbf{b}_9$ & $-\mathbf{e}^\Gamma_+ + \mathbf{e}^\Gamma_-  $ & 4\\
\hline
\hline
\multicolumn{3}{|c|}{Generators of \{AI\}} \\
\hline
Basis & \multicolumn{2}{l|}{Symmetry data} \\
\hline
$\mathbf{a}_1 (G_+^{1a})$ & \multicolumn{2}{l|}{$\mathbf{e}^\Gamma_+ + \mathbf{e}^\mathrm{R}_+ + \mathbf{e}^\mathrm{T}_+ + \mathbf{e}^\mathrm{U}_+ + \mathbf{e}^\mathrm{V}_+ + \mathbf{e}^\mathrm{X}_+ + \mathbf{e}^\mathrm{Y}_+ + \mathbf{e}^\mathrm{Z}_+ $} \\
\hline
$\mathbf{a}_2 (G_-^{1a})$ & \multicolumn{2}{l|}{$\mathbf{e}^\Gamma_- + \mathbf{e}^\mathrm{R}_- + \mathbf{e}^\mathrm{T}_- + \mathbf{e}^\mathrm{U}_- + \mathbf{e}^\mathrm{V}_- + \mathbf{e}^\mathrm{X}_- + \mathbf{e}^\mathrm{Y}_- + \mathbf{e}^\mathrm{Z}_- $} \\
\hline
$\mathbf{a}_3 (G_+^{1b})$ & \multicolumn{2}{l|}{$\mathbf{e}^\Gamma_+ + \mathbf{e}^\mathrm{R}_- + \mathbf{e}^\mathrm{T}_- + \mathbf{e}^\mathrm{U}_- + \mathbf{e}^\mathrm{V}_+ + \mathbf{e}^\mathrm{X}_+ + \mathbf{e}^\mathrm{Y}_+ + \mathbf{e}^\mathrm{Z}_- $} \\
\hline
$\mathbf{a}_4 (G_+^{1c})$ & \multicolumn{2}{l|}{$\mathbf{e}^\Gamma_+ + \mathbf{e}^\mathrm{R}_- + \mathbf{e}^\mathrm{T}_- + \mathbf{e}^\mathrm{U}_+ + \mathbf{e}^\mathrm{V}_- + \mathbf{e}^\mathrm{X}_+ + \mathbf{e}^\mathrm{Y}_- + \mathbf{e}^\mathrm{Z}_+ $} \\
\hline
$\mathbf{a}_5 (G_+^{1d})$ & \multicolumn{2}{l|}{$\mathbf{e}^\Gamma_+ + \mathbf{e}^\mathrm{R}_- + \mathbf{e}^\mathrm{T}_+ + \mathbf{e}^\mathrm{U}_- + \mathbf{e}^\mathrm{V}_- + \mathbf{e}^\mathrm{X}_- + \mathbf{e}^\mathrm{Y}_+ + \mathbf{e}^\mathrm{Z}_+ $} \\
\hline
$\mathbf{a}_6 (G_+^{1e})$ & \multicolumn{2}{l|}{$\mathbf{e}^\Gamma_+ + \mathbf{e}^\mathrm{R}_+ + \mathbf{e}^\mathrm{T}_- + \mathbf{e}^\mathrm{U}_- + \mathbf{e}^\mathrm{V}_+ + \mathbf{e}^\mathrm{X}_- + \mathbf{e}^\mathrm{Y}_- + \mathbf{e}^\mathrm{Z}_+ $} \\
\hline
$\mathbf{a}_7 (G_+^{1f})$ & \multicolumn{2}{l|}{$\mathbf{e}^\Gamma_+ + \mathbf{e}^\mathrm{R}_+ + \mathbf{e}^\mathrm{T}_- + \mathbf{e}^\mathrm{U}_+ + \mathbf{e}^\mathrm{V}_- + \mathbf{e}^\mathrm{X}_- + \mathbf{e}^\mathrm{Y}_+ + \mathbf{e}^\mathrm{Z}_- $} \\
\hline
$\mathbf{a}_8 (G_+^{1g})$ & \multicolumn{2}{l|}{$\mathbf{e}^\Gamma_+ + \mathbf{e}^\mathrm{R}_+ + \mathbf{e}^\mathrm{T}_+ + \mathbf{e}^\mathrm{U}_- + \mathbf{e}^\mathrm{V}_- + \mathbf{e}^\mathrm{X}_+ + \mathbf{e}^\mathrm{Y}_- + \mathbf{e}^\mathrm{Z}_- $} \\
\hline
$\mathbf{a}_9 (G_+^{1h})$ & \multicolumn{2}{l|}{$\mathbf{e}^\Gamma_+ + \mathbf{e}^\mathrm{R}_- + \mathbf{e}^\mathrm{T}_+ + \mathbf{e}^\mathrm{U}_+ + \mathbf{e}^\mathrm{V}_+ + \mathbf{e}_X^- + \mathbf{e}_Y^- + \mathbf{e}_Z^- $} \\
\hline
\end{tabular}
\caption{\label{tab:P-1} The generators of \{BS\} and \{AI\} of the space group \#2. Here $\mathbf{e}^\mathbf{K}_{\pm}$ represents the basis where all entries are zero except $n^\mathbf{K}_{\pm}=1$, $\kappa_i$ represents the order of $\mathbf{b}_i$, i.e., the minimal positive integer making $\kappa_i\mathbf{b}_i\in\{\mathrm{AI}\}$, and the symbol $G^{w}_{\pm}$ in parenthesis represents that the corresponding atomic symmetry data is induced from the orbital having the $P$ eigenvalue $\pm1$ at the Wyckoff position $w$. The notations for momenta and positions are defined as $\Gamma(000)$, $\rm R(\pi\pi\pi)$, $\rm T(0\pi\pi)$, $\rm U(\pi 0\pi)$, $\rm V(\pi\pi 0)$, $\rm X(\pi 00)$, $\rm Y(0\pi 0)$, $\rm Z(00\pi)$, and $1a(000)$, $1b(00\frac12)$, $1c(0\frac12 0)$,  $1d(\frac12 00)$, $1e(\frac12\frac12 0)$, $1f(\frac12 0\frac12)$, $1g(0\frac12\frac12)$, $1h(\frac12\frac12\frac12)$, respectively.}
\end{table}

According to Ref. \cite{Bradlyn2017}, to generate $\{\mathrm{AI}\}$ only the orbitals at the ``maximal Wyckoff positions'' need to be considered.
Here maximal Wyckoff position refers to position whose on-site symmetry group is a maximal finite subgroup of the space group.
For space group \#2 there are only eight such positions, i.e., $(x_1,x_2,x_3)$ with $x_{1,2,3}=0,\frac12$.
Since all of these positions are inversion-invariant (modulo a lattice), on each position we have only two kinds of orbitals: the even one and the odd one. 
We denote the atomic orbital with parity $\chi=\pm1$ at the inversion-invariant position $\mathbf{t}$ in the lattice $\mathbf{R}$ as $|\chi\mathbf{t+R}\rangle$, and define the corresponding Bl\"och wave function as $|\phi_{\chi\mathbf{t}}(\mathbf{k})\rangle = \frac{1}{\sqrt{N}}\sum_\mathbf{R} e^{i(\mathbf{t+R})\cdot\mathbf{k}} |\chi\mathbf{t+R}\rangle$.
Under inversion operation, $P$, the atomic orbital firstly get an inversion eigenvalue and then move to the inverted position, i.e., $\hat{P}|\chi\mathbf{t+R}\rangle=\chi|\chi\mathbf{-t-R}\rangle$.
(We use unhatted symbols for the symmetries themselves, and hatted ones for the corresponding operators.)
Using this property it is direct to obtain the inversion eigenvalues of $|\phi_{\chi\mathbf{t}}(\mathbf{k})\rangle$ for $\mathbf{k}\in\mathrm{TRIMs}$.
By this method we generate sixteen symmetry data (two orbitals for each position and eight positions in total) and find that only nine of them are linearly independent, which is consistent with the statement that $\{\mathrm{AI}\}$ has same rank with $\{\mathrm{BS}\}$.
We choose the nine listed in table \ref{tab:P-1} as the generators of $\{\mathrm{AI}\}$.
A crucial observation then follows: $\mathbf{b}_{i=1\cdots5}$ can be generated from $\mathbf{a}_{i}$'s, for example, $\mathbf{b}_1 = \mathbf{a}_1 + \mathbf{a}_2 - \mathbf{a}_6$, whereas $\mathbf{b}_{i=6,7,8,9}$ can not.
Nevertheless the less, $2\mathbf{b}_6$, $2\mathbf{b}_7$, $2\mathbf{b}_8$, and $4\mathbf{b}_9$ can be generated from $\mathbf{a}_{i}$'s, for example, $4\mathbf{b}_{9}=3\mathbf{a}_1+4\mathbf{a}_2-\sum_{i=3}^9\mathbf{a}_i$.
Therefore, we have $\kappa_{i=1\cdots5}=1$, $\kappa_{i=6,7,8}=2$, $\kappa_{i=9}=4$, and the indicator group $\{\mathrm{BS}\}/\{\mathrm{AI}\}$ is $\mathbb{Z}_2\times\mathbb{Z}_2\times\mathbb{Z}_2\times\mathbb{Z}_4$.

By counting the numbers of even and odd states at particular TRIMs, we find the following formulae give a successful mapping from \{BS\} to $\mathbb{Z}_2\times\mathbb{Z}_2\times\mathbb{Z}_2\times\mathbb{Z}_4$.
\begin{equation}
\label{eq:p-1}
\begin{aligned}
z_{2,1}\equiv&\sum_{\substack{\mathbf{K}\in\mathrm{TRIM} \\ \text{at } \{k_1=\pi\}}}\frac{N_-(\mathbf{K})-N_+(\mathbf{K})}{2}\;\mathrm{mod}\;2\\
z_{2,2}\equiv&\sum_{\substack{\mathbf{K}\in\mathrm{TRIM} \\ \text{at } \{k_2=\pi\}}}\frac{N_-(\mathbf{K})-N_+(\mathbf{K})}{2}\;\mathrm{mod}\;2\\
z_{2,3}\equiv&\sum_{\substack{\mathbf{K}\in\mathrm{TRIM} \\ \text{at } \{k_3=\pi\}}}\frac{N_-(\mathbf{K})-N_+(\mathbf{K})}{2}\;\mathrm{mod}\;2\\
z_{4}\equiv&\sum_{\mathbf{K}\in\mathrm{TRIM}}\frac{N_-(\mathbf{K})-N_+(\mathbf{K})}{2}\;\mathrm{mod}\;4
\end{aligned},
\end{equation}
where $N_\pm(\mathbf{K})$ is the number of valence bands having positive (negative) parity.
Here we use the notation $N_{\pm}(\mathbf{K})$ instead of $n^\mathbf{K}_\pm$ to emphasize that these equations are applicable for all centrosymmetric space groups, where $N_{\pm}(\mathbf{K})$ is obtained from more general symmetry data.
For a general centrosymmetric space group, one ignores all symmetries but inversion and translation, and calculate $N_{\pm}(\mathbf{K})$ as if it were a symmetry data of space group \#2.
As shown in Sec. \ref{sec:centro} most indicators in centrosymmetric space groups are indeed induced from the above four equations.

As mentioned in Sec. \ref{sub:SI}, the indcators are in general convention-dependent due to the many choices of \{BS\} generators.
Here we take the $z_4$ indicator as an example to show the convention dependence and discuss the physical interpretation of this convention-dependence.
Below we proceed with another choice of $\mathbf{b}_8$, but the discussion applies for all \{BS\} generators.
Upon redefining the $8$-th generator as $\mathbf{b}_8-2\mathbf{b}_9$, which leaves the order $\kappa_8=2$ invariant, the $z_{2,i=1,2,3}$ indicators keep invariant, whereas the $z_4$ indicator changes to $ z_4+2z_{2,3}\;\mathrm{mod}\;4$.
Specifically, upon the redefining, the indicator set $(0011)$ interchanges with $(0013)$, and the indicator set $(0010)$ interchanges with $(0012)$.
Then, we find that, from the physical point of view, redefining indicators in this way corresponds to a different choice of inversion center from $(000)$ to $(00\frac12)$.
Choosing a different inversion center redefines the inversion operation as $\{\bar1|001\}$, i.e., inversion centered at $(000)$ followed by a translation $(001)$.
(Following BILBAO \cite{Aroyo2009}, we use $\{p|\mathbf{t}\}$ to represent the space group operation composed of point group operation $p$ followed by translation $\mathbf{t}$.)
The translation $(001)$ leads to additional ``$-$'' signs in the inversion eigenvalues at the $k_3=\pi$ momenta, such that the generator $\mathbf{b}_8=-\mathbf{e}^\Gamma_{+} + \mathbf{e}^\Gamma_{-} + \mathbf{e}^\mathrm{Z}_{+} - \mathbf{e}^\mathrm{Z}_{-}$, which have indicator set $(0010)$ due to Eq. (\ref{eq:p-1}), changes to $\mathbf{b}_8=-\mathbf{e}^\Gamma_{+}+\mathbf{b}_8 + \mathbf{e}^\Gamma_{-} + \mathbf{e}^\mathrm{Z}_{-} - \mathbf{e}^\mathrm{Z}_{+} $, which have the indicator set $(0012)$ due to Eq. (\ref{eq:p-1}). (See table \ref{tab:P-1} for definitions of the nation $\mathbf{e}^\mathbf{K}_\pm$.)
Generalizing the discussion for changing inversion center from $(000)$ to $(\frac{i}{2}\frac{j}{2}\frac{k}{2})$ ($i,j,k=0,1$), one can easily find that the $z_{2,i=1,2,3}$ indicators keep invariant whereas the $z_4$ indicator changes to $z_{4}+2iz_{2,1}+2jz_{2,2}+2kz_{2,3}\;\mathrm{mod}\;4$.
It should be noticed that, whichever inversion center is chosen, the parity of $z_4$ keeps the same, \emph{i.~e.}, convention independent. 
In fact odd $z_4$ corresponds to an odd number of nodal loops centering at TRIMs, as discussed in Sec. \ref{sec:centro}.

\subsection{Space group \#10: example showing indicator induction and more} \label{sub:SI-P2/m}

Space group \#10 is generated by three lattice translations along $\mathbf{a}_{1,2,3}$, an inversion, $P$, and an rotation, $C_2$, wherein the rotation axis passes through the inversion center and is parallel to $\mathbf{a}_2$ \mbox{\cite{Aroyo2009}}. 
%\#10 also has a mirror symmetry, $M=PC_2$, generated by the 2-fold rotation followed by the  inversion.
In the absence of SOC, these operators satisfy $\hat{P}^2=\hat{C}_2^2=1$ and $[\hat{P},\hat{C}_2]=0$.
There are two types of high symmetry momenta: (i) the eight TRIMs which are invariant under both $P$ and the $C_2$, (ii) the four high symmetry lines $(k_1,k_2,k_3)$ ($k_1,k_3=0,\pi$, $k_2\neq 0,\pi$) which are invariant under only $C_2$.
The TRIMs have four one-dimensional irreducible representations, which have the $C_2$ and $P$ eigenvalues $(1,1)$, $(1,-1)$, $(-1,1)$, and $(-1,-1)$, respectively.
Thus the corresponding entries in symmetry data are given by $n^{\mathbf{K}}_{\pm,\pm}$, representing the number of states having the $C_2$ and $P$ eigenvalues $\pm1,\pm1$ at $\mathbf{K}$.
On the other hand, the $C_2$-invariant lines have only two kinds of one-dimensional irreducible representations, i.e., the one having the $C_2$ eigenvalue 1 and the one having the $C_2$ eigenvalue -1.
Thus the corresponding entries in symmetry data are given by $n^{\mathbf{K}}_{\pm}$,  where $\pm$ represents the rotation eigenvalue $\pm1$.

Now we turn to the compatibility relations.
The first kind of compatibility relations simply require the same occupation numbers at high symmetry momenta.
The second kind require that the $C_2$ eigenvalues of the occupied bands remain invariant along the $C_2$-invariant lines.
On one hand, the second kind relations make $n_{\pm}^{\mathbf{K}}$ ($\mathbf{K}\notin \mathrm{TRIM}$) a constant along a $C_2$-invariant line.
On the other hand, viewing TRIMs as particular points in the $C_2$-invariant lines,  these relations require $n_{\pm,+}^{\mathbf{K}}+n_{\pm,-}^{\mathbf{K}} = n_{\pm}^{\mathbf{K}+(0,k_2,0)}$ for $\mathbf{K}\in \mathrm{TRIM}$ and  $k_2\neq0,\pi$, implying that the symmetry data on the $C_2$-invariant lines are completely determined by the symmetry data on the TRIMs.
Therefore in the following we will keep only the symmetry data on the eight TRIMs and denote the eight TRIMs as  $\mathbf{K}_{i=1\cdots 8}$.
The two kinds of compatibility relations are then given by
\begin{equation}
\sum_{\zeta=\pm1,\chi=\pm1} n^{\mathbf{K}_i}_{\zeta,\chi} - n^{\mathbf{K}_j}_{\zeta,\chi}=0,\qquad(i,j=1\cdots 8),
\end{equation}
and
\begin{equation}
\sum_{\chi=\pm1} n^{\mathbf{K}_i}_{1,\chi} - n^{\mathbf{K}_i+(0,\pi,0)}_{1,\chi}=0,\qquad(i=1\cdots 8),
\end{equation}
respectively.       
We tabulate the 15 generators of the solutions in table \ref{tab:P2/m}.

\begin{table}
{\scriptsize
\begin{tabular}{|C{0.61cm}|L{7.1cm}|C{0.3cm}|}
\hline
\multicolumn{3}{|c|}{Generators of \{BS\}} \\
\hline
Basis & Symmetry data & $\kappa_i$ \\
\hline
$\mathbf{b}_{1}$ & $\mathbf{e}^\mathrm{A}_{+,+}  +\mathbf{e}^\mathrm{B}_{+,+}  +\mathbf{e}^\mathrm{C}_{+,+}  +\mathbf{e}^\mathrm{D}_{+,+}  +\mathbf{e}^\mathrm{E}_{+,+}  +\mathbf{e}^\mathrm{G}_{+,+}  +\mathbf{e}^\mathrm{Y}_{+,+}  +\mathbf{e}^\mathrm{Z}_{+,+}  $ & 1 \\
\hline
$\mathbf{b}_{2}$ & $\mathbf{e}^\mathrm{A}_{+,-}  +\mathbf{e}^\mathrm{B}_{+,-}  +\mathbf{e}^\mathrm{C}_{+,-}  +\mathbf{e}^\mathrm{D}_{+,-}  +\mathbf{e}^\mathrm{E}_{+,-}  +\mathbf{e}^\mathrm{G}_{+,-}  +\mathbf{e}^\mathrm{Y}_{+,-}  +\mathbf{e}^\mathrm{Z}_{+,-}  $ & 1 \\
\hline
$\mathbf{b}_{3}$ & $\mathbf{e}^\mathrm{A}_{-,+}  +\mathbf{e}^\mathrm{B}_{-,+}  +\mathbf{e}^\mathrm{C}_{-,+}  +\mathbf{e}^\mathrm{D}_{-,+}  +\mathbf{e}^\mathrm{E}_{-,+}  +\mathbf{e}^\mathrm{G}_{-,+}  +\mathbf{e}^\mathrm{Y}_{-,+}  +\mathbf{e}^\mathrm{Z}_{-,+}  $ & 1 \\
\hline
$\mathbf{b}_{4}$ & $\mathbf{e}^\mathrm{A}_{-,-}  +\mathbf{e}^\mathrm{B}_{-,-}  +\mathbf{e}^\mathrm{C}_{-,-}  +\mathbf{e}^\mathrm{D}_{-,-}  +\mathbf{e}^\mathrm{E}_{-,-}  +\mathbf{e}^\mathrm{G}_{-,-}  +\mathbf{e}^\mathrm{Y}_{-,-}  +\mathbf{e}^\mathrm{Z}_{-,-}  $ & 1 \\
\hline
$\mathbf{b}_{5}$ & $\mathbf{e}^\mathrm{C}_{+,+}  -\mathbf{e}^\mathrm{C}_{+,-}  +\mathbf{e}^\mathrm{D}_{+,+}  -\mathbf{e}^\mathrm{D}_{+,-}  +\mathbf{e}^\mathrm{E}_{+,+}  -\mathbf{e}^\mathrm{E}_{+,-}  +\mathbf{e}^\mathrm{Z}_{+,+}  -\mathbf{e}^\mathrm{Z}_{+,-}  $ & 1 \\
\hline
$\mathbf{b}_{6}$ & $\mathbf{e}^\mathrm{C}_{-,+}  -\mathbf{e}^\mathrm{C}_{-,-}  +\mathbf{e}^\mathrm{D}_{-,+}  -\mathbf{e}^\mathrm{D}_{-,-}  +\mathbf{e}^\mathrm{E}_{-,+}  -\mathbf{e}^\mathrm{E}_{-,-}  +\mathbf{e}^\mathrm{Z}_{-,+}  -\mathbf{e}^\mathrm{Z}_{-,-}  $ & 1 \\
\hline
$\mathbf{b}_{7}$ & $\mathbf{e}^\mathrm{C}_{+,-}  -\mathbf{e}^\mathrm{C}_{-,-}  -\mathbf{e}^\mathrm{D}_{+,+}  +\mathbf{e}^\mathrm{D}_{+,-}  -\mathbf{e}^\mathrm{E}_{+,+}  +\mathbf{e}^\mathrm{E}_{+,-}  +\mathbf{e}^\mathrm{G}_{+,+}  -\mathbf{e}^\mathrm{G}_{-,-}  +\mathbf{e}^\mathrm{Y}_{+,+}  -\mathbf{e}^\mathrm{Y}_{-,-}  +\mathbf{e}^\mathrm{Z}_{+,-}  -\mathbf{e}^\mathrm{Z}_{-,-}  $ & 1 \\
\hline
$\mathbf{b}_{8}$ & $\mathbf{e}^\mathrm{D}_{+,+}  -\mathbf{e}^\mathrm{D}_{+,-}  +\mathbf{e}^\mathrm{D}_{-,+}  -\mathbf{e}^\mathrm{D}_{-,-}  +\mathbf{e}^\mathrm{E}_{+,+}  -\mathbf{e}^\mathrm{E}_{+,-}  +\mathbf{e}^\mathrm{E}_{-,+}  -\mathbf{e}^\mathrm{E}_{-,-}  -\mathbf{e}^\mathrm{G}_{+,+}  +\mathbf{e}^\mathrm{G}_{+,-}  -\mathbf{e}^\mathrm{G}_{-,+}  +\mathbf{e}^\mathrm{G}_{-,-}  -\mathbf{e}^\mathrm{Y}_{+,+}  +\mathbf{e}^\mathrm{Y}_{+,-}  -\mathbf{e}^\mathrm{Y}_{-,+}  +\mathbf{e}^\mathrm{Y}_{-,-}  $ & 1 \\
\hline
$\mathbf{b}_{9}$ & $\mathbf{e}^\mathrm{B}_{+,+}  -\mathbf{e}^\mathrm{B}_{-,-}  +\mathbf{e}^\mathrm{D}_{+,-}  -\mathbf{e}^\mathrm{D}_{-,+}  -\mathbf{e}^\mathrm{E}_{+,+}  +\mathbf{e}^\mathrm{E}_{+,-}  -\mathbf{e}^\mathrm{E}_{-,+}  +\mathbf{e}^\mathrm{E}_{-,-}  +2\mathbf{e}^\mathrm{G}_{+,+}  -\mathbf{e}^\mathrm{G}_{+,-}  +\mathbf{e}^\mathrm{G}_{-,+}  -2\mathbf{e}^\mathrm{G}_{-,-}  +\mathbf{e}^\mathrm{Y}_{+,+}  -\mathbf{e}^\mathrm{Y}_{+,-}  +\mathbf{e}^\mathrm{Y}_{-,+}  -\mathbf{e}^\mathrm{Y}_{-,-}  +\mathbf{e}^\mathrm{Z}_{+,+}  -\mathbf{e}^\mathrm{Z}_{-,-}  $ & 1 \\
\hline
$\mathbf{b}_{10}$ & $-\mathbf{e}^\mathrm{B}_{+,-}  +\mathbf{e}^\mathrm{B}_{-,+}  -\mathbf{e}^\mathrm{D}_{+,-}  +\mathbf{e}^\mathrm{D}_{-,+}  -\mathbf{e}^\mathrm{G}_{+,-}  +\mathbf{e}^\mathrm{G}_{-,+}  -\mathbf{e}^\mathrm{Z}_{+,-}  +\mathbf{e}^\mathrm{Z}_{-,+}  $ & 1 \\
\hline
$\mathbf{b}_{11}$ & $\mathbf{e}^\mathrm{E}_{+,+}  -\mathbf{e}^\mathrm{E}_{+,-}  +\mathbf{e}^\mathrm{E}_{-,+}  -\mathbf{e}^\mathrm{E}_{-,-}  -\mathbf{e}^\mathrm{G}_{+,+}  +\mathbf{e}^\mathrm{G}_{+,-}  -\mathbf{e}^\mathrm{G}_{-,+}  +\mathbf{e}^\mathrm{G}_{-,-}  -\mathbf{e}^\mathrm{Y}_{+,+}  +\mathbf{e}^\mathrm{Y}_{+,-}  -\mathbf{e}^\mathrm{Y}_{-,+}  +\mathbf{e}^\mathrm{Y}_{-,-}  -\mathbf{e}^\mathrm{Z}_{+,+}  +\mathbf{e}^\mathrm{Z}_{+,-}  -\mathbf{e}^\mathrm{Z}_{-,+}  +\mathbf{e}^\mathrm{Z}_{-,-}  $ & 1 \\
\hline
$\mathbf{b}_{12}$ & $\mathbf{e}^\mathrm{G}_{+,+}  -\mathbf{e}^\mathrm{G}_{+,-}  +\mathbf{e}^\mathrm{G}_{-,+}  -\mathbf{e}^\mathrm{G}_{-,-}  +\mathbf{e}^\mathrm{Y}_{+,+}  -\mathbf{e}^\mathrm{Y}_{+,-}  +\mathbf{e}^\mathrm{Y}_{-,+}  -\mathbf{e}^\mathrm{Y}_{-,-}  $ & 1 \\
\hline
$\mathbf{b}_{13}$ & $\mathbf{e}^\mathrm{G}_{+,-}  - \mathbf{e}^\mathrm{G}_{-,+} +\mathbf{e}^\mathrm{Z}_{+,+}  -\mathbf{e}^\mathrm{Z}_{-,-}$ & 2 \\
\hline
$\mathbf{b}_{14}$ & $\mathbf{e}^\mathrm{G}_{-,-}  - \mathbf{e}^\mathrm{G}_{+,+} +\mathbf{e}^\mathrm{Z}_{-,+}  -\mathbf{e}^\mathrm{Z}_{+,-}$ & 2 \\
\hline
$\mathbf{b}_{15}$ & $\mathbf{e}^\mathrm{G}_{+,+}  -\mathbf{e}^\mathrm{G}_{+,-}  +\mathbf{e}^\mathrm{G}_{-,+}  -\mathbf{e}^\mathrm{G}_{-,-}  $ & 2 \\
\hline
\end{tabular}
\begin{tabular}{|C{1.4cm}|L{6.805cm}|}
\hline
\multicolumn{2}{|c|}{Generators of \{AI\}} \\
\hline
Basis & Symmetry data \\
\hline
$\mathbf{a}_{1}$($G^{1a}_{+,+}$) & $\mathbf{e}^\mathrm{A}_{+,+}  +\mathbf{e}^\mathrm{B}_{+,+}  +\mathbf{e}^\mathrm{C}_{+,+}  +\mathbf{e}^\mathrm{D}_{+,+}  +\mathbf{e}^\mathrm{E}_{+,+}  +\mathbf{e}^\mathrm{G}_{+,+}  +\mathbf{e}^\mathrm{Y}_{+,+}  +\mathbf{e}^\mathrm{Z}_{+,+}  $ \\
\hline
$\mathbf{a}_{2}$($G^{1a}_{+,-}$) & $\mathbf{e}^\mathrm{A}_{+,-}  +\mathbf{e}^\mathrm{B}_{+,-}  +\mathbf{e}^\mathrm{C}_{+,-}  +\mathbf{e}^\mathrm{D}_{+,-}  +\mathbf{e}^\mathrm{E}_{+,-}  +\mathbf{e}^\mathrm{G}_{+,-}  +\mathbf{e}^\mathrm{Y}_{+,-}  +\mathbf{e}^\mathrm{Z}_{+,-}  $ \\
\hline
$\mathbf{a}_{3}$($G^{1a}_{-,+}$) & $\mathbf{e}^\mathrm{A}_{-,+}  +\mathbf{e}^\mathrm{B}_{-,+}  +\mathbf{e}^\mathrm{C}_{-,+}  +\mathbf{e}^\mathrm{D}_{-,+}  +\mathbf{e}^\mathrm{E}_{-,+}  +\mathbf{e}^\mathrm{G}_{-,+}  +\mathbf{e}^\mathrm{Y}_{-,+}  +\mathbf{e}^\mathrm{Z}_{-,+}  $ \\
\hline
$\mathbf{a}_{4}$($G^{1a}_{-,-}$) & $\mathbf{e}^\mathrm{A}_{-,-}  +\mathbf{e}^\mathrm{B}_{-,-}  +\mathbf{e}^\mathrm{C}_{-,-}  +\mathbf{e}^\mathrm{D}_{-,-}  +\mathbf{e}^\mathrm{E}_{-,-}  +\mathbf{e}^\mathrm{G}_{-,-}  +\mathbf{e}^\mathrm{Y}_{-,-}  +\mathbf{e}^\mathrm{Z}_{-,-}  $ \\
\hline
$\mathbf{a}_{5}$($G^{1b}_{+,+}$) & $\mathbf{e}^\mathrm{A}_{+,+}  +\mathbf{e}^\mathrm{B}_{+,+}  +\mathbf{e}^\mathrm{C}_{+,-}  +\mathbf{e}^\mathrm{D}_{+,-}  +\mathbf{e}^\mathrm{E}_{+,-}  +\mathbf{e}^\mathrm{G}_{+,+}  +\mathbf{e}^\mathrm{Y}_{+,+}  +\mathbf{e}^\mathrm{Z}_{+,-}  $ \\
\hline
$\mathbf{a}_{6}$($G^{1b}_{-,+}$) & $\mathbf{e}^\mathrm{A}_{-,+}  +\mathbf{e}^\mathrm{B}_{-,+}  +\mathbf{e}^\mathrm{C}_{-,-}  +\mathbf{e}^\mathrm{D}_{-,-}  +\mathbf{e}^\mathrm{E}_{-,-}  +\mathbf{e}^\mathrm{G}_{-,+}  +\mathbf{e}^\mathrm{Y}_{-,+}  +\mathbf{e}^\mathrm{Z}_{-,-}  $ \\
\hline
$\mathbf{a}_{7}$($G^{1c}_{+,+}$) & $\mathbf{e}^\mathrm{A}_{-,-}  +\mathbf{e}^\mathrm{B}_{-,-}  +\mathbf{e}^\mathrm{C}_{+,+}  +\mathbf{e}^\mathrm{D}_{-,-}  +\mathbf{e}^\mathrm{E}_{-,-}  +\mathbf{e}^\mathrm{G}_{+,+}  +\mathbf{e}^\mathrm{Y}_{+,+}  +\mathbf{e}^\mathrm{Z}_{+,+}  $ \\
\hline
$\mathbf{a}_{8}$($G^{1c}_{+,-}$) & $\mathbf{e}^\mathrm{A}_{-,+}  +\mathbf{e}^\mathrm{B}_{-,+}  +\mathbf{e}^\mathrm{C}_{+,-}  +\mathbf{e}^\mathrm{D}_{-,+}  +\mathbf{e}^\mathrm{E}_{-,+}  +\mathbf{e}^\mathrm{G}_{+,-}  +\mathbf{e}^\mathrm{Y}_{+,-}  +\mathbf{e}^\mathrm{Z}_{+,-}  $ \\
\hline
$\mathbf{a}_{9}$($G^{1d}_{+,+}$) & $\mathbf{e}^\mathrm{A}_{-,-}  +\mathbf{e}^\mathrm{B}_{+,+}  +\mathbf{e}^\mathrm{C}_{-,-}  +\mathbf{e}^\mathrm{D}_{+,+}  +\mathbf{e}^\mathrm{E}_{-,-}  +\mathbf{e}^\mathrm{G}_{+,+}  +\mathbf{e}^\mathrm{Y}_{-,-}  +\mathbf{e}^\mathrm{Z}_{+,+}  $ \\
\hline
$\mathbf{a}_{10}$($G^{1d}_{+,-}$) & $\mathbf{e}^\mathrm{A}_{-,+}  +\mathbf{e}^\mathrm{B}_{+,-}  +\mathbf{e}^\mathrm{C}_{-,+}  +\mathbf{e}^\mathrm{D}_{+,-}  +\mathbf{e}^\mathrm{E}_{-,+}  +\mathbf{e}^\mathrm{G}_{+,-}  +\mathbf{e}^\mathrm{Y}_{-,+}  +\mathbf{e}^\mathrm{Z}_{+,-}  $ \\
\hline
$\mathbf{a}_{11}$($G^{1e}_{+,+}$) & $\mathbf{e}^\mathrm{A}_{-,-}  +\mathbf{e}^\mathrm{B}_{+,+}  +\mathbf{e}^\mathrm{C}_{-,+}  +\mathbf{e}^\mathrm{D}_{+,-}  +\mathbf{e}^\mathrm{E}_{-,+}  +\mathbf{e}^\mathrm{G}_{+,+}  +\mathbf{e}^\mathrm{Y}_{-,-}  +\mathbf{e}^\mathrm{Z}_{+,-}  $ \\
\hline
$\mathbf{a}_{12}$($G^{1f}_{+,+}$) & $\mathbf{e}^\mathrm{A}_{-,-}  +\mathbf{e}^\mathrm{B}_{-,-}  +\mathbf{e}^\mathrm{C}_{+,-}  +\mathbf{e}^\mathrm{D}_{-,+}  +\mathbf{e}^\mathrm{E}_{-,+}  +\mathbf{e}^\mathrm{G}_{+,+}  +\mathbf{e}^\mathrm{Y}_{+,+}  +\mathbf{e}^\mathrm{Z}_{+,-}  $ \\
\hline
$\mathbf{a}_{13}$($G^{1g}_{+,+}$) & $\mathbf{e}^\mathrm{A}_{+,+}  +\mathbf{e}^\mathrm{B}_{-,-}  +\mathbf{e}^\mathrm{C}_{-,-}  +\mathbf{e}^\mathrm{D}_{-,-}  +\mathbf{e}^\mathrm{E}_{+,+}  +\mathbf{e}^\mathrm{G}_{+,+}  +\mathbf{e}^\mathrm{Y}_{-,-}  +\mathbf{e}^\mathrm{Z}_{+,+}  $ \\
\hline
$\mathbf{a}_{14}$($G^{1g}_{+,-}$) & $\mathbf{e}^\mathrm{A}_{+,-}  +\mathbf{e}^\mathrm{B}_{-,+}  +\mathbf{e}^\mathrm{C}_{-,+}  +\mathbf{e}^\mathrm{D}_{-,+}  +\mathbf{e}^\mathrm{E}_{+,-}  +\mathbf{e}^\mathrm{G}_{+,-}  +\mathbf{e}^\mathrm{Y}_{-,+}  +\mathbf{e}^\mathrm{Z}_{+,-}  $ \\
\hline
$\mathbf{a}_{15}$($G^{1h}_{+,+}$) & $\mathbf{e}^\mathrm{A}_{+,+}  +\mathbf{e}^\mathrm{B}_{-,-}  +\mathbf{e}^\mathrm{C}_{-,+}  +\mathbf{e}^\mathrm{D}_{-,+}  +\mathbf{e}^\mathrm{E}_{+,-}  +\mathbf{e}^\mathrm{G}_{+,+}  +\mathbf{e}^\mathrm{Y}_{-,-}  +\mathbf{e}^\mathrm{Z}_{+,-}  $ \\
\hline
\end{tabular}
}
\caption{\label{tab:P2/m} The generators of \{BS\} and \{AI\} of the space group \#10. Here $\mathbf{e}^\mathbf{K}_{\pm,\pm}$ represents the basis where all entries are zero except $n^\mathbf{K}_{\pm,\pm}=1$, $\kappa_i$ represents the order of $\mathbf{b}_i$, i.e., the minimal positive integer making $\kappa_i\mathbf{b}_i\in\{\mathrm{AI}\}$, and the symbol $G^{w}_{\pm,\pm}$ in parenthesis represents that the corresponding atomic symmetry data is induced from the orbital having the $C_2$ and $P$ eigenvalues $\pm1,\pm1$ at the Wyckoff position $w$. The notations for momenta and positions are defined as $\rm A(\pi 0 \pi)$, $\rm B(00\pi)$, $\rm C(\pi\pi 0)$, $\rm D(0\pi\pi)$, $\rm E(\pi\pi\pi)$, $\rm \Gamma(000)$, $\rm Y(\pi 0 0)$, $\rm Z(0\pi 0)$, and $1a(000)$, $1b(0\frac12 0)$, $1c(00\frac12)$,  $1d(\frac12 00)$, $1e(\frac12\frac12 0)$, $1f( 0\frac12\frac12)$, $1g(\frac12 0\frac12)$, $1h(\frac12\frac12\frac12)$, respectively.}
\end{table}

Generators of \{AI\} can be calculated by the same method described in Sec. \ref{sub:SI-P-1}.
Space group \#10 has eight maximal Wyckoff positions\cite{Bradlyn2017}, i.e., $(x_1,x_2,x_3)$ ($x_{1,2,3}=0,\frac12$), each of which has four kinds of atomic orbitals with the $C_2$ and $P$ eigenvalues $(1,1)$, $(1,-1)$, $(-1,1)$, and $(-1,-1)$, respectively.
By a direct Fourier transformation of these atomic orbitals we get the 15 independent atomic symmetry data, which are chosen as generators of \{AI\}, shown in table \ref{tab:P2/m}.
Then it follows that $\mathbf{b}_{i=1\cdots 12} \in \{\mathrm{AI}\}$ and $\mathbf{b}_{i=13,14,15} \notin \{\mathrm{AI}\}$, whereas $2\mathbf{b}_{i=13,14,15} \in \{\mathrm{AI}\}$, implying that the indicator group is $\mathbb{Z}_2\times\mathbb{Z}_2\times\mathbb{Z}_2$.

To find explicit formulae for the indicators, we first notice that \#2 is a subgroup of \#10 and thus we can induce the indicator formulae from \#2 by the method described in Sec. \ref{sub:SI}.
Substituting $\mathbf{b}_{i}$'s into Eq. (\ref{eq:p-1}), we find that the \#2 indicator set $(z_{2,1}z_{2,2}z_{2,3}z_{4})$ of the three nontrivial generators, $\mathbf{b}_{13,14,15}$, are $(0100)$, $(0100)$, and $(0002)$, respectively, generating a $\mathbb{Z}_2\times\mathbb{Z}_2$ group.
Therefore two of the three $\mathbb{Z}_2$ indicators are induced from $z_{2,2}$ and $z_{4}$ of space group \#2, while the left $\mathbb{Z}_2$ indicator is not induced from \#2.
Since $z_4=1,3$ are no more realizable in space group \#10, $z_4$ reduce to a $\mathbb{Z}_2$ number and we introduce $z_2^\prime=z_4/2$ to represent this $\mathbb{Z}_2$ number.
To find the left indicators, we notice that states in the $k_2=\pi$-plane can be divided into two sectors due to their mirror ($M=PC_2$) eigenvalues and thus the parities in this plane can be counted for the two sectors separately as 
\begin{equation}
\label{eq:zpm}
z^{(\pm)}_{2}\equiv\sum_{ \substack{ \mathbf{K}\in\mathrm{TRIM} \\ \mathrm{at}\;\{k_2=\pi\} }} \frac{N^{(\pm)}_-(\mathbf{K})-N^{(\pm)}_+(\mathbf{K})}{2}\;\mathrm{mod}\;2,
\end{equation}
where $N^{(\pm)}_{+}(\mathbf{K})$ is the number of states having parity $+1$ in the $M=\pm1$ sector, and $N^{(\pm)}_{-}(\mathbf{K})$ is the number of states having parity $-1$ in the $M=\pm1$ sector. 
Apparently, sum of $z_{2}^{(+)}$ and $z_{2}^{(-)}$ just gives $z_{2,2}$ (mod 2), whereas $z_2^{(+)}$ and $z_2^{(-)}$ themself are new indicators, either of which realizes a successful mapping from \{BS\} to the missing $\mathbb{Z}_2$ indicator.
Therefore, we choose the independent indicators of space group \#10 as $z_{2}^{(+)}$,$z_2^{(-)}$, $z_2^\prime$ such that the nontrivial generators, $\mathbf{b}_{13,14,15}$, have the indicators $(100)$, $(010)$, and $(001)$, respectively.

\section{Centrosymmetric space groups}
\label{sec:centro}
In the absence of spin-orbital coupling, the time-reversal operator, $\hat{T}$, satisfies $\hat{T}^2=+1$.
Hamiltonians having the $\hat{T}^2=+1$ symmetry belong to the orthogonal Wigner-Dyson class \cite{Wigner1955}, also known as class AI in the new Altland-Zirnbauer system \cite{Altland1997}.
When inversion symmetry, $\hat{P}$, is also present, the composite symmetry $PT$ satisfies $(\hat{P}\hat{T})^2=+1$. 
In the presence of $PT$, generic band crossings are nodal lines, which are robust against symmetric perturbations, due to the $\pi$-Berry phase associated with any loop that links with a nodal line \cite{Kim2015,Fang2015a}.

First we introduce a lemma in a 2D system with $P$ and $T$, which later we will see have quite a few useful variations.
Consider a closed path in the 2D BZ that encircles exact one half of the BZ, such that the inside and the outside of the loop are mapped to each other under inversion (see Fig. \ref{fig:BerryPhase1}(a)).
\begin{figure}
\begin{centering}
\includegraphics[width=1\linewidth]{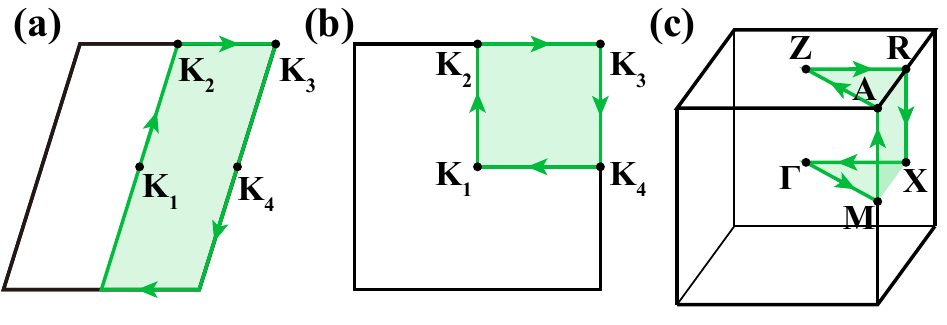}
\par\end{centering}
\protect\caption{\label{fig:BerryPhase1}The loops on which Berry phases are considered in this paper. The plane in (a) can be any 2D sub-manifold of the 3D BZ in a system having inversion and the plane in (a, b) can be any $k$-slice in a system having a twofold or fourfold axis. (a) A loop enclosing half of the BZ, passing four TRIM on its way, denoted by $\mathbf{K}_{1,2,3,4}$. (b) A loop enclosing a quarter of the BZ, passing all four TRIM. (c) A loop particularly considered in a simple tetragonal lattice for space group \#130.}
\end{figure}
Following Ref. [\onlinecite{Fang2012}], one can prove that the Berry phase of the loop is given by\\
\textit{Lemma}
\begin{equation}\label{eq:lemma}
e^{i\Phi_B}=\prod_{n\in{occ.},\mathrm{K}\in\mathrm{TRIM}}\chi_n(\mathrm{K}),
\end{equation}
where $occ.$ represents the set of valence bands and $\chi_n(\mathrm{K})$ the inversion eigenvalue of the $n$-th band at $\mathrm{K}$.
When the product of inversion eigenvalues at TRIM of all valence bands is $-1$, the Berry phase is $\pi$, so that there must be one (or an odd number of) Dirac point inside the area enclosed by the loop.
Suppose the Dirac point has momenta $\mathbf{k}_0$, then due to time-reversal there is another at $-\mathbf{k}_0$, so that there are at least two Dirac points in the 2D BZ.

This lemma helps diagnose the presence and the configuration of nodal loops in the BZ.
In a 3D system having $P$ and $T$, there are eight TRIMs that are invariant under parity.
The above lemma applies to any 2D slice in the 3D Brillouin zone as long as the slice is invariant under inversion.
It is straightforward to confirm that any such 2D slice passes exactly four out of the eight TRIM, and if for that slice $\Phi_B=\pi$, there must be at least two points where the nodal line passes through this 2D slice.
Therefore, the inversion eigenvalues at all eight TRIM indicate how many times (modulo four) a given 2D slice in the 3D BZ is passed by nodal lines.

In this section, we also need to diagnose nodal lines in the presence of other symmetries in addition to inversion, where a simple extension of the lemma is incurred.
In the presence of mirror symmetry (denoted by $M$) in addition to inversion symmetry, there are special 2D planes in the BZ where $M\mathbf{k}=\mathbf{k}$.
On these mirror-invariant planes, thanks to $\hat{M}^2=+1$ and $[\hat{M},\hat{P}]=[\hat{M},\hat{T}]=0$, we can separate all bands into the $\hat{M}=+1$-sector and $\hat{M}=-1$-sector, and the bands in each sector have its respective inversion eigenvalues at the four TRIM, as well as its respective Berry phase associate with the loop in Fig. \ref{fig:BerryPhase1}(a), namely\\
\textit{Lemma: first variation}
\begin{equation}
\label{eq:lemma1}
e^{i\Phi^{(\pm)}_B}=\prod_{n\in{occ.},\mathrm{K}\in\mathrm{TRIM}}\chi^\pm_n(\mathrm{K}),
\end{equation}
If both $\Phi_B^{(+)}$ and $\Phi_B^{(-)}$ are $\pi$, then there are two Dirac points in the $+1$-sector and two in the $-1$ sector.
The Dirac points in opposite sectors may appear at the same momenta via fine tuning, but cannot pairwise annihilate each other as their hybridization is disallowed by mirror symmetry.

We now apply the lemma and its variations (more to be introduced later) to study the band topology of each centrosymmetric space group that has a nontrivial indicator group.
These 41 space groups \cite{Po2017} are further divided into six classes according to the presence of rotation and/or screw axes.
Within each class, we first write down the explicit expressions of all indicators, followed by analyzing the band crossings for all combinations of nonzero indicators.
The minimal possible configurations of nodal lines for each nonzero set of indicators for each of the 41 space groups are tabulated in table \ref{tab:1}, where ``minimal'' means that we have tried to minimize the number of lines among all possible configurations.

\LTcapwidth=0.97\textwidth
\begin{longtable*}{C{\figwids}|C{\figwids}|C{\figwids}|C{\figwid}|C{\figwid}|c|c|c|cc}
\caption{\label{tab:1}Minimal configurations of nodal lines in centrosymmetric space groups, given any nonzero set of symmetry-based indicators. First column is the class of the space groups by which the text is broken into subsections. Second column is the indicators we choose for the generators of the indicator group. Third column is for the numbers of space groups. The fourth column gives the Brillouin zone. The reciprocal lattice ($\mathbf{g}_{i=1,2,3}$) setting and convention for Brillouin zone follow Ref. \cite{aroyo_BZ_2014}. The fifth column contains all possible combinations of nonzero indicators and their corresponding configurations, where blue lines represent nodal rings without $\mathbb{Z}_2$-monopole charge, yellow lines represent nodal rings with $\mathbb{Z}_2$-monopole charge, gray planes are mirror planes at which nodal rings are centered and red lines the rotation or screw axes surrounded by one or two nodal rings.  Note here that if two rings or lines are related by a reciprocal vector, only one of them will be drawn.}\\
\hline
\hline
\hline
{\textbf{Symm.}} & {\textbf{Indicator}} & {\textbf{SGs}} & {\textbf{BZ}} & \multicolumn{ 4 }{c}{\textbf{Configurations}} \\
\hline
\endfirsthead
\multicolumn{10}{c} {{\bfseries \tablename\ \thetable{} -- continued}}\\
\hline
{\textbf{Symm.}} & {\textbf{Indicator}} & {\textbf{SGs}} & {\textbf{BZ}} & \multicolumn{ 4 }{c}{\textbf{Configurations}} \\
\hline
\endhead
\hline \multicolumn{10}{r}{{Continued on next page}} \\
\endfoot
\hline \hline
\endlastfoot
\hline
%
% P-1 =========================================================================
%
\multirow{2}{\figwids}{Inversion only} & \multirow{2}{*}{$z_{2,i=123}$, $z_{4}$} & \multirow{2}{*}{2} & \textit{aP} &
 0001, 0003 & 1001, 1003 & 0002 & 1000, 1002 & 1100, 1102 \\%1110, 1112\\
\hliness
& & &   \raisebox{-0.9\height}{\includegraphics[width=\figwid]{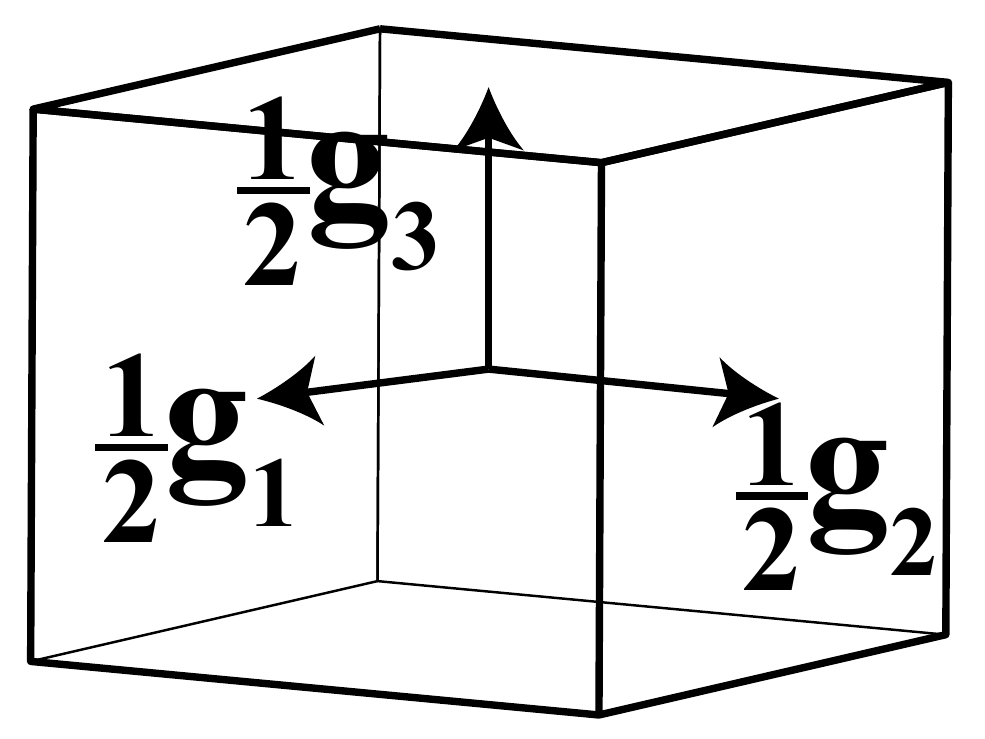}} & 
        \raisebox{-0.9\height}{\includegraphics[width=\figwid]{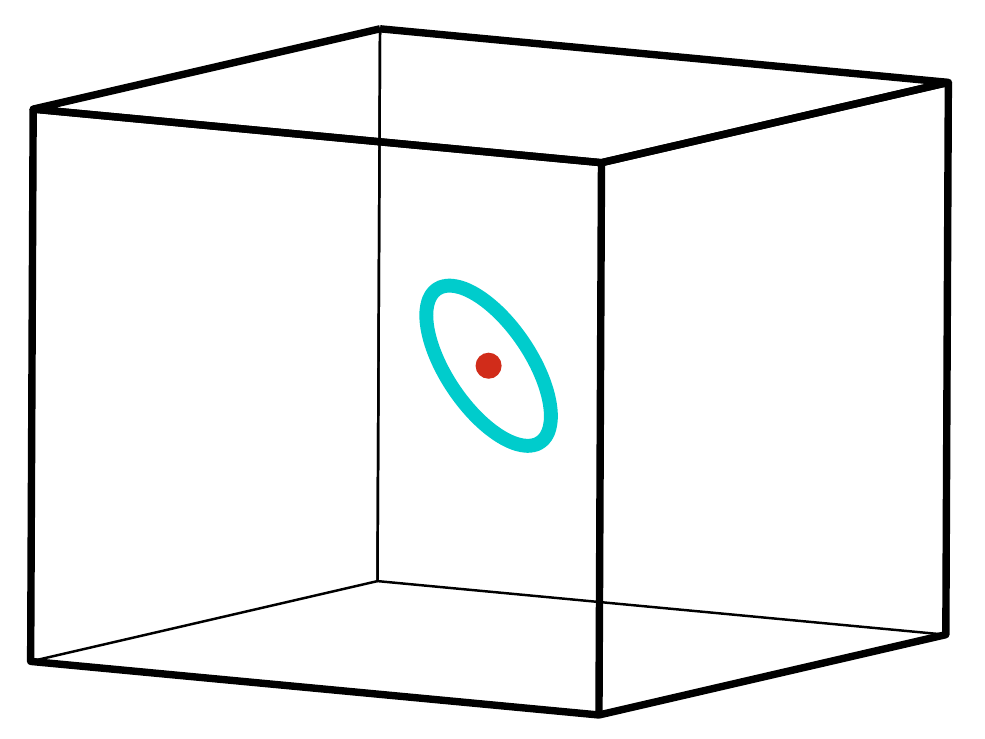}} &
        \raisebox{-0.9\height}{\includegraphics[width=\figwid]{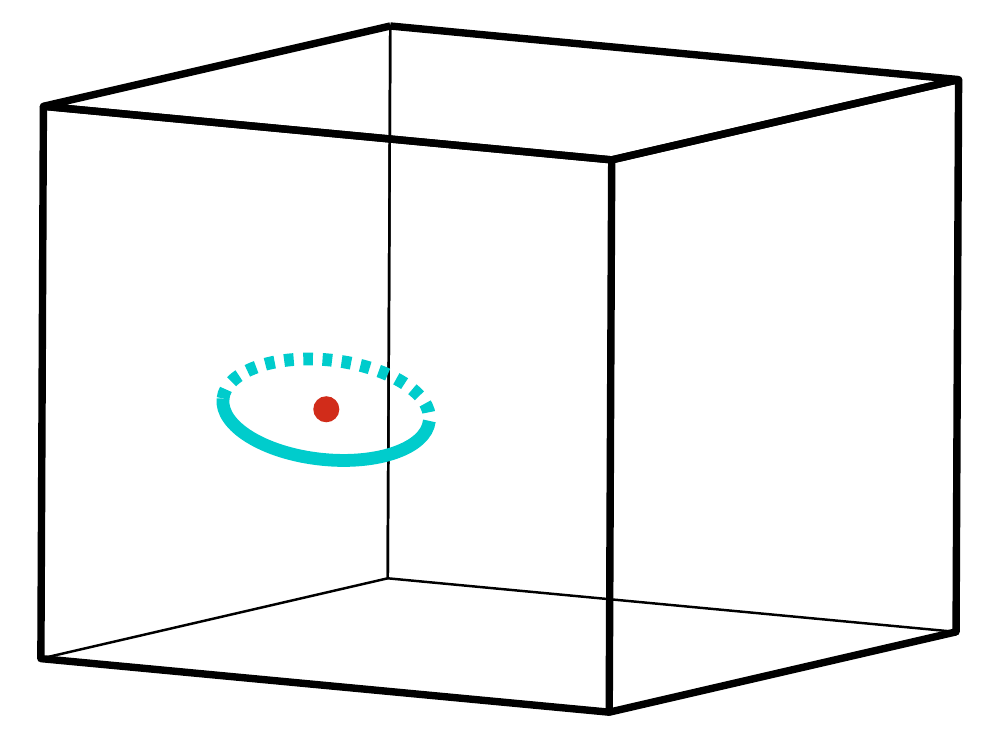}} &
        \raisebox{-0.9\height}{\includegraphics[width=\figwid]{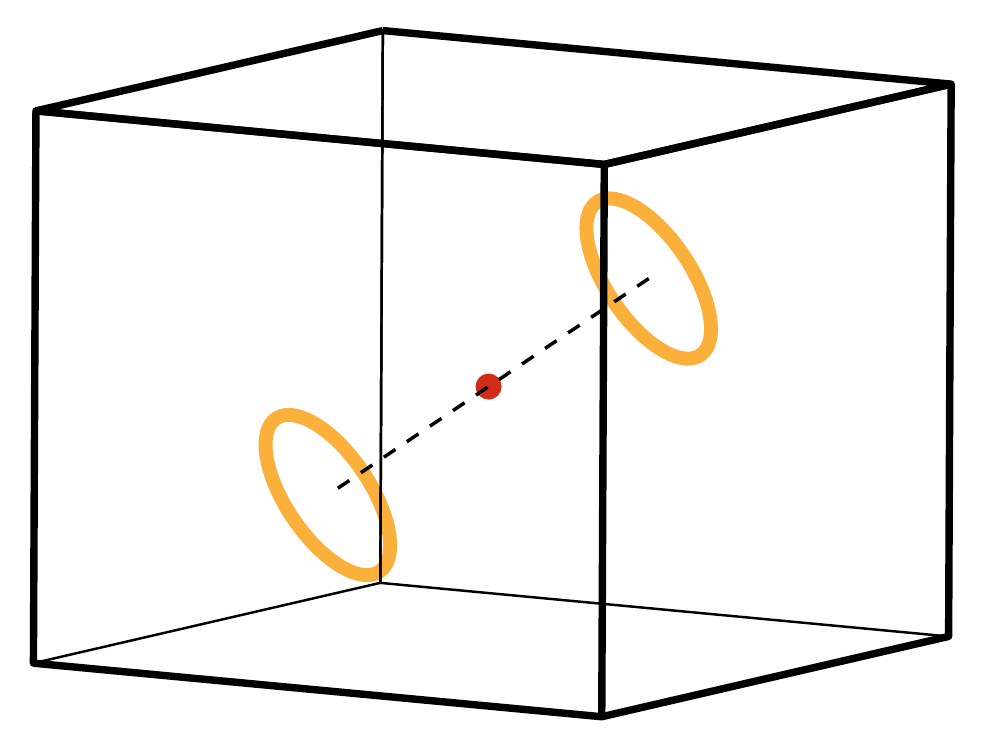}} &
        \raisebox{-0.9\height}{\includegraphics[width=\figwid]{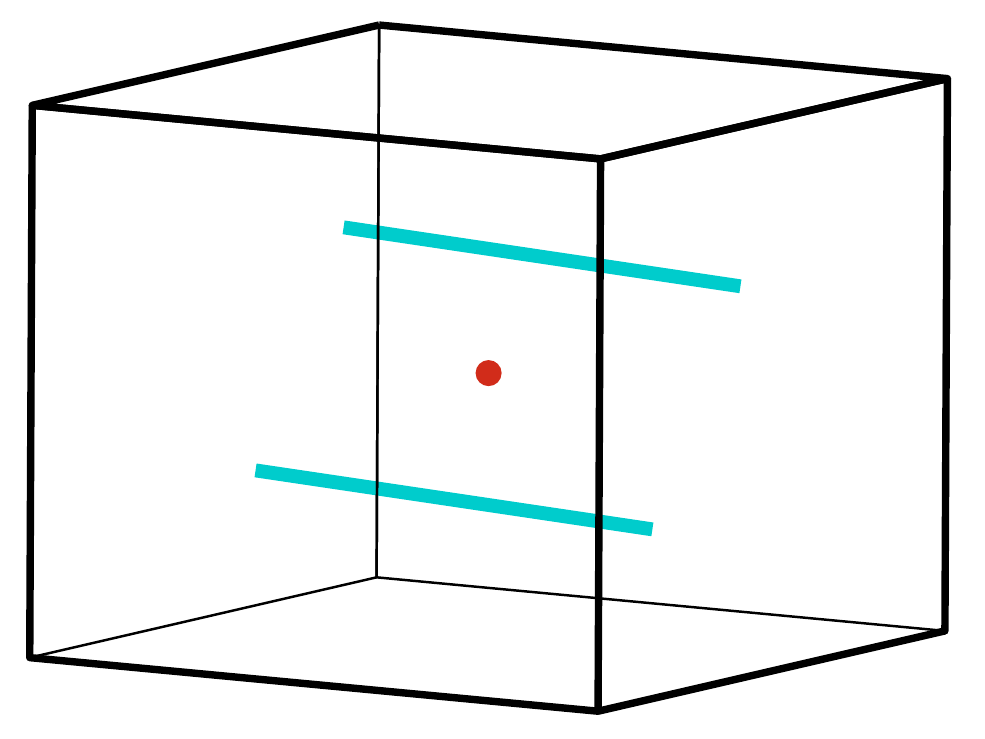}} &
        \raisebox{-0.9\height}{\includegraphics[width=\figwid]{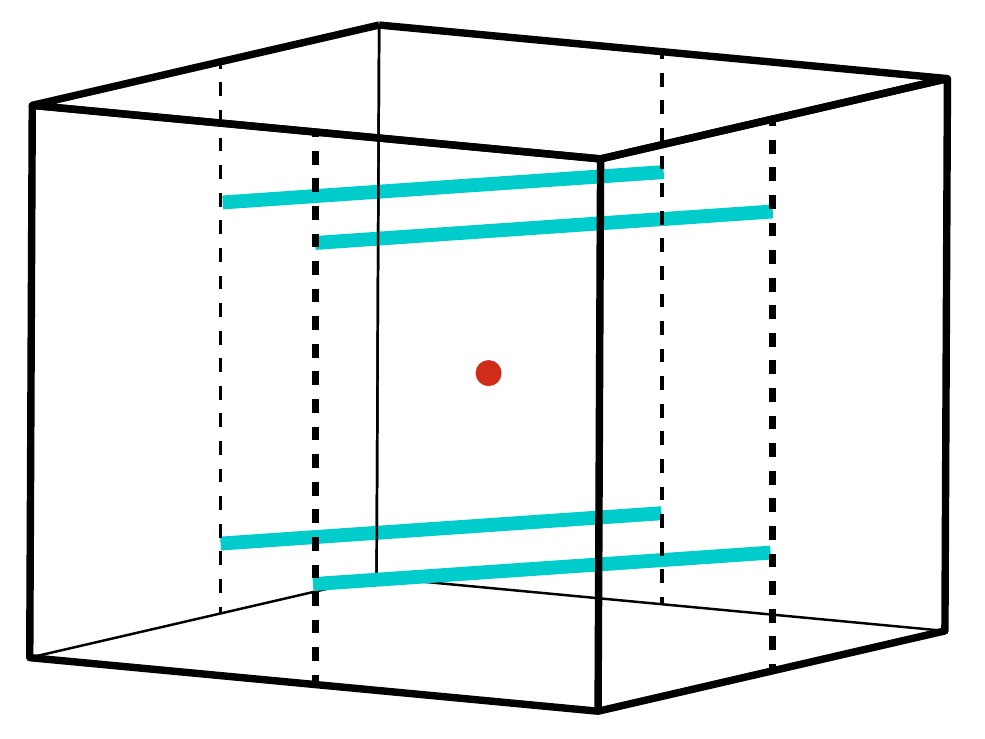}} \\
        %\raisebox{-0.9\height}{\includegraphics[width=\figwid]{2_111.pdf}} \\
\hline
%%
%% C2 ==========================================================================
%%
\multirow{16}{\figwids}{Inversion plus twofold rotation or screw} & \multirow{2}{*}{$z_2^\prime$} & \multirow{2}{*}{11} & \textit{mP} & 
 1 \\
\hliness
& & & \raisebox{-0.9\height}{\includegraphics[width=\figwid]{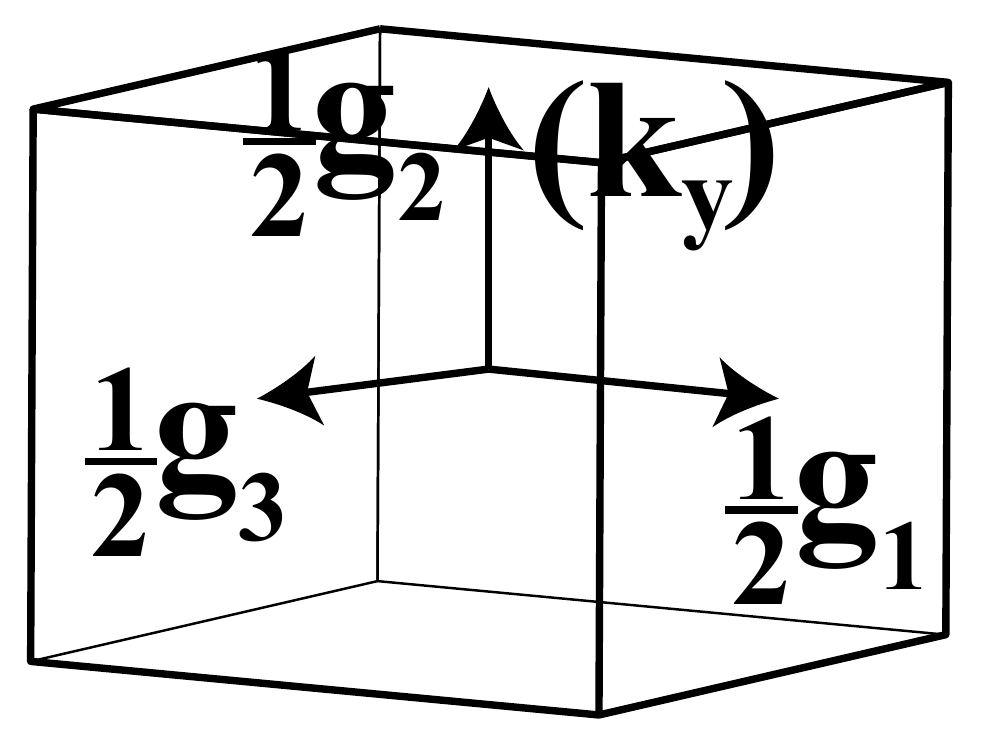}} & 
      \raisebox{-0.9\height}{\includegraphics[width=\figwid]{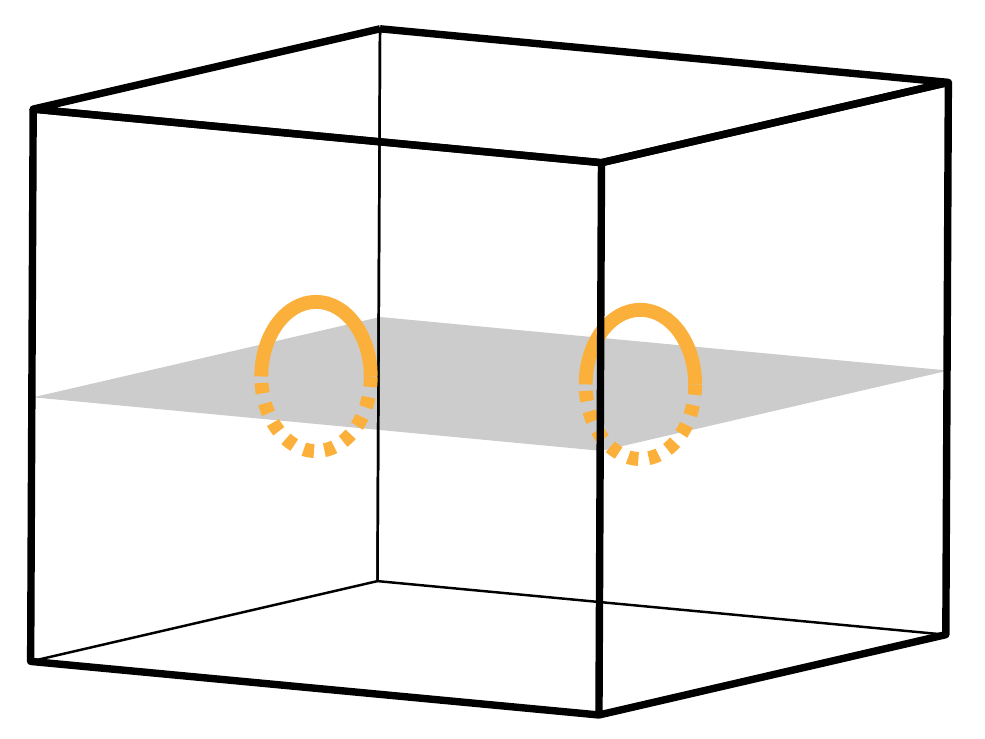}} \\
\hlines
& \multirow{2}{*}{$z_2^\prime$} & \multirow{2}{*}{14} & \textit{mP} &
 1 \\
\hliness
& & & \raisebox{-0.9\height}{\includegraphics[width=\figwid]{BZ_mP.pdf}} & 
      \raisebox{-0.9\height}{\includegraphics[width=\figwid]{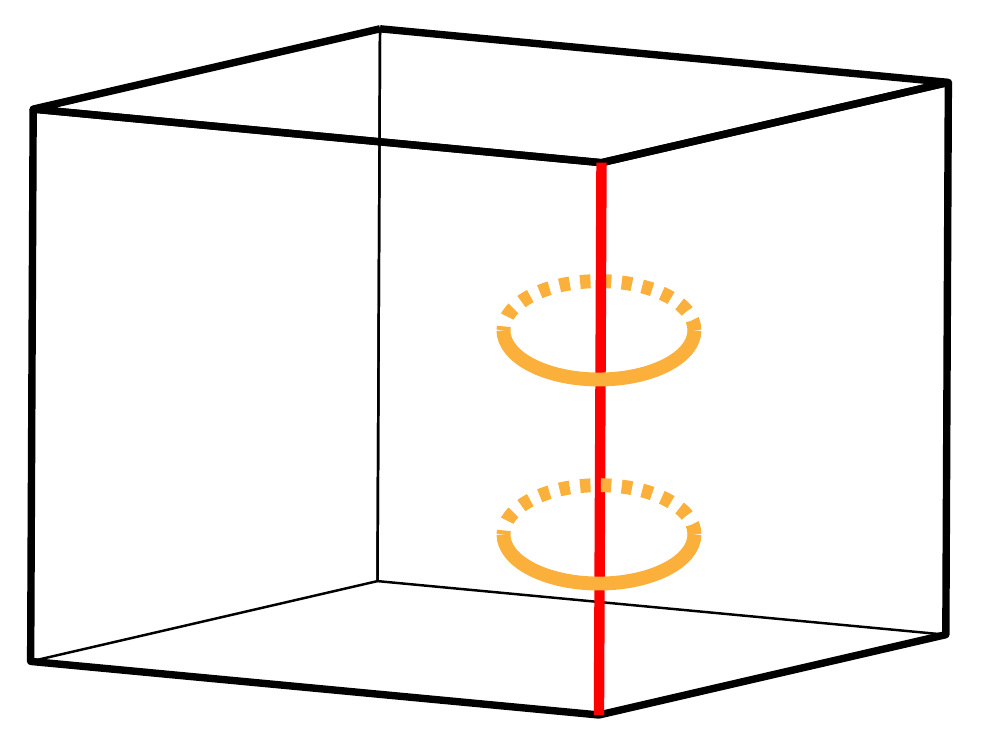}} \\
\hlines
& \multirow{2}{*}{$z_2^\prime$} & \multirow{2}{\figwids}{48, 49, 50, 52, 53, 54, 56, 58, 60} & \textit{oP} &
 \multicolumn{2}{c|}{1} \\
\hliness
& & & \raisebox{-0.9\height}{\includegraphics[width=\figwid]{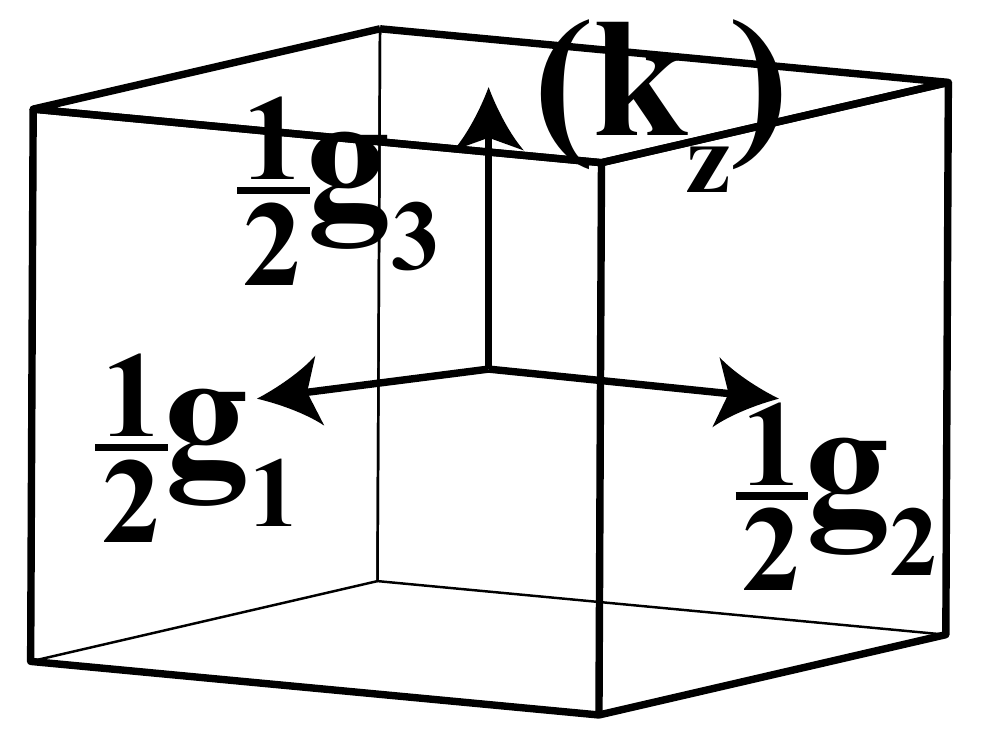}} & \multicolumn{2}{c|}{\raisebox{-0.9\height}{\includegraphics[width=\figwid]{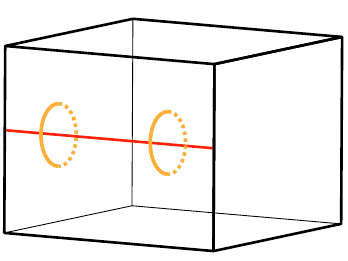}} or \raisebox{-0.9\height}{\includegraphics[width=\figwid]{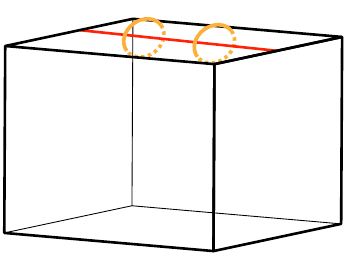}}} \\
\hlines
& \multirow{2}{*}{$z_2^\prime$} & \multirow{2}{*}{66, 68} & \textit{oC} &
 1 \\
\hliness
& & & \raisebox{-0.9\height}{\includegraphics[width=\figwid]{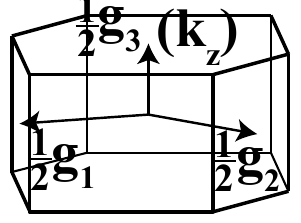}} & \raisebox{-0.9\height}{\includegraphics[width=\figwid]{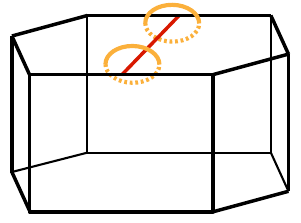}} \\
\hlines
& \multirow{2}{*}{$z_2^\prime$} & \multirow{2}{*}{70} & \textit{oF} &
 1 \\
\hliness
& & & \raisebox{-0.9\height}{\includegraphics[width=\figwid]{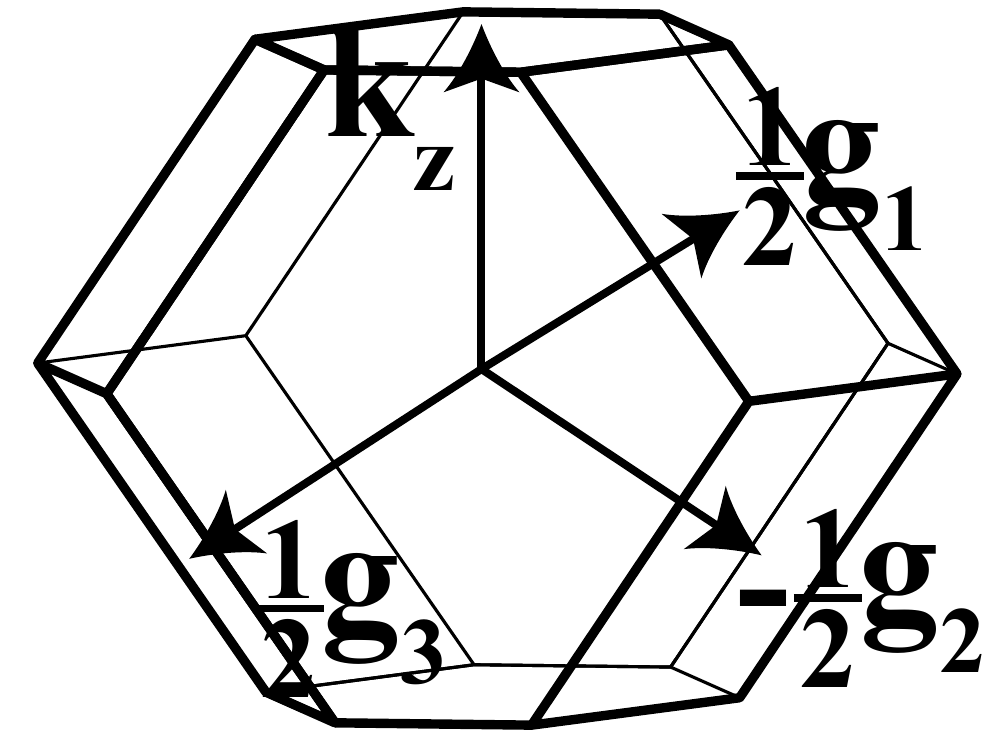}} & 
      \raisebox{-0.9\height}{\includegraphics[width=\figwid]{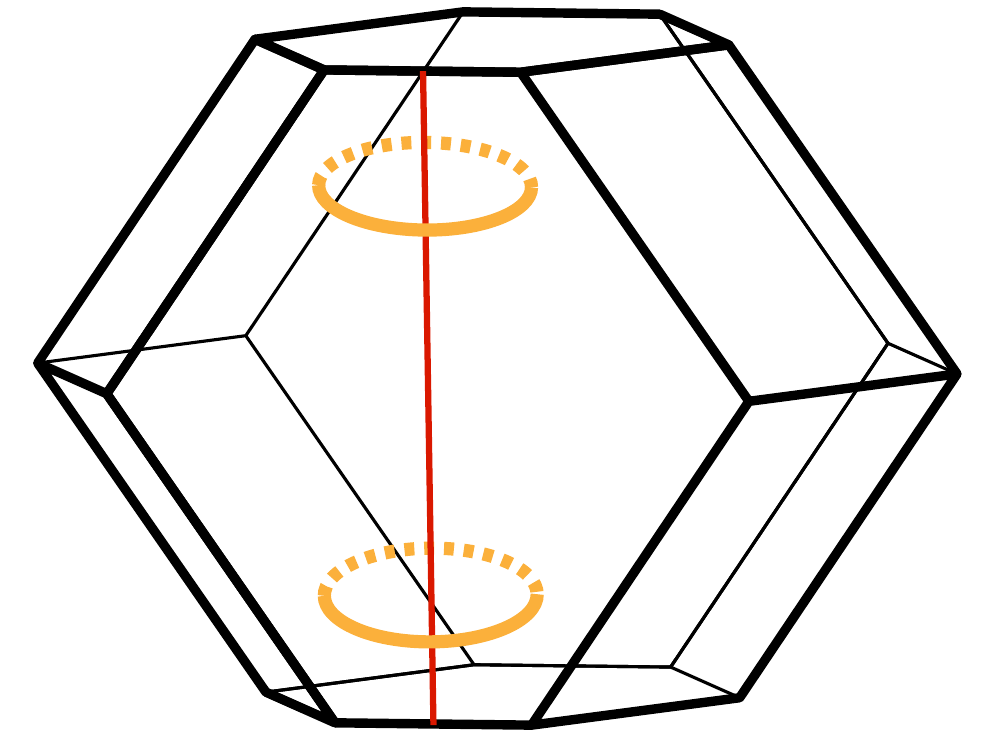}} \\
\hlines
& \multirow{2}{*}{$z_{2,2}$, $z_2^\prime$} & \multirow{2}{*}{12} & \textit{mC} &
 01 & 10, 11 \\
\hliness
& &   & \raisebox{-0.9\height}{\includegraphics[width=\figwid]{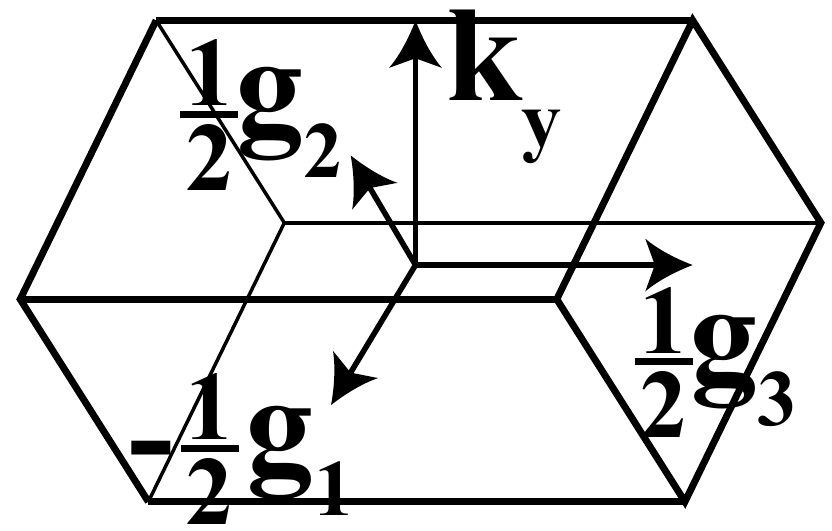}} &
        \raisebox{-0.9\height}{\includegraphics[width=\figwid]{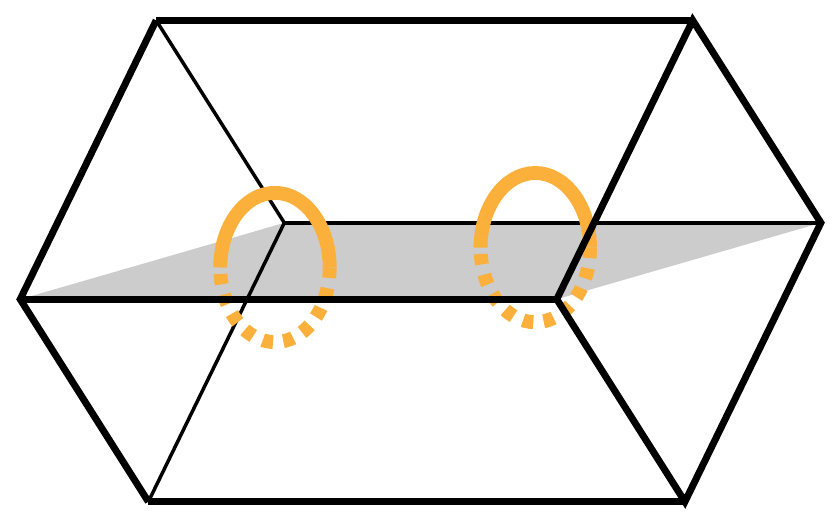}} &
        \raisebox{-0.9\height}{\includegraphics[width=\figwid]{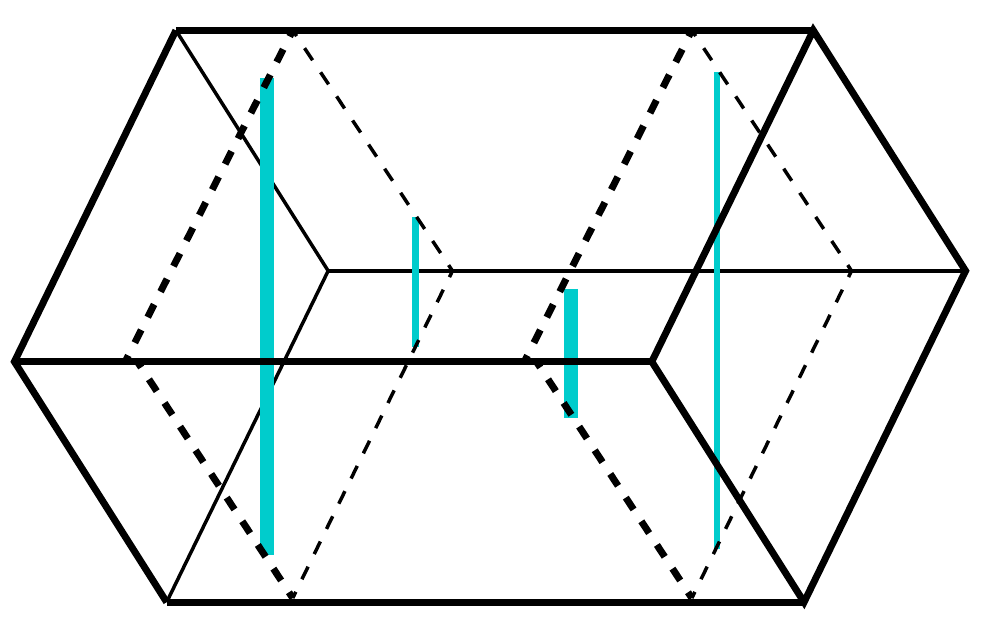}} \\
\hlines
& \multirow{2}{*}{$z_{2,2}$, $z_2^\prime$} & \multirow{2}{*}{13} & \textit{mP} &
 01 & 10, 11 \\
\hliness
& & &  \raisebox{-0.9\height}{\includegraphics[width=\figwid]{BZ_mP.pdf}} & 
       \raisebox{-0.9\height}{\includegraphics[width=\figwid]{14.pdf}} &
       \raisebox{-0.9\height}{\includegraphics[width=\figwid]{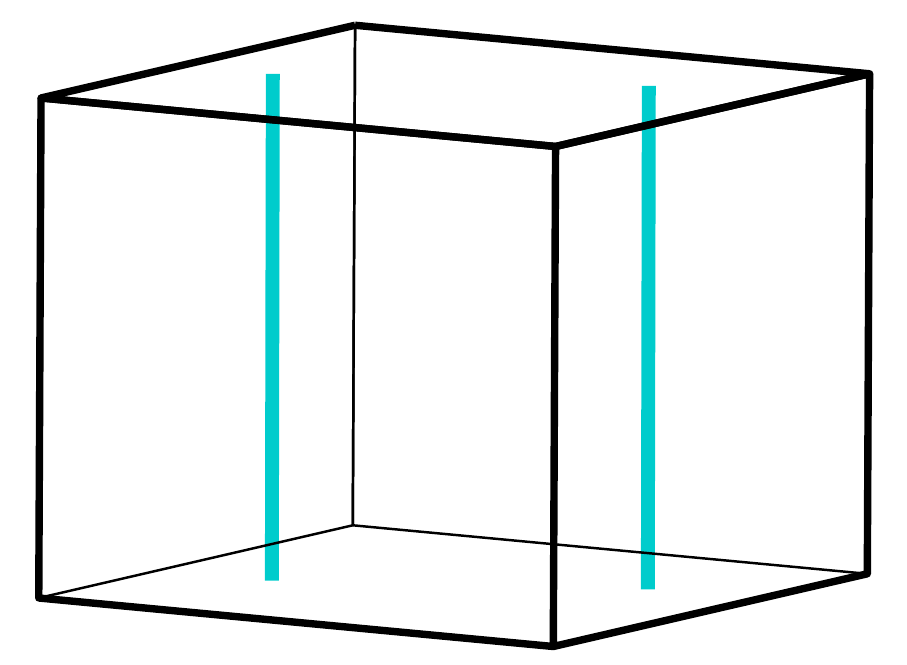}} \\
\hlines
& \multirow{2}{*}{$z_{2,2}$, $z_2^\prime$} & \multirow{2}{*}{15} & \textit{mC} &
 01 & 10, 11 \\
\hliness
& & & \raisebox{-0.9\height}{\includegraphics[width=\figwid]{BZ_mC.pdf}} &
      \raisebox{-0.9\height}{\includegraphics[width=\figwid]{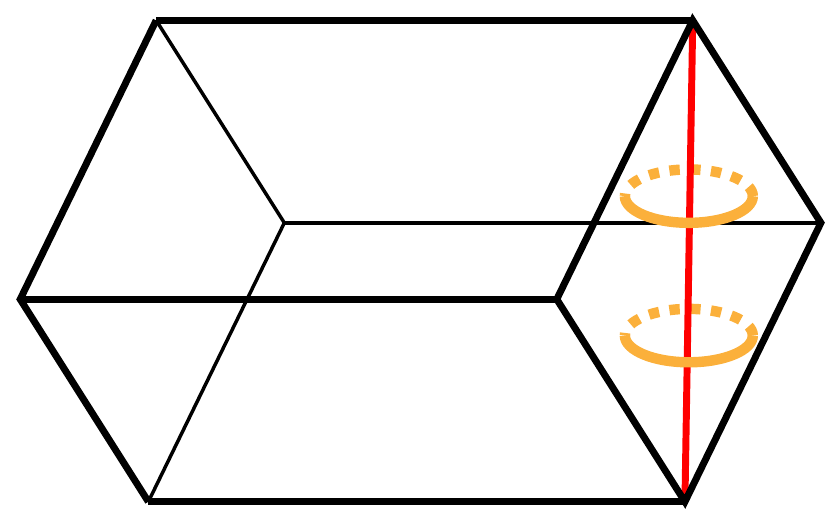}} &
      \raisebox{-0.9\height}{\includegraphics[width=\figwid]{12_11.pdf}} \\
\hlines
& \multirow{2}{\figwids}{$z_2^{(+)}$, $z_2^{(-)}$, $z_2^\prime$} & \multirow{2}{*}{10} & \textit{mP} &
 100, 010, 101, 011 & 001 & 110 & 111\\
\hliness
& & & \raisebox{-0.9\height}{\includegraphics[width=\figwid]{BZ_mP.pdf}} &
      \raisebox{-0.9\height}{\includegraphics[width=\figwid]{13_11.pdf}} &
      \raisebox{-0.9\height}{\includegraphics[width=\figwid]{11.pdf}} &
      \raisebox{-0.9\height}{\includegraphics[width=\figwid]{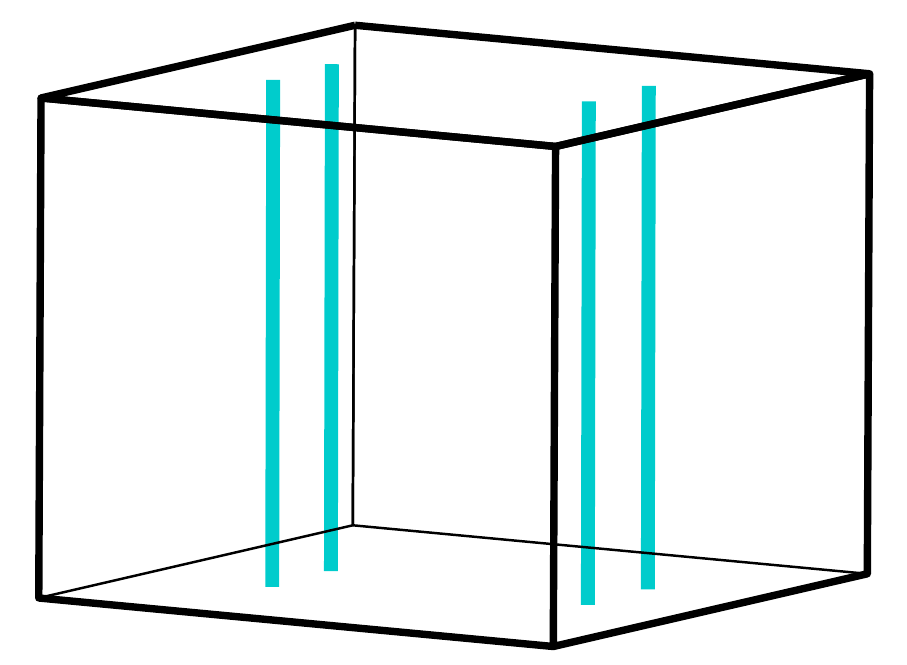}} &
      \raisebox{-0.9\height}{\includegraphics[width=\figwid]{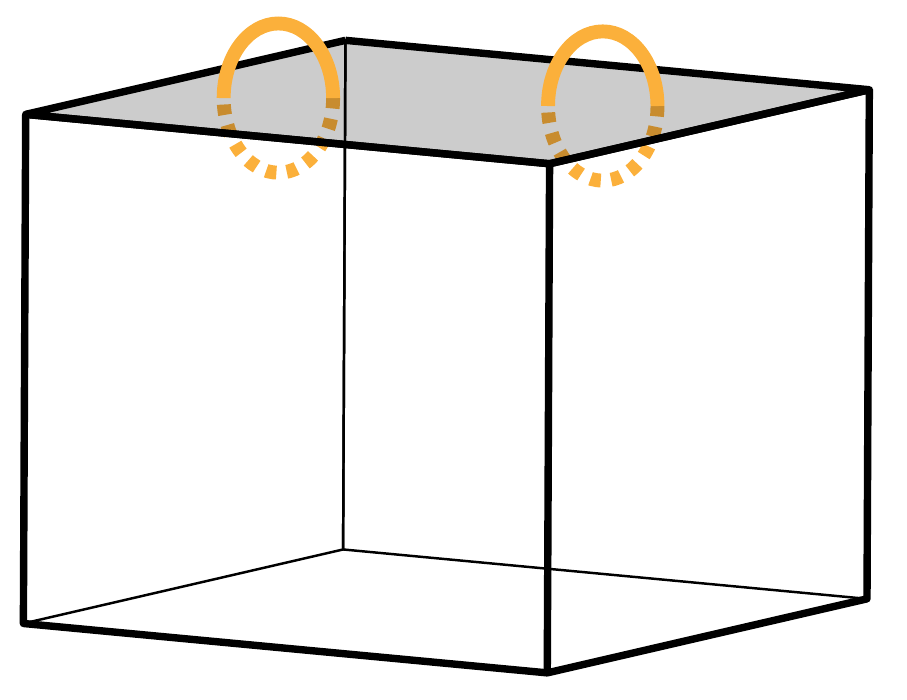}} \\
\hline
%
% C3 ==========================================================================
%
\multirow{8}{\figwids}{Inversion plus threefold rotation or screw} & \multirow{2}{\figwids}{$z_2^\prime$} & \multirow{2}{\figwids}{162, 163, 164, 165} & \textit{hP} &
 1 \\
\hliness
& & & \raisebox{-0.9\height}{\includegraphics[width=\figwid]{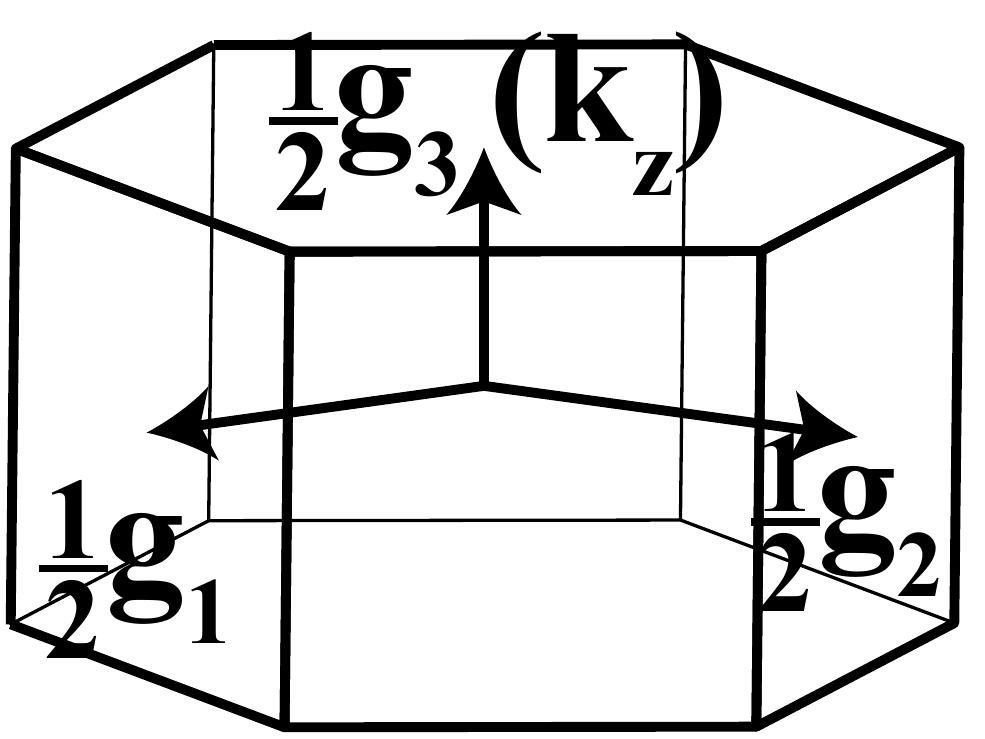}} 
    & \raisebox{-0.9\height}{\includegraphics[width=\figwid]{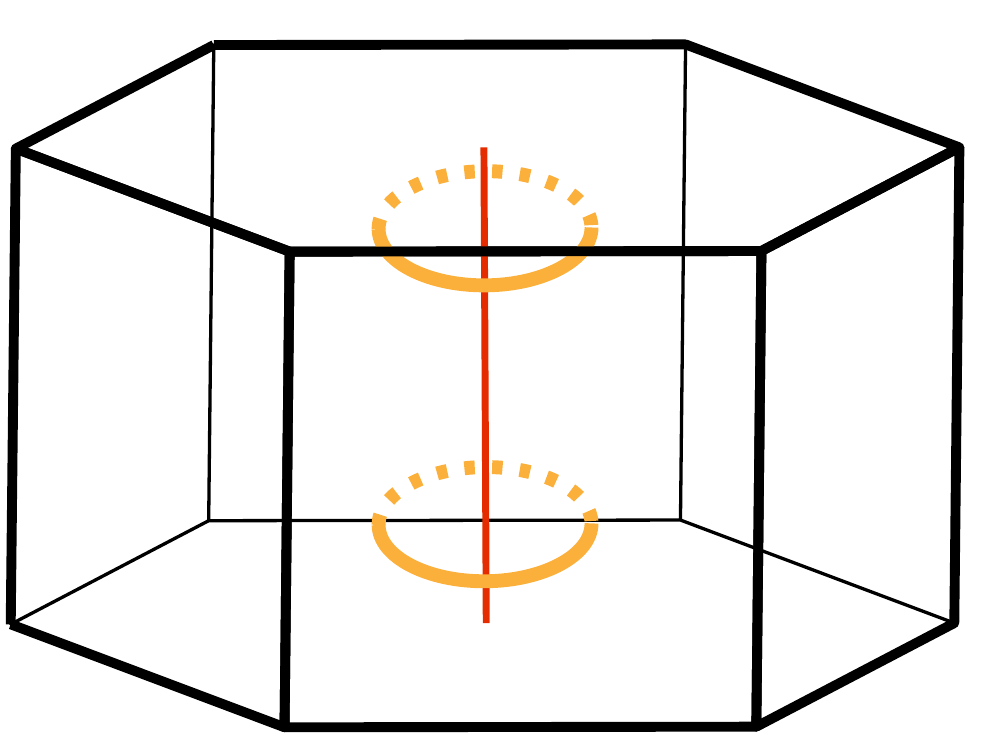}} \\
\hlines
& \multirow{2}{*}{$z_2^\prime$} & \multirow{2}{*}{166, 167} & \textit{hR} &
 1 \\
\hliness
& & & \raisebox{-0.9\height}{\includegraphics[width=\figwid]{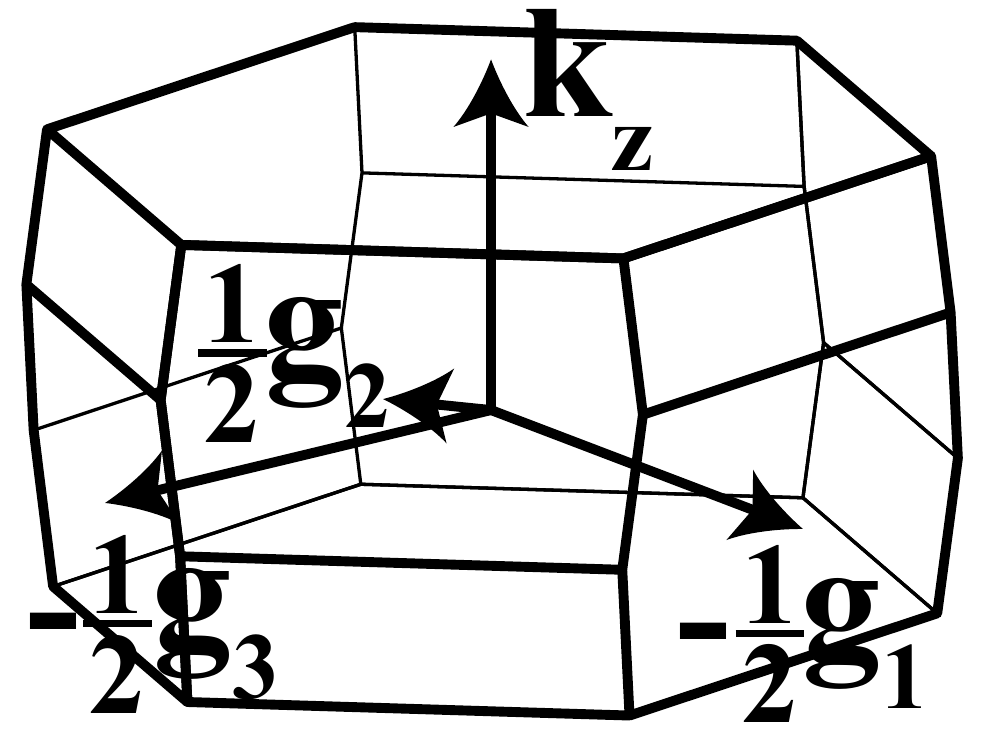}}
    & \raisebox{-0.9\height}{\includegraphics[width=\figwid]{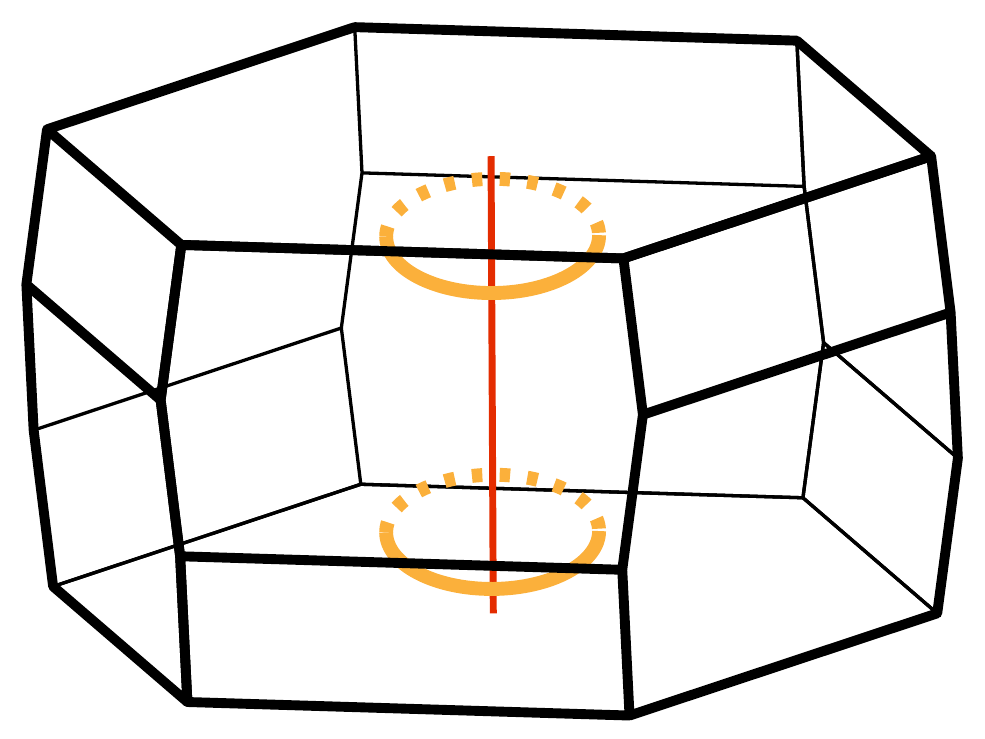}} \\
\hlines
& \multirow{2}{*}{$z_{2,3}$, $z_4$} & \multirow{2}{*}{147} & \textit{hP} &
 01, 03 &  11, 13 & 02 & 10, 12 \\
\hliness
& & & \raisebox{-0.9\height}{\includegraphics[width=\figwid]{BZ_hP.pdf}} &
      \raisebox{-0.9\height}{\includegraphics[width=\figwid]{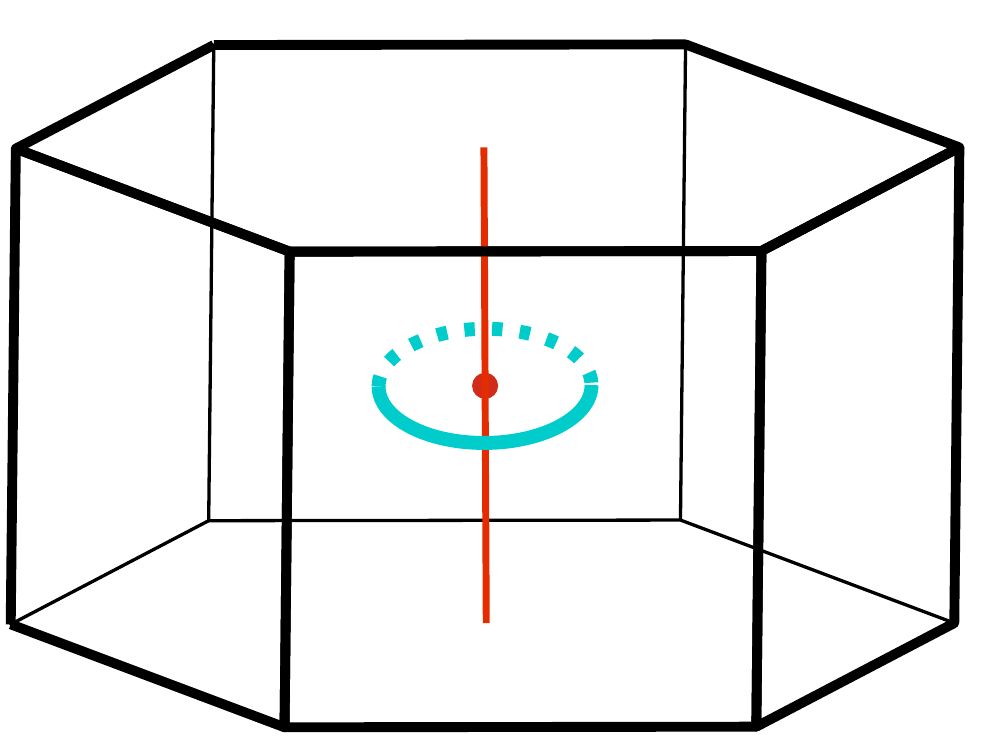}} &
      \raisebox{-0.9\height}{\includegraphics[width=\figwid]{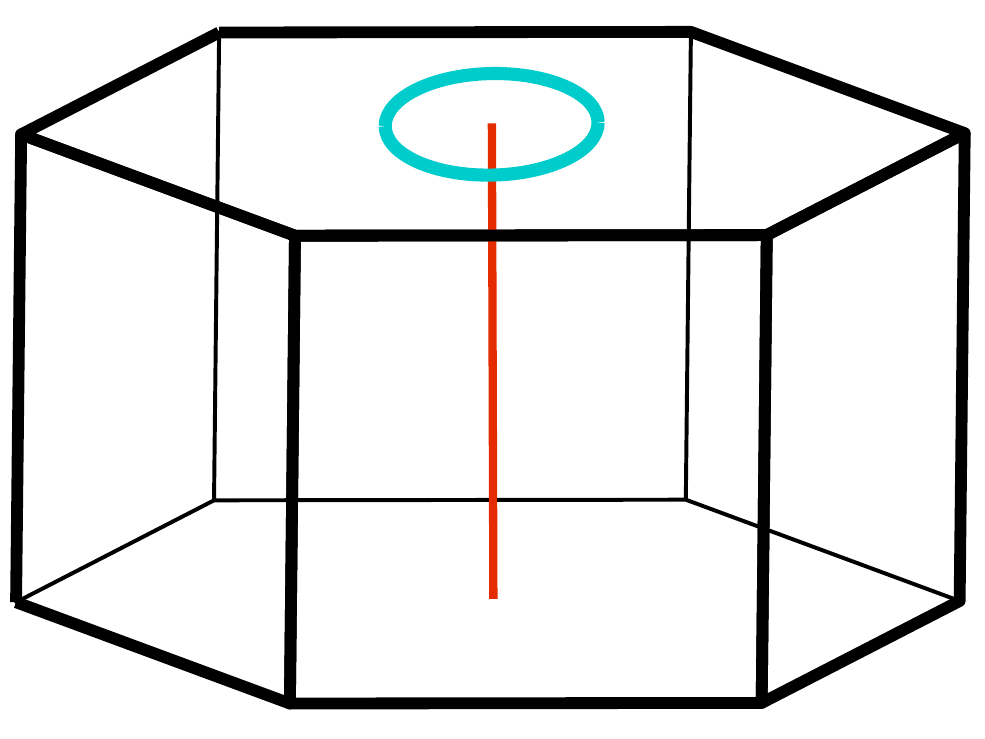}} &
      %\raisebox{-0.9\height}{\includegraphics[width=\figwid]{147_02.pdf}} &
      \raisebox{-0.9\height}{\includegraphics[width=\figwid]{162.pdf}} &
      \raisebox{-0.9\height}{\includegraphics[width=\figwid]{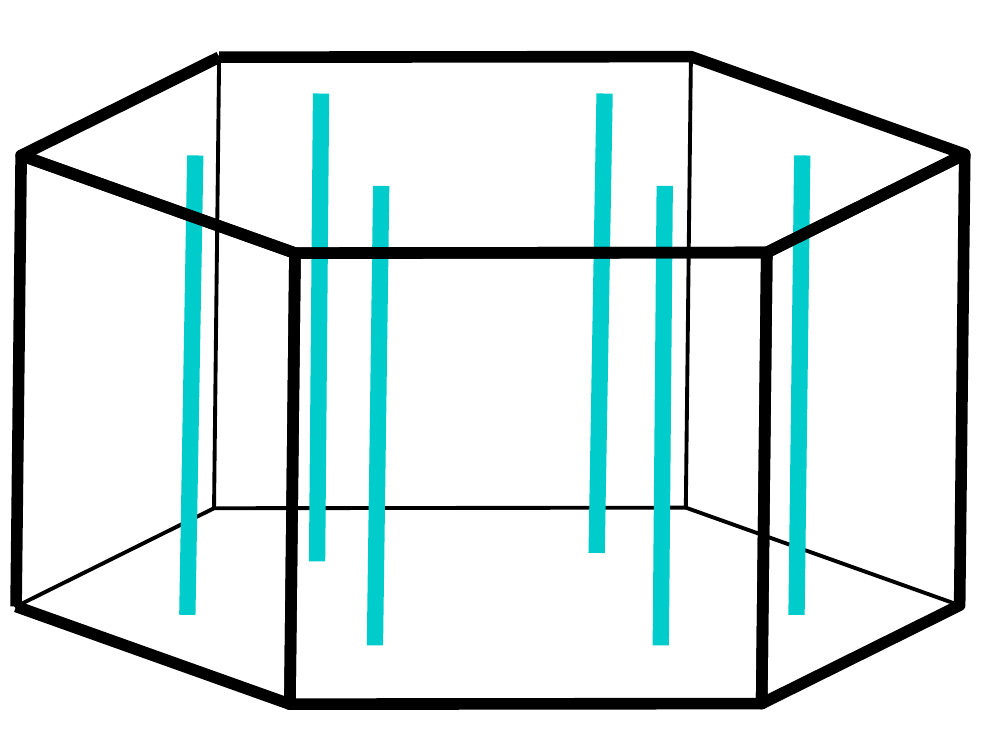}} \\
\hlines

& \multirow{2}{*}{$z_{2,3}$, $z_4$} & \multirow{2}{*}{148} & \textit{hR} & 
 01, 03 &  11, 13 & 02 & 10, 12 \\
\hliness
& & & \raisebox{-0.9\height}{\includegraphics[width=\figwid]{BZ_hR.pdf}} &
        \raisebox{-0.9\height}{\includegraphics[width=\figwid]{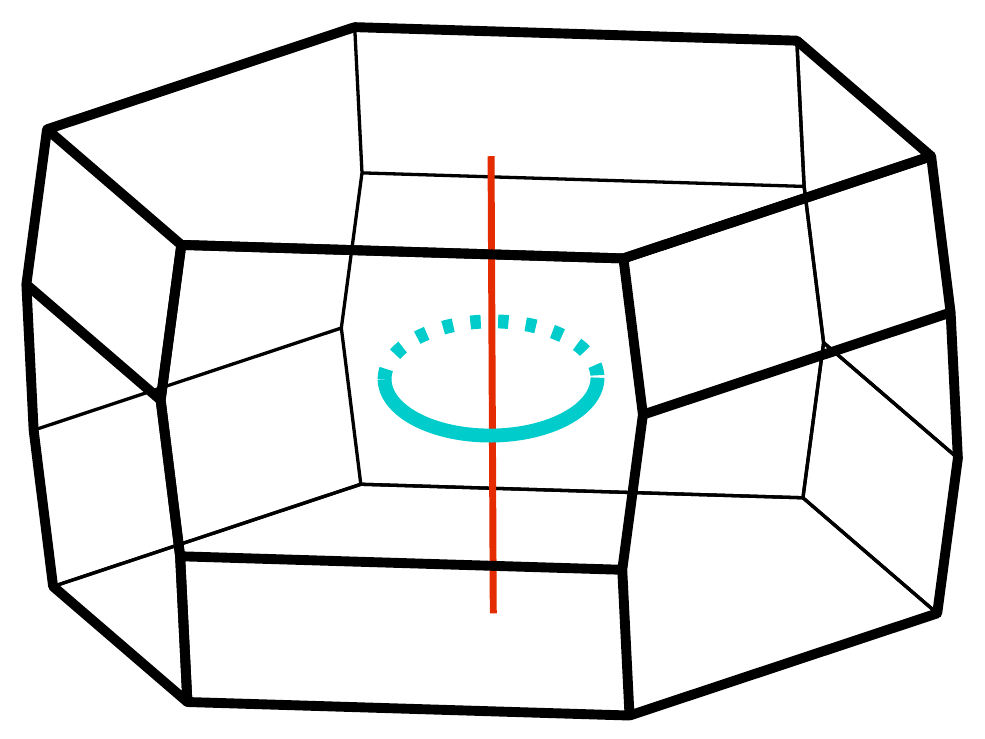}} &
        \raisebox{-0.9\height}{\includegraphics[width=\figwid]{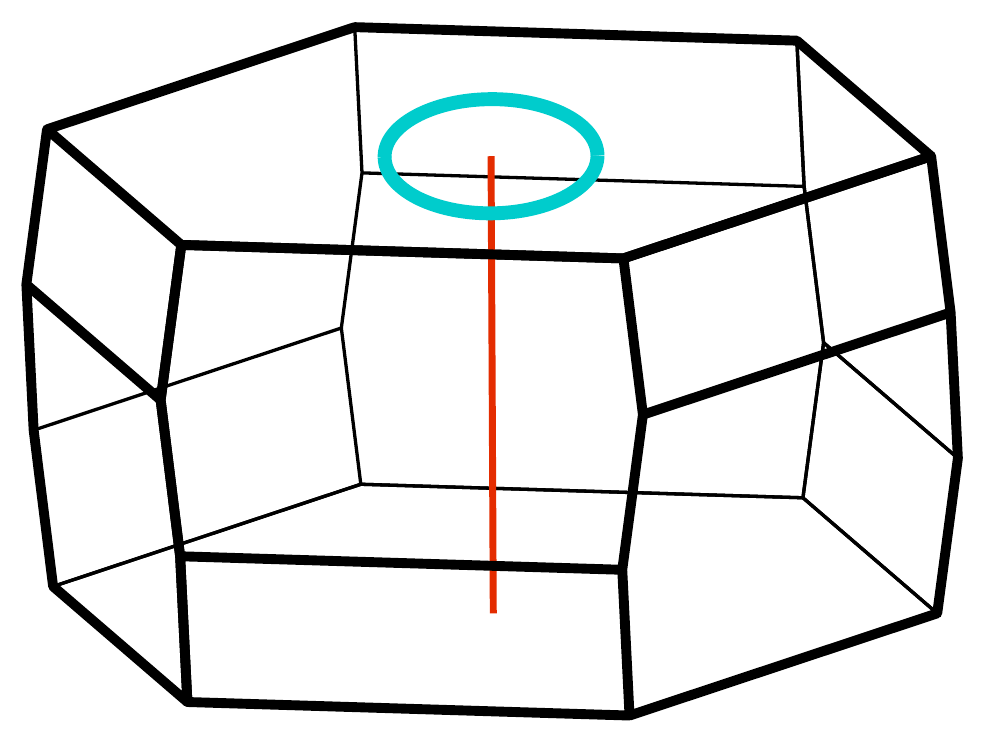}} &
        %\raisebox{-0.9\height}{\includegraphics[width=\figwid]{148_02.pdf}} &
        \raisebox{-0.9\height}{\includegraphics[width=\figwid]{166.pdf}} &
        \raisebox{-0.9\height}{\includegraphics[width=\figwid]{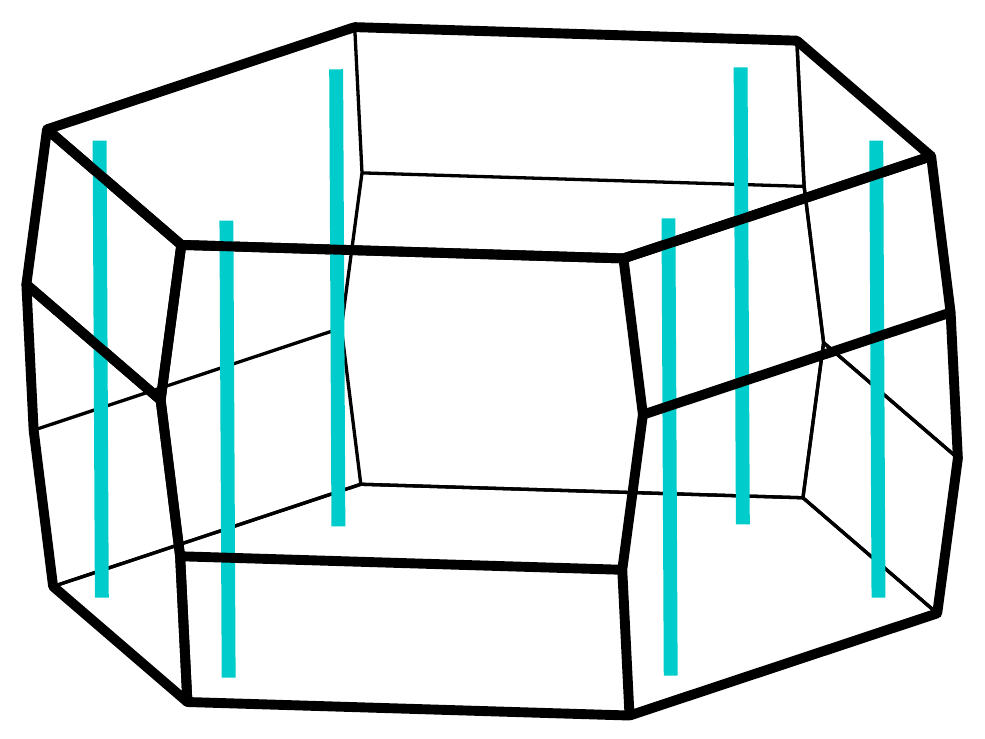}} \\
\hline
%
% C6 ==========================================================================
%
\multirow{6}{\figwids}{Inversion plus sixfold rotation or screw} & \multirow{2}{*}{$z_2^\prime$} & \multirow{2}{*}{176} & \textit{hP} &
 1 \\
\hliness
& & & \raisebox{-0.9\height}{\includegraphics[width=\figwid]{BZ_hP.pdf}} & \raisebox{-0.9\height}{\includegraphics[width=\figwid]{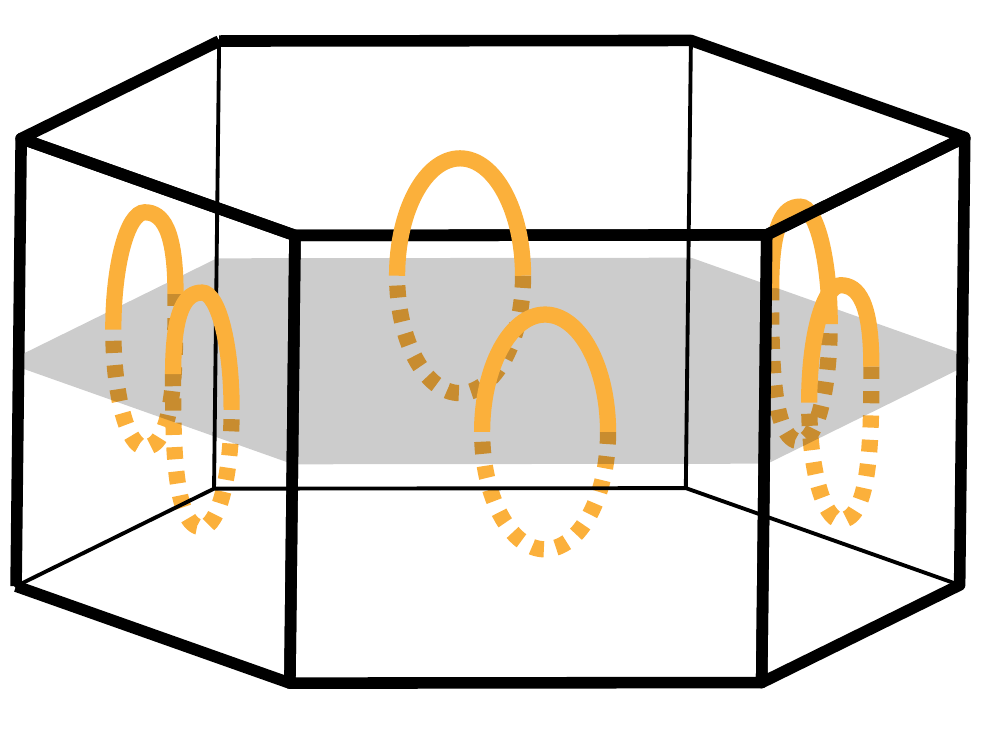}} \\
\hlines
& \multirow{2}{*}{$z_2^\prime$} & \multirow{2}{*}{192} &  \textit{hP} &
 1 \\
\hliness
& & & \raisebox{-0.9\height}{\includegraphics[width=\figwid]{BZ_hP.pdf}} & \raisebox{-0.9\height}{\includegraphics[width=\figwid]{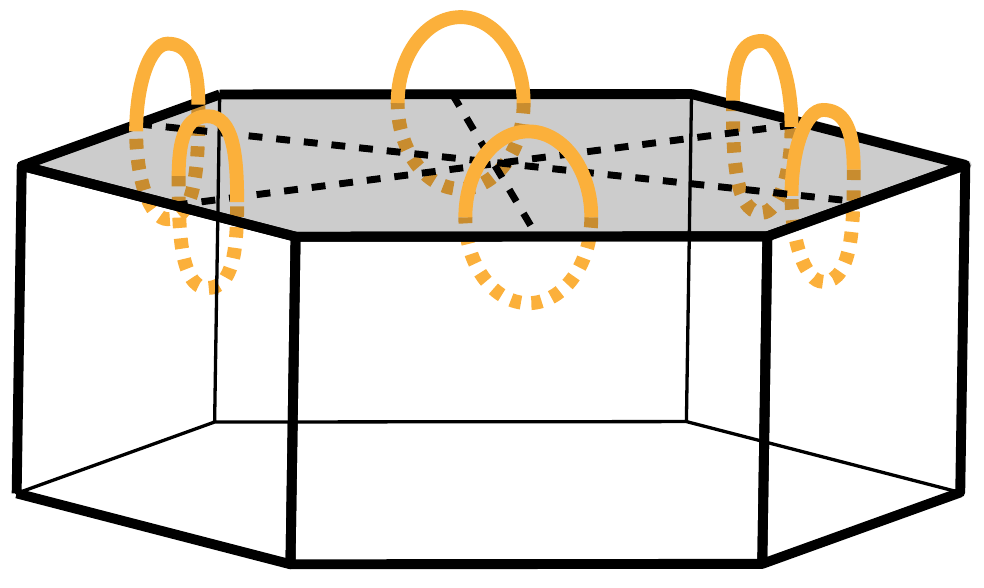}} \\
\hlines
& \multirow{2}{\figwids}{$z_2^{(+)}$, $z_2^{(-)}$, $z_2^\prime$} & \multirow{2}{*}{175} &  \textit{hP} &
 100, 010, 101, 011 & 001 & 110 & 111 \\
\hliness
& & & \raisebox{-0.9\height}{\includegraphics[width=\figwid]{BZ_hP.pdf}} & 
        \raisebox{-0.9\height}{\includegraphics[width=\figwid]{147_10.pdf}} &
        \raisebox{-0.9\height}{\includegraphics[width=\figwid]{176.pdf}} &
        \raisebox{-0.9\height}{\includegraphics[width=\figwid]{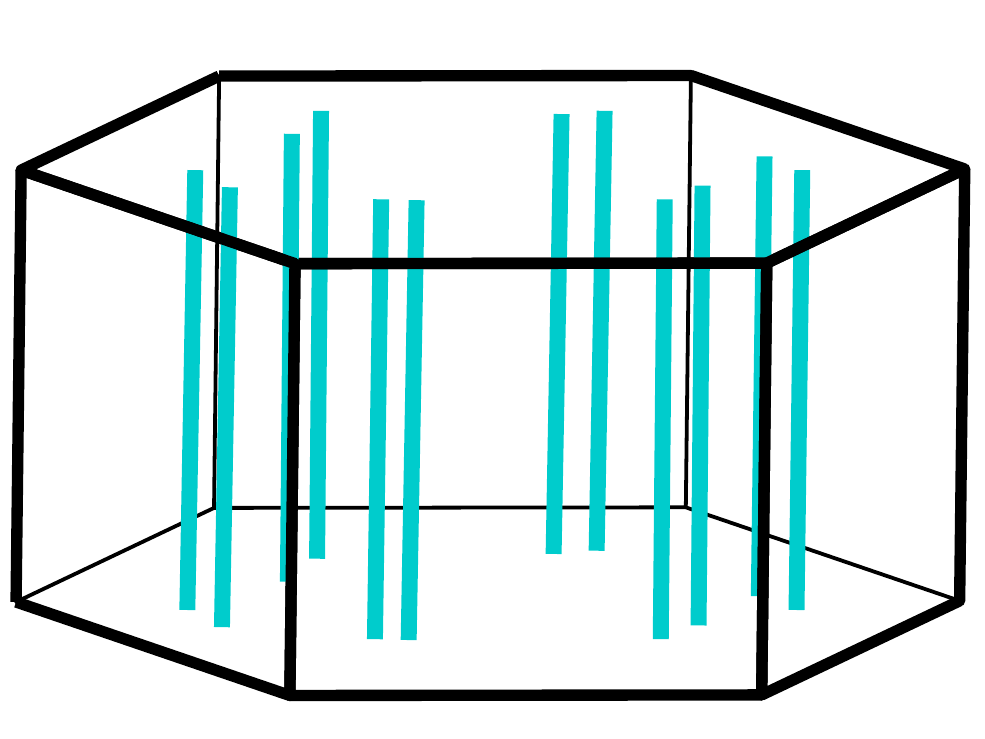}} &
        \raisebox{-0.9\height}{\includegraphics[width=\figwid]{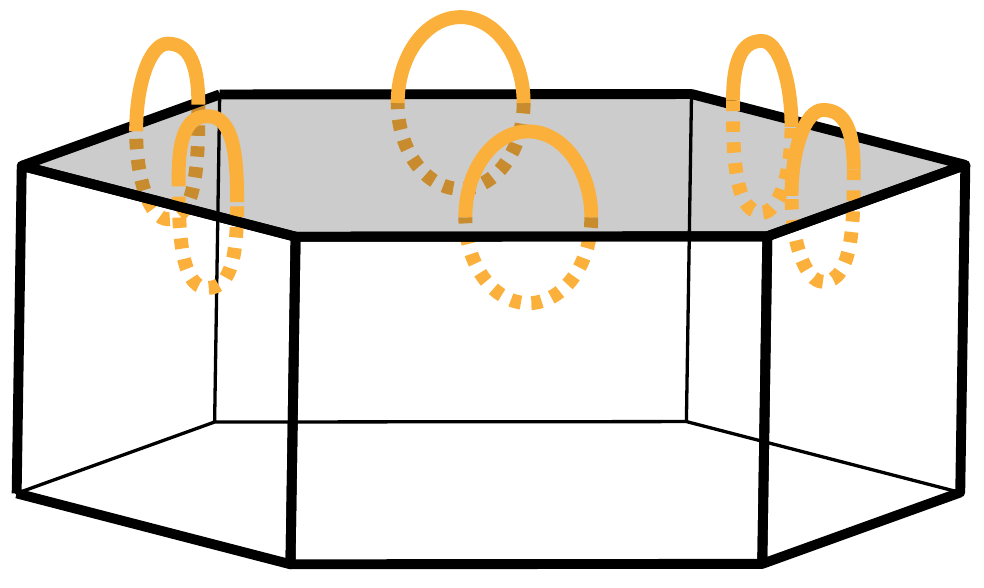}} \\
\hline
%
% C4 ==========================================================================
%
\multirow{8}{\figwids}{Inversion plus fourfold rotation or screw} & \multirow{2}{*}{$\delta_2$} & \multirow{2}{*}{85, 86} & \textit{tP} &
 1 \\
\hliness
& & &  \raisebox{-0.9\height}{\includegraphics[width=\figwid]{BZ_oP.pdf}}  & \raisebox{-0.9\height}{\includegraphics[width=\figwid]{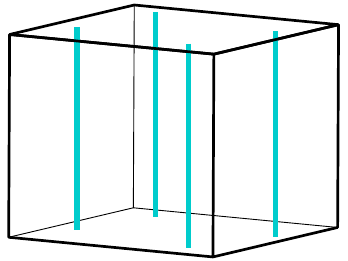}} \\
\hlines
& \multirow{2}{*}{$z'_2$} & \multirow{2}{*}{88} & \textit{tI} &
 1 \\
\hliness
& & & \raisebox{-0.9\height}{\includegraphics[width=\figwid]{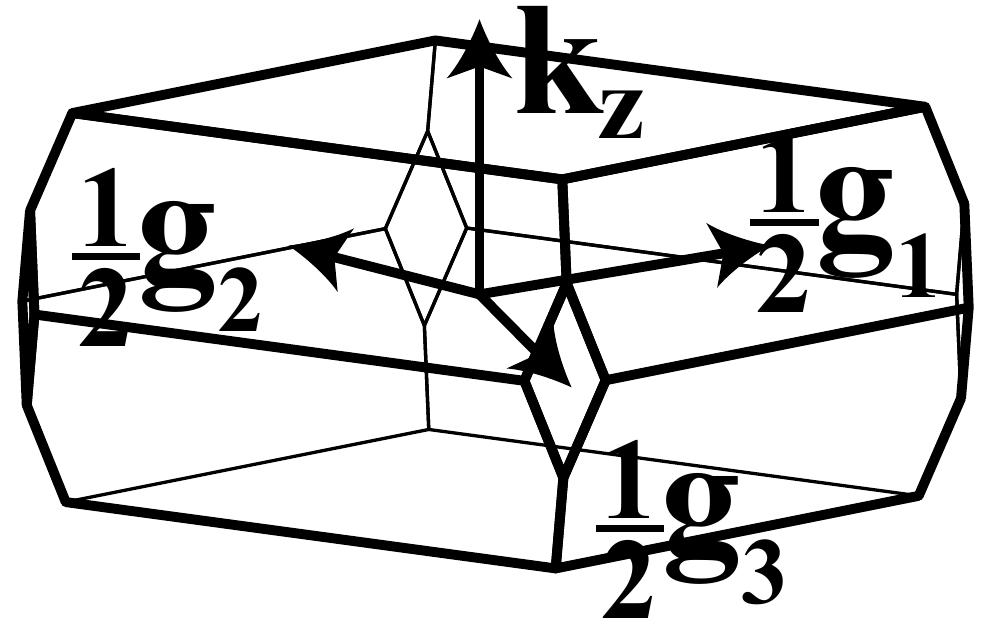}}  & \raisebox{-0.9\height}{\includegraphics[width=\figwid]{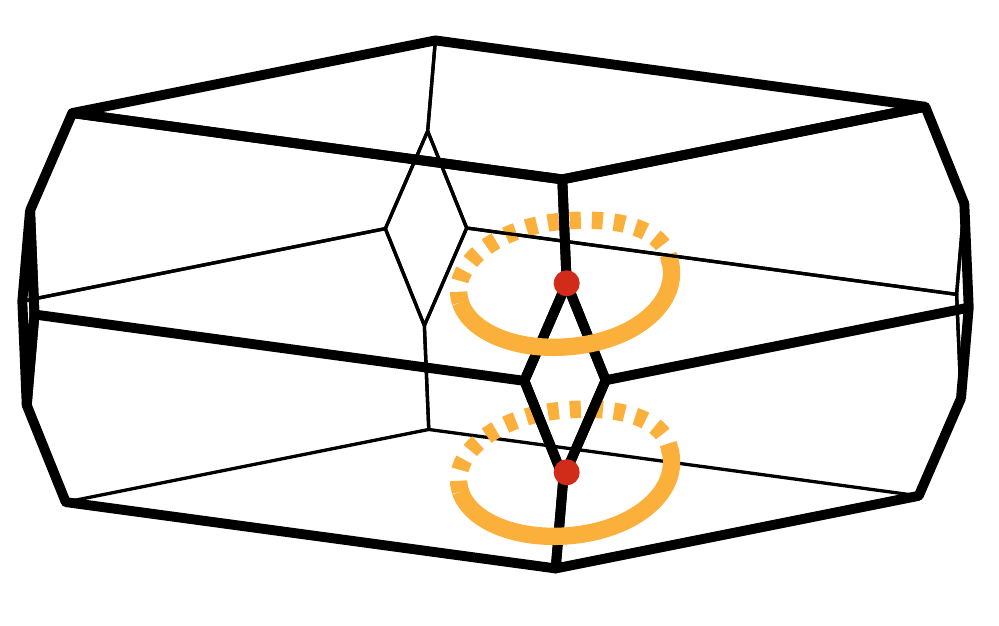}} \\
\hlines
& \multirow{2}{*}{$\delta_2^{(+),\pi}$} & \multirow{2}{*}{124, 128} & \textit{tP} &
 1 \\
\hliness
& & & \raisebox{-0.9\height}{\includegraphics[width=\figwid]{BZ_oP.pdf}} \ & \raisebox{-0.9\height}{\includegraphics[width=\figwid]{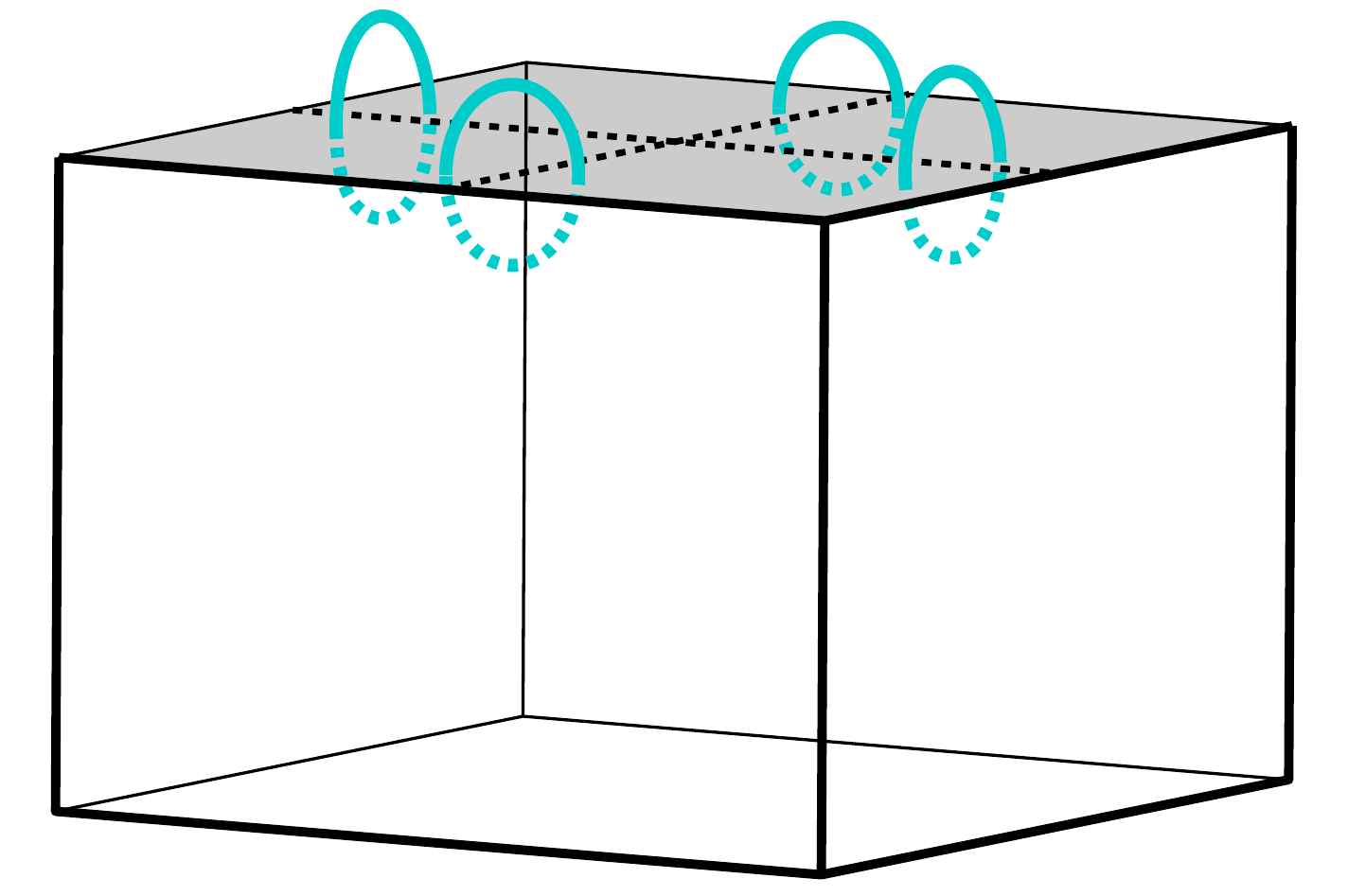}} \\
\hlines
& \multirow{2}{*}{$\theta_2$} & \multirow{2}{*}{130} & \textit{tP} &
 1 \\
\hliness
& & & \raisebox{-0.9\height}{\includegraphics[width=\figwid]{BZ_oP.pdf}}  & \raisebox{-0.9\height}{\includegraphics[width=\figwid]{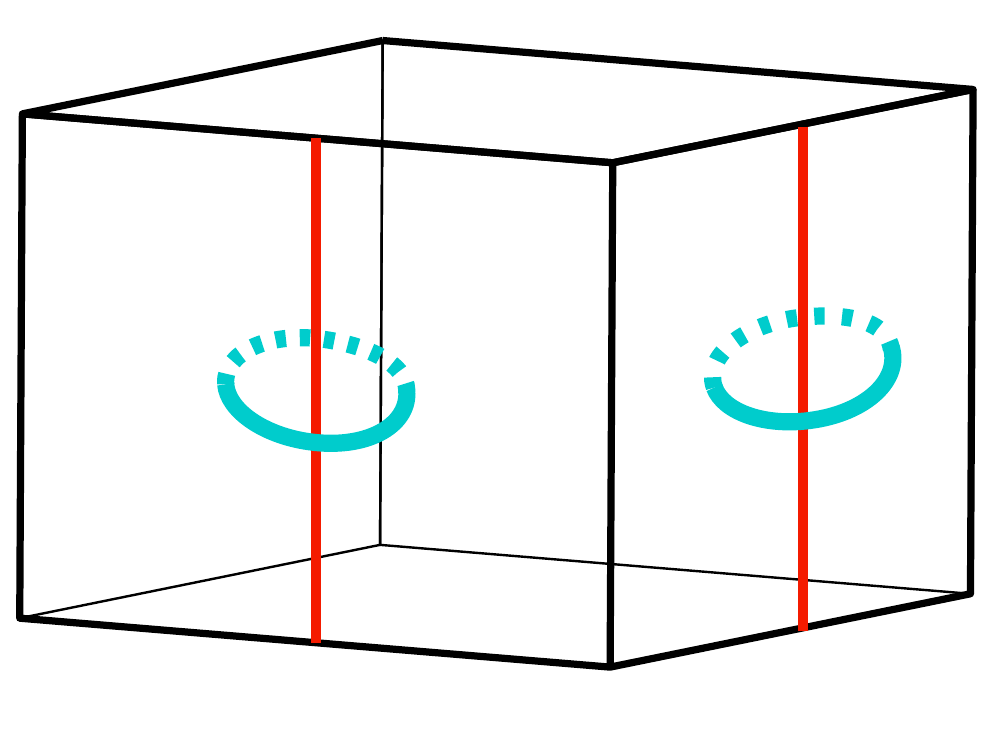}} \\
\hlines
& \multirow{2}{\figwids}{$\delta_2^{(+),0}$, $\delta_2^{(-),0}$} & \multirow{2}{*}{84} & \textit{tP} &
 10, 01 & 11 \\
\hliness
& & & \raisebox{-0.9\height}{\includegraphics[width=\figwid]{BZ_oP.pdf}} & 
        \raisebox{-0.9\height}{\includegraphics[width=\figwid]{84_10.pdf}} &
        \raisebox{-0.9\height}{\includegraphics[width=\figwid]{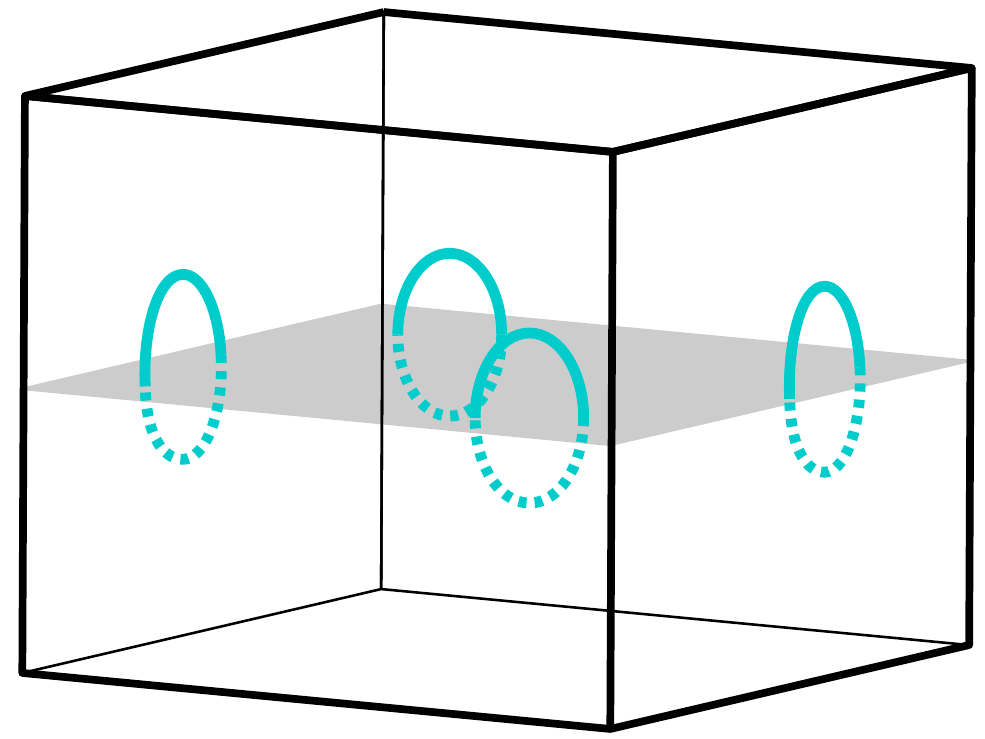}} \\
\hlines
& \multirow{2}{*}{$\phi_2$, $\delta_2^{(+),0}$} & \multirow{2}{*}{87} & \textit{tI} &
 10 & 01 \\
\hliness
& & &  \raisebox{-0.9\height}{\includegraphics[width=\figwid]{BZ_tI.pdf}} & 
       \raisebox{-0.9\height}{\includegraphics[width=\figwid]{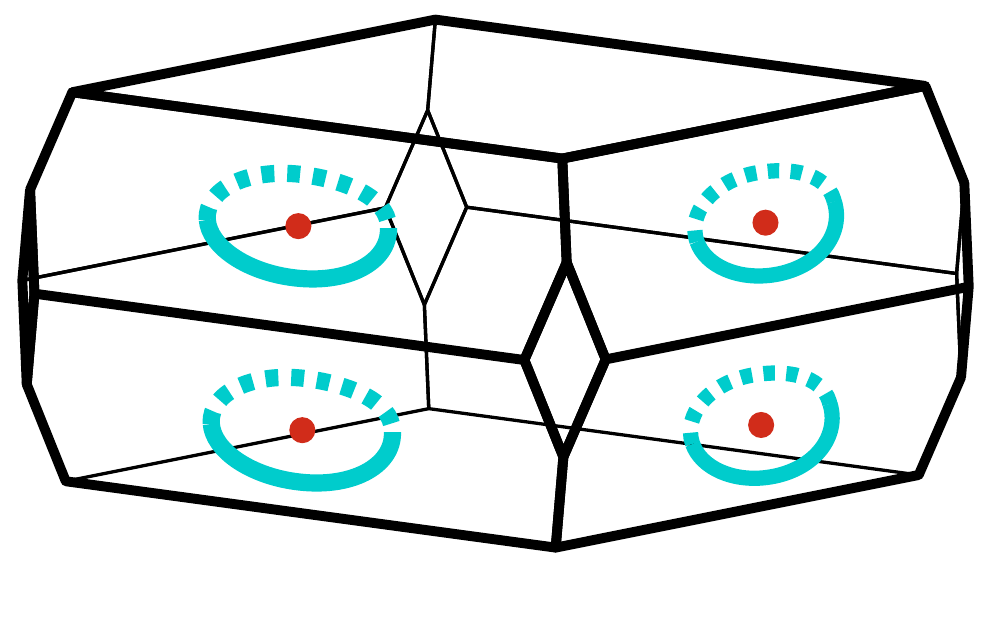}} &
        \raisebox{-0.9\height}{\includegraphics[width=\figwid]{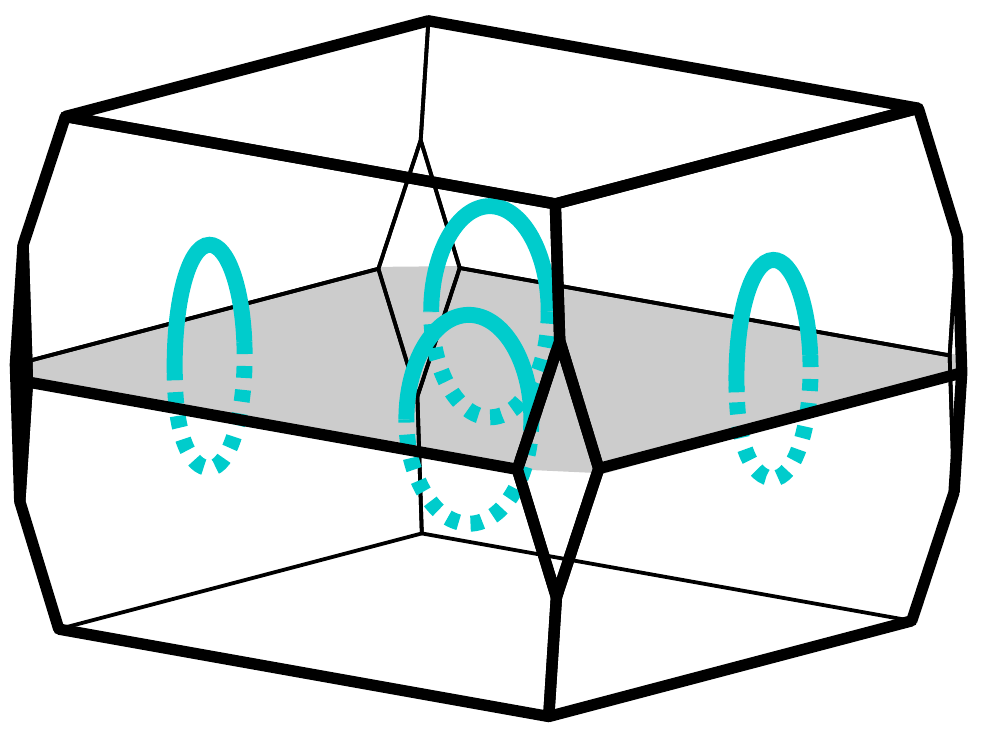}} \\
\hlines
& \multirow{2}{\figwids}{$\delta_2^{(+),\pi}$, $\delta_{2}^{(-),\pi}$, $\delta'_2$} & \multirow{2}{*}{83} & \textit{tP} &
100, 010, 101, 011 & 001 & 111 & 110 \\
\hliness
& & & \raisebox{-0.9\height}{\includegraphics[width=\figwid]{BZ_oP.pdf}}  & 
        \raisebox{-0.9\height}{\includegraphics[width=\figwid]{84_10.pdf}} &
        \raisebox{-0.9\height}{\includegraphics[width=\figwid]{84_11.pdf}} &
        \raisebox{-0.9\height}{\includegraphics[width=\figwid]{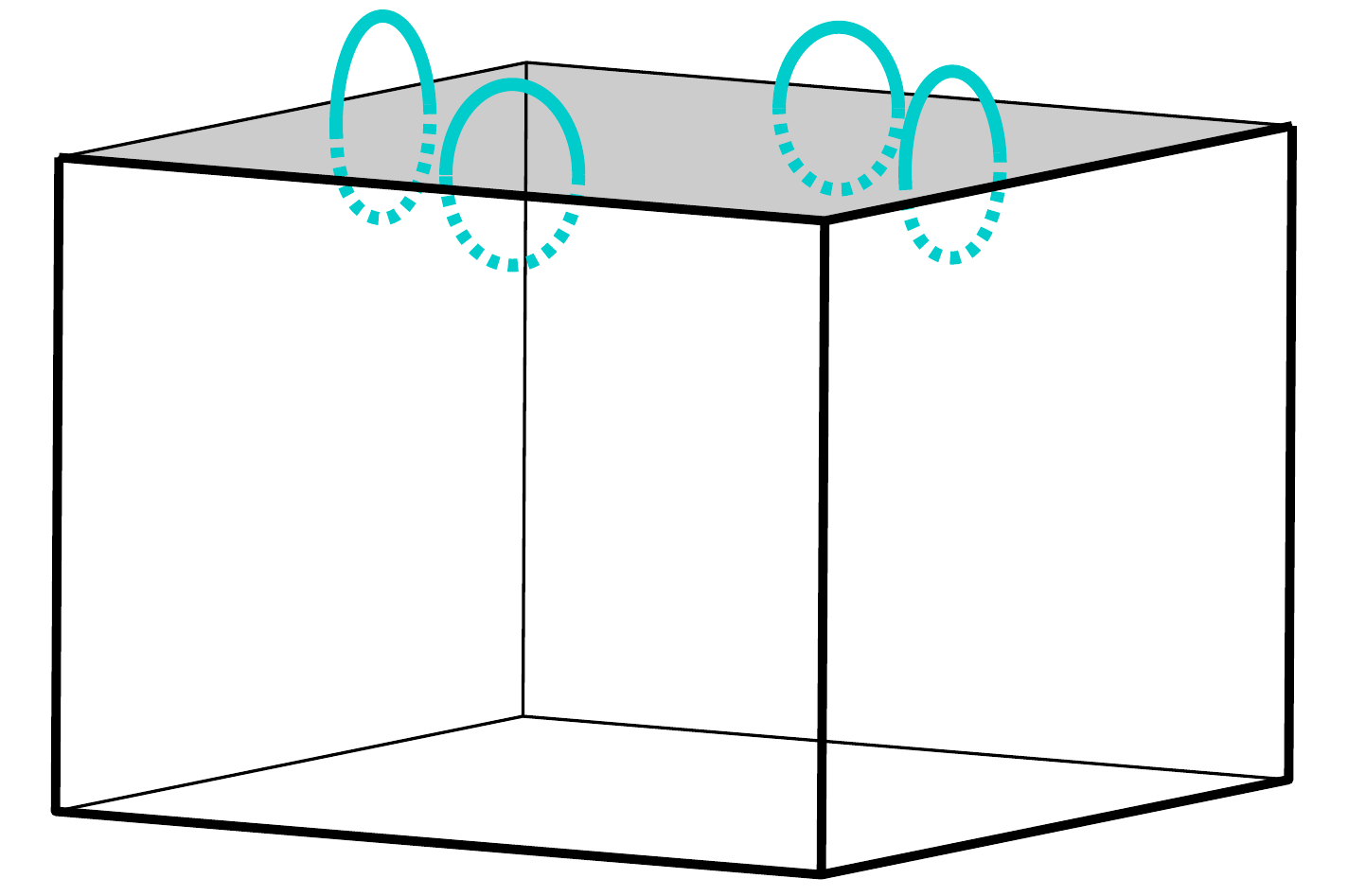}} &
        \raisebox{-0.9\height}{\includegraphics[width=\figwid]{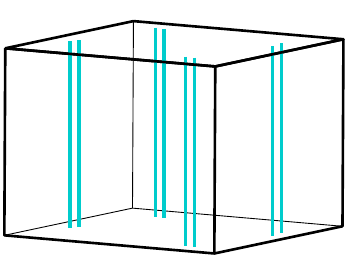}} \\
\hline
%
% Cubic =======================================================================
%
\multirow{4}{\figwids}{Inversion plus cubic symmetry} & \multirow{2}{*}{$z_2^\prime$} & \multirow{2}{*}{201} & \textit{cP} &
 1 \\
\hliness
& & & \raisebox{-0.9\height}{\includegraphics[width=\figwid]{BZ_oP.pdf}} & \raisebox{-0.9\height}{\includegraphics[width=\figwid]{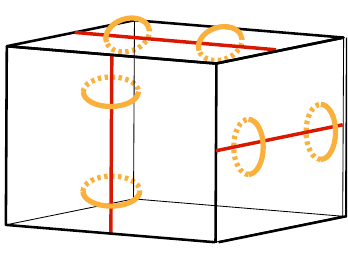}} \\
\hlines
& \multirow{2}{*}{$z_2^\prime$} & \multirow{2}{*}{203} & \textit{cF} &
 1 \\
\hliness
& & & \raisebox{-0.9\height}{\includegraphics[width=\figwid]{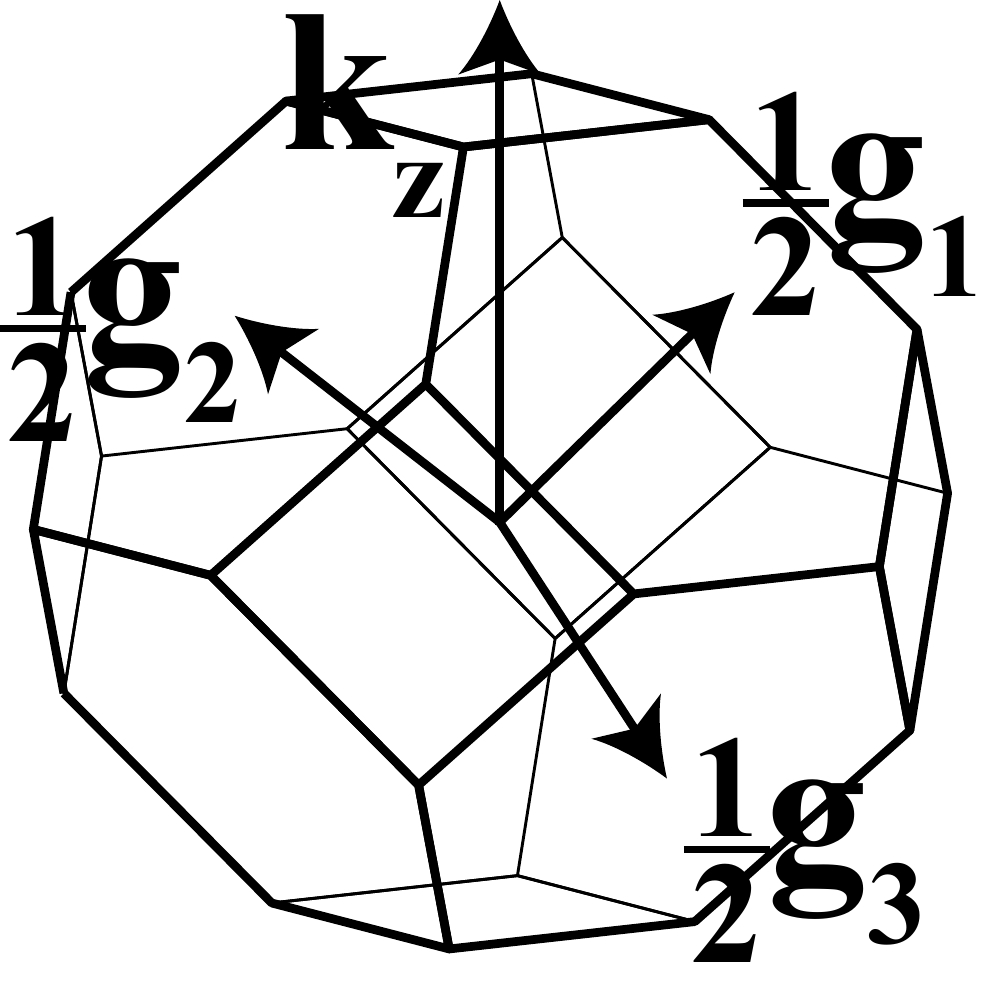}} & \raisebox{-0.9\height}{\includegraphics[width=\figwid]{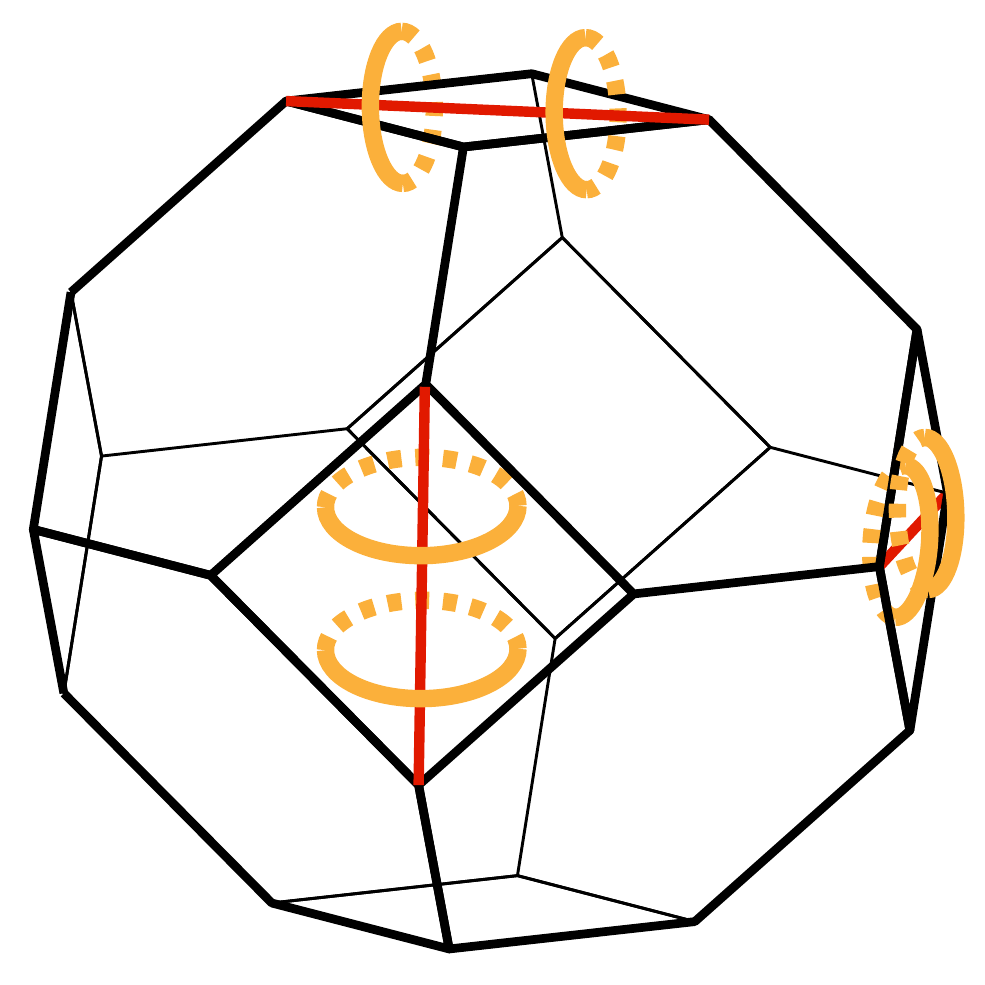}} \\
\hline
\end{longtable*}
\subsection{Space group \#2}
\label{sec:SPG2}

The indicator group for space group \#2 is $\mathbb{Z}_2\times\mathbb{Z}_2\times\mathbb{Z}_2\times\mathbb{Z}_4$, and the corresponding indicator set, $(z_{2,1}z_{2,2}z_{2,3}z_4)$, are defined in Eq. (\ref{eq:p-1}).
The definitions of $z_{2,i=1,2,3}$ are identical to the expressions for the weak indicators in the original Fu-Kane formula, only that in the original context, their nonzero values imply weak topological insulators.
The definition of $z_4$ may look both familiar and strange: $z_4$ mod 2 is the familiar strong indicator in the original Fu-Kane formula, whose nonzero value implies a strong topological insulator if SOC is present, so it is unsurprising if $z_4=1,3$ gives us some topological semimetal without SOC, while the physical meaning of $z_4=2$ remains far from clear at this point.

Before analyzing the implications of nonzero $z_{2,i}$ and $z_4$, we remark that $z_{2,i}$'s are ``weak'', in the sense that they change value under translation-breaking perturbations that preserve inversion symmetry; but that $z_4$ is a strong indicator.
Translation-breaking perturbations, such as density wave order parameters, effectively fold the BZ, such that all eight TRIM in the unfolded BZ are now all at $\Gamma$, and that each new TRIM (which were interior points of the original BZ) has the same number of bands having positive parity as that of bands having negative parity.
Therefore, after this folding, it is easy to confirm that $z_{2,i}=0$ in the folded BZ. 
It is also easy to check that $z_4$ remains the same after this folding.
Intuitively, this is because in its expression, all TRIM contribute equally.

Invoking the lemma in Eq. (\ref{eq:lemma}), we can establish the following statements regarding which planes are crossed by nodal lines for how many times.
(i) If $z_4=0,2$ and $z_{2,i}=1$, both the slice at $k_i=0$ and $k_i=\pi$ are crossed by nodal lines for 2 mod 4 times, and the crossing points are pairwise related by time-reversal, where $k_i$ is the component of the momentum when decomposed into three reciprocal lattice vectors, i.~e., $\mathbf{k}=\sum_{i=1,2,3}k_i\mathbf{b}_i$.
(ii) If $z_4=1,3$ and $z_{2,i}=0$, the slice at $k_i=0$ is crossed by nodal lines for 2 mod 4 times, so that a possible configuration of nodal line is a single nodal ring that is symmetric about $\Gamma$.
(iii) If $z_4=1,3$ and $z_{2,i}=1$, the slice at $k_i=\pi$ is crossed by nodal lines for 2 mod 4 times.
To summarize, if $z_4=0,2$ and some $z_{2,i}\neq 0$, a possible configuration of nodal lines has two lines, related by time-reversal, that run along the direction of $\sum_iz_{2,i}\mathbf{b}_i$, and if $z_4=1,3$, a possible configuration is a single nodal ring around the TRIM at $\sum_i z_{2,i}\mathbf{b}_i/2$.

Analysis above shows that $z_4=0$, $z_{2,i}=1$ and $z_4=2$, $z_{2,i}=1$ correspond to the same nodal line configuration.
Indeed, this must be true because the indicator sets $z_4=0$, $z_{2,i}=1$ and $z_4=2$, $z_{2,i}=1$ change to each other upon changing the inversion center used to define parity (Sec. \ref{sub:SI-P-1}), which leaves all physical observables invariant.
However, it should be noted that, given fixed inversion center, state with $z_4=0$, $z_{2,i}=1$ and state with $z_4=2$, $z_{2,i}=1$ have a relative difference, since by definition we cannot tune one to another without closing gaps at high symmetry momenta.
This relative difference can be detected by additional gapless modes (in addition to the bulk modes) on a domain wall of the two states, provided that the configuration keeps inversion symmetry.

There is one and only one remaining nonzero set of indicators, where $(z_{2,1}z_{2,2}z_{2,3}z_4)=(0002)$.
Using the lemma, we find that there is no plane that must be crossed by nodal lines, so that the natural question becomes: could this state be a (topological) gapped state?
To answer this question, we first look at a related question: is a double-copy of the $(0001)$-state a gapped state? 
According to the above results, the $(0001)$-state has one nodal ring around $\Gamma$, crossing $k_3=0$-plane at $\pm\mathbf{k}_0$.
By doubling the state of $(0001)$, we have also doubled the nodal rings.
Then we introduce coupling between the two copies while preserving $P$ and $T$.
Illustrated in Fig. \ref{fig:2}, the two points on the plane $k_3=0$ where the two rings cross the plane can be gapped, and now we have two disconnected loops on each side of the plane.
We will now show that each loop has \textit{nonzero $\mathbb{Z}_2$-monopole charge} \cite{Fang2015a}, and is therefore topologically stable against any inversion-preserving perturbations.
\begin{figure}
\begin{centering}
\includegraphics[width=1\linewidth]{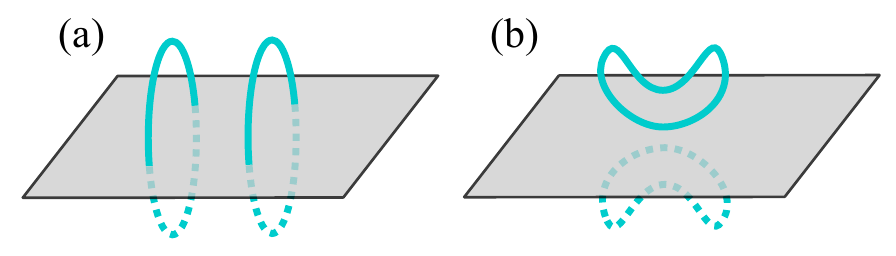}
\par\end{centering}
\protect\caption{\label{fig:2} (a) Two nodal rings that are contributed by the two copies of the $(0001)$-state have two loops at the same position (deliberately separated for distinction), each crossing the $k_3=0$-plane. (b) Adding a hybridization between the two copies can open a full gap on the $k_3=0$-plane, and this results in a reconnection of the nodal structure, such that on each side of $k_3=0$, there is a nodal ring with nontrivial $\mathbb{Z}_2$-monopole charge.}
\end{figure}
The $\mathbb{Z}_2$-monopole charge is defined on a closed 2D manifold, e. g., a sphere, that encloses the nodal loop, originating from the nontrivial second homotopy group of real Grassmanian manifolds.
There has not been any prediction of electronic materials hosting this new type of nodal line, probably due to the involved calculation of its topological invariant.
To show that the nodal loops in Fig. \ref{fig:2}(b) carry nonzero $\mathbb{Z}_2$-charge, one should closely study how the crossing points on the $k_3=0$-plane are gapped out.
Near $\mathbf{k}_0$, where the nodal loop crosses $k_3=0$ plane, the effective theory of a $z_4=1$ state on the $k_3=0$-plane takes the Dirac form
\begin{equation}
h(q_1,q_2)=q_1\sigma_x+q_2\sigma_z,
\end{equation}
where $\mathbf{q}\equiv\mathbf{k}-\mathbf{k}_0$ and $\sigma_i$'s are Pauli matrices; and we have implicitly chosen the symmetry representation $\hat{P}\hat{T}=K$, where $K$ means complex conjugation.
Doubling the whole system automatically doubles $h(\mathbf{q})$ to $H(\mathbf{q})=h(\mathbf{q})\oplus{h}(\mathbf{q})$. 
For the four-band model $H(\mathbf{q})$, there is only one term that gaps the spectrum while preserving $PT$: $m\tau_y\sigma_y$, where $\tau_z=\pm1$ is the flavor index. 
Due to time-reversal symmetry, the two Dirac points at $-\mathbf{k}_0$ also gap out each other, such that the plane $k_3=0$ becomes fully gapped, where the $\mathbb{Z}_2$-invariant, $\nu$, protected by $PT$-symmetry can be defined. 
For four-band models of the form
\begin{equation}
H(\mathbf{k})=d_x(\mathbf{k})\tau_0\sigma_x+d_y(\mathbf{k})\tau_0\sigma_z+d_z(\mathbf{k})\tau_y\sigma_y,
\end{equation}
the $\mathbb{Z}_2$ invariant is given by \cite{Fang2015a}
\begin{equation}
\label{eq:kappa}
\nu=\frac{1}{4\pi}\int\int_{BZ}dk^2\hat{d}\cdot\times\partial_{k_1}\hat{d}\times\partial_{k_2}\hat{d}\;\mathrm{mod} 2,
\end{equation}
where $d_i(\mathbf{k})$ are real functions and $\hat{d}$ the unit vector of $(d_x,d_y,d_z)$.
Inserting $H(\mathbf{q})$ into the integral in Eq.(\ref{eq:kappa}), one finds that $H(\mathbf{q})$ contributes $1/2$ to $\nu$. 
To obtain the integral near $-\mathbf{k}_0$, we first notice that the Bloch wavefunctions at $-\mathbf{k}_0$ can be obtained by acting $T$ on the wavefunctions at $\mathbf{k}_0$, so as to fix the basis vectors for the k.p-theory near $-\mathbf{k}_0$. 
In this basis, the effective theory at $-\mathbf{k}_0$ is simply $H'(\mathbf{q})=H(-\mathbf{q})$. 
Explicit calculation shows that the contribution of $H'(\mathbf{q})$ is also $1/2$, and $\nu$ for entire the $k_3=0$-slice is $\nu(k_z=0)=1$.
On the other hand $\nu(k_z=\pi)$ for $k_3=\pi$-slice is trivial:
Before doubling, the $k_3=\pi$-plane is gapped, and may possess either a trivial or nontrivial $\nu$, but due to its $\mathbb{Z}_2$-nature, after doubling $\nu$ simply vanishes. 
Therefore, the difference of $\nu$ between $k_3=0$- and $k_3=\pi$-planes implies that there is a nodal loop with nontrivial $\mathbb{Z}_2$ charge on each side of the $k_3=0$-plane.

While the doubled $(0001)$-state is a $(0002)$-state, the inverse is unnecessarily true.
Two states with same indicators $(0002)$ may in principle still be topologically distinct, and we cannot exclude the possibility that there be $(0002)$-states that are fully gapped.
Nevertheless, we argue that this is highly unlikely based on the following observation.
Suppose we start from the $(0002)$-state built from doubling a $(0001)$-state.
In order to gap out the two nodal loops, they have to pairwise annihilate each other at a TRIM, which we assume to be $\Gamma$ without loss of generality.
Consider a four-band k$\cdot$p-model near $\Gamma$ that describes this annihilation process:
\begin{equation} \label{eq:z4=2-kp}
H=k_1\tau_0\sigma_x+k_2\tau_0\sigma_z+k_3\tau_z\sigma_0+(m-k_z^2)\tau_y\sigma_y,
\end{equation}
where the symmetry operators are represented by $\hat{P}=\tau_y\sigma_y$ and $\hat{T}=K\tau_y\sigma_y$.
The two loops with $\mathbb{Z}_2$ charge are, if $m>0$,
\begin{equation}
L_\pm\equiv\{(k_1,k_2,k_3)|k_3=\pm\sqrt{m},\sqrt{k_1^2+k_2^2}=m\}.
\end{equation}
At $m=0$ both loops shrink to a point and vanish at $\Gamma$, and the model is fully gapped for $m<0$.
Therefore the model indeed describes a topological transition from $m>0$ to $m<0$, where two loops having $\mathbb{Z}_2$ charge annihilate at $m=0$.
For $m>0$ ($m<0$), the two occupied bands at $\Gamma$ have negative (positive) parity, such that after the band inversion, the strong index $z_4$ changes by two.
Therefore, within a four-band model, it is impossible for a $(0002)$-state to be fully gapped.
For a system of more than four bands, we can reasonably assume that when the two nodal loops are sufficiently close to each other, the bands away from the Fermi energy can be energetically separated from the four bands involved in the annihilation without changing band topology, so that the above analysis remains valid.

To further confirm the correspondence between $z_4=2$ and nodal loops with $\mathbb{Z}_2$ charge, we generalize the k$\cdot$p model in Eq. (\ref{eq:z4=2-kp}) to a tight-binding model given by
\begin{align}
H(\mathbf{k})& = \sin k_1\tau_0\sigma_x + \sin k_2\tau_0\sigma_z + \sin k_3 \tau_z\sigma_0 \nonumber \\
             & + \Big(\Delta-\sum_i\cos k_i\Big)\tau_y\sigma_y,
\end{align}
and the symmetry operators are still $\hat{P}=\tau_y\sigma_y$ and $\hat{T}=K\tau_y\sigma_y$.
On one hand, the $\mathbb{Z}_2$ invariant (Eq. ({\ref{eq:kappa}})) of a $k_z$-plane can be explicitly derived as 
\begin{equation}
\nu(k_z)=\mathrm{sgn}((\Delta-\cos k_3-2)(\Delta-\cos k_3+2))\;\mathrm{mod}\;2,
\end{equation}
The $\mathbb{Z}_2$ charge in half of BZ, given by the difference of $\nu(k_3=0)$ and $\nu(k_3=\pi)$, can be expressed as $\mathrm{sgn}(\Delta-3)(\Delta-1)(\Delta+1)(\Delta+3)\;\mathrm{mod}\;2$ \cite{Fang2015a}.
Therefore, in half of BZ exits one (mod 2) nodal loop with  $\mathbb{Z}_2$ charge  when $1<|\Delta|<3$.
On the other hand, it is direct to verify that for $1<|\Delta|<3$ the indicator set of the lower two bands is $(0002)$, whereas for $|\Delta|>3$ or $|\Delta|<1$ the indicator set is $(0000)$.
Therefore the correspondence between $z_4=2$ and nodal loops with $\mathbb{Z}_2$ charge still holds in this tight-binding model.

Up to this point, we have used both mathematics and physical arguments to support the conclusion that a state with indicators $(0002)$ must have one (or an odd number of) nodal ring(s) in each half of the BZ, and each nodal ring has nonzero $\mathbb{Z}_2$-monopole charge.
This concludes the analysis of each nonzero set of indicators for space group \#2.
Before going to the next part, we briefly comment on the difficulty in finding a rigorous proof for the relation between indicator set $(0002)$ and the presence of $Z_2$-charged nodal ring. One may attempt at relating the $Z_2$ invariant of a 2D plane with $P$ and $T$ to the inversion eigenvalues, and then show that if $k_z=0$ and $k_z=\pi$ have different $Z_2$, there must be one nodal ring having $Z_2$ monopole charge. The same proof worked for the relation between the number of Weyl points and the inversion eigenvalues in 3D with inversion but not time-reversal. However, in our case, one may easily prove that 2D systems having $P$ and $T$ do \emph{not} have any indicator, so that there does not exist any relation between the $Z_2$ invariant and the inversion eigenvalues.

\subsection{Inversion plus twofold rotation/screw axis}

From this point, we start adding one additional rotation or screw axis to enhance the space group \#2 to a higher space group.
As convention, the direction of the highest-order rotation (screw) axis is defined as the $z$-direction.

There are 18 space groups that have (i) at least one twofold rotation/screw axis, (ii) no higher-order rotation/screw axis and (iii) nontrivial indicator group.
Two observations relate their indicators to the indicators of \#2:
(i) Among the 18, space group \#14 is their common subgroup and a study of the compatibility relation in \#14 reveals that $z_4\in{even}$; and
(ii) All 18 space groups are subgroups of space group \#192, and the compatibility relations of \#192 allows at least one solution having $z_4=2$.
These two facts together give at least one strong $\mathbb{Z}_2$-indicator shared by all 18 space groups: $z'_2\equiv{z}_4/2$.

The subsection hence is further divided into parts by the number of independent $\mathbb{Z}_2$-indicators other than $z'_2$.

\subsubsection{$\mathbb{Z}_2$-indicators in space groups \#11, 14, 48, 49, 50, 52, 53, 54, 56, 58, 60, 66, 68, 70}
\label{sec:twofold1}

Since $z'_2$ is always an indicator and the 14 space groups in this part have only one indicator, this indicator must be $z'_2$.
As \#2 is their common subgroup, when $z'_2=1$, there is one (or an odd number of) nodal ring in each half of the BZ, but their configurations now are further constrained by rotation/screw axes.

Finding the constraints requires a detailed study of the k$\cdot$p-model of a nodal ring having $\mathbb{Z}_2$-monopole charge.
The k$\cdot$p-model expanded about the center of a nodal ring with $\mathbb{Z}_2$-monopole charge is \cite{Fang2015a}
\begin{equation}
\label{eq:singlering}
h(\mathbf{q})=q_1\sigma_x+q_2\sigma_z+q_3\tau_y\sigma_y+m\tau_z\sigma_0,
\end{equation}
where the only assumed symmetry $PT$ is represented by $\hat{P}\hat{T}=K$.
Hamiltonian in Eq. (\ref{eq:singlering}) gives a nodal ring of radius $\sqrt{|m|}$ perpendicular to the $q_3$-direction.
It is easy to confirm that $h(\mathbf{q})$ is consistent with an additional twofold rotation $\hat{C}_2=\sigma_y$ satisfying $C_2^2=1$, such that
\begin{equation}
\label{eq:C2}
\hat{C}_2^{-1}h(q_1,q_2,q_3)\hat{C}_2=h(-q_1,-q_2,q_3).
\end{equation}
It is important to notice that $\hat{C}_2$ thus defined \textit{anticommutes} with $\hat{P}\hat{T}$, and that it is impossible to find any Dirac matrix that commutes with $PT$ while satisfying Eq. (\ref{eq:C2}).
Therefore, the nodal ring may be centered at a $C_2$-axis where 
\begin{equation}
\label{eq:rotation}
\hat{C}_2\hat{P}\hat{T}=-\hat{P}\hat{T}\hat{C}_2.
\end{equation}
This is satisfied only if (i) in real space, the $C_2$-axis does not pass any inversion center, and (ii) at the $C_2$-invariant momentum, commuting $P$ and $C_{2}$ generates a minus sign.
In any lattice having twofold rotation and inversion, we can set the inversion center at $(000)$, and the rotation axis as the line passing through $\mathbf{t}/2$,  where $\mathbf{t}$ is a vector perpendicular to the rotation axis, such that $C_2=\{2|\mathbf{t}\}$ and $C_2P=\{1|2\mathbf{t}\}PC_2$.
Then for a $C_2$-invariant line in momentum space, if it satisfies $2\mathbf{k}\cdot\mathbf{t}=\pi\;\mathrm{mod}\;2\pi$, Eq. (\ref{eq:rotation}) holds. 
Similar results apply for twofold screw axis $C_{2_1}=\{2|\mathbf{t+s}\}$, where $\mathbf{s}$ the screw-vector parallel with screw axis, but the anti-commutation is replaced by
\begin{equation}
\label{eq:screw}
\hat{C}_{2_1}\hat{P}\hat{T}=-e^{2i\mathbf{s}\cdot\mathbf{k}}\hat{P}\hat{T}\hat{C}_{2_1}.
\end{equation}
Applying these results to the listed space groups, we find that space groups \#14, 48, 49, 50, 52, 53, 54, 56, 58, 60, 66, 68 and 70 have twofold axes not passing through any inversion center, and we plot in Table \ref{tab:1} nodal rings around one $C_2$- or $C_{2_1}$-invariant line where either Eq. (\ref{eq:rotation}) or Eq. (\ref{eq:screw}) is met.
In space group \#11, where the screw axis contains an inversion center, each of the two nodal rings are centered and symmetric about the $k_2=0$-plane, and the two rings are related to each other under the twofold screw rotation. (For \#11, the screw axis is along the $k_2$-direction, as shown in table \ref{tab:1}.)

\subsubsection{$\mathbb{Z}_2\times\mathbb{Z}_2$-indicators in space groups \#12, 13, 15}
\label{sec:twofold2}

Other than the $\mathbb{Z}_2$-indicator $z'_2$, there is another $\mathbb{Z}_2$-indicator for these three monoclinic space groups.
Analysis of the compatibility relations reveals that this is nothing but the second weak index $z_{2,2}$.
When $z_{2,2}=1$, the strong indicator $z'_2$ becomes convention-dependent (see discussion in Sec. \ref{sub:SI-P-1} and \ref{sec:SPG2}), and the nodal lines are two lines running along the $k_2$-direction, related to each other under time-reversal.
When $z_{2,2}=0$ and $z'_2=1$, there are again two nodal rings having $\mathbb{Z}_2$-monopole charge, the positions of which depend on if the rotation axis contains any inversion center, as explained in Sec. \ref{sec:twofold1}.

\subsubsection{$\mathbb{Z}_2\times\mathbb{Z}_2\times\mathbb{Z}_2$-indicators in space group \#10}
\label{sec:twofold3}

Space group \#10 has a simple monoclinic lattice with a twofold rotation axis, so that $P$ and $C_2$ implies a mirror plane $M\equiv{P}C_2$ that also contains the inversion center.
In this case, the 2D slice at $k_2=\pi$ is a mirror invariant plane in the BZ, on which we can define two $\mathbb{Z}_2$-indicators for the $M=+1$- and $M=-1$-sectors as $z_2^{(+)}$ and $z_2^{(-)}$ (Eq. (\ref{eq:zpm})), respectively.
From Eq. (\ref{eq:zpm}) and Eq. (\ref{eq:p-1}), we see that the previously defined index $z_{2,2}$ satisfies
\begin{equation}
z_{2,2}=z^{(+)}_2+z^{(-)}_2\;\mathrm{mod}\;2.
\end{equation}
Therefore, if $z^{(+)}_2=1$ and $z^{(-)}_2=0$ (or vice versa), there are two nodal lines running along the $k_2$-direction, related to each other by time-reversal.
We are now left with three cases (a) $z_2^{(\pm)}=0,z'_2=1$, (b) $z_2^{(\pm)}=1,z'_2=0$ and (c) $z_2^{(\pm)}=1,z'_2=1$. 
\begin{enumerate}[fullwidth,itemsep=0cm,label=(\alph*)]
\item $z_2^{(\pm)}=0$ means that $k_2=\pi$-plane is crossed by 0 mod 4 nodal lines, and $z'_2=1$ means that there are in the entire BZ 2 mod 4 nodal rings.
According to the discussion in Sec. \ref{sec:twofold1}, since the rotation axis contains at least one inversion center, a nodal ring cannot be symmetric about any $C_2$-invariant line.
The only possibility is that both nodal rings are vertical, symmetric about the $k_2=0$-plane but related to each other under the twofold rotation.
\item $z_2^{(\pm)}=1$ means that the nodal lines cross the $k_z=\pi$-plane four times, two in the $M=+1$- and two in the $M=-1$-sectors, respectively, while $z'_2=0$ means that the same happens on the $k_0=0$-plane.
A typical configuration has in total four nodal lines running along the $k_2$-direction.
\item $z_2^{(\pm)}=1$ means that the nodal lines cross the $k_z=\pi$-plane four times, two in the $M=+1$- and two in the $M=-1$-sectors, respectively, while $z'_2=1$ means there are in total 2 mod 4 nodal rings in the entire BZ.
A typical configuration has two vertical nodal rings, related to each other by twofold rotation, but now centered and symmetric about the $k_2=\pi$-plane.
\end{enumerate}

\subsection{Inversion plus threefold rotation/screw axis}

Now we consider, instead of a twofold axis, adding a threefold/screw rotation axis to space group \#2.
Among all space groups having these two symmetries, the following eight have nontrivial indicator group: \#147, 148 and \#162-167.
All eight are either subgroups of \#166 or \#192, both having at least one band structure that gives $z_4=2$.
Therefore, they at least have a strong $\mathbb{Z}_2$-indicator $z'_2=z_4/2$.
In fact, it is the only indicator for space group \#162-167, where the indicator group is $\mathbb{Z}_2$.
When $z'_2=1$, there are 2 mod 4 nodal rings, and there their configuration must observe both $C_3$ and time-reversal symmetries.
Unlike twofold axes, threefold axes always contain inversion centers in these space groups.
A study of the model in Eq. (\ref{eq:singlering}) shows that we can define the
\begin{equation}
\hat{C}_3=-\exp(i\sigma_y\pi/3)
\end{equation}
which satisfies $\hat{C}_3^3=1$, $[\hat{C}_3,\hat{P}\hat{T}]=0$ and
\begin{equation}
\hat{C}^{-1}_3h(\mathbf{k})\hat{C}_3=h(C_3\mathbf{k}).
\end{equation}
Therefore, while a nodal ring having a $\mathbb{Z}_2$-monopole charge is inconsistent with a twofold rotation through its center if $[\hat{C}_2,\hat{P}\hat{T}]=0$, it is \textit{consistent} with a threefold rotation, so that a ring may be surrounding a $C_3$-invariant line in the BZ.

\subsubsection{$\mathbb{Z}_2$-indicators in space groups \#162-167}

As is said, $z'_2$ is the only indicator for these groups, and if $z'_2=1$, there are two nodal rings, each of which symmetric about a $C_3$-invariant line in the BZ, and are related to each other by time-reversal.

\subsubsection{$\mathbb{Z}_2\times\mathbb{Z}_4$-indicators in space groups \#147, 148}

These two space groups have higher symmetry indicators.
In fact, they allow $z_4$ defined in Eq.(\ref{eq:p-1}) to take all four possible values, and also allow the three weak indicators to be same and nonzero, i.~e., $z_{2,1}=z_{2,2}=z_{2,3}=1$.
The topological information indicated by their nonzero combinations is identical to those discussed in Sec.\ref{sec:SPG2}, with the only difference that, due to the threefold axis, the state having $z_{2,3}=1$ and $z_4=0,2$ should have 6 mod 12 nodal lines running in the third direction.

\subsection{Inversion plus six rotation/screw axis}

There are only three space groups that have sixfold axis and nontrivial indicator group, all of which are subgroups of space group \#192.
\#192 allows at least one band structure having $z_4=2$, such that the three groups have at least one strong $\mathbb{Z}_2$-indicator, $z'_2\equiv{z}_4/2$.
When $z'_2=1$, there must be 2 mod 4 nodal rings protected by $P$ and $T$, and the additional symmetries put further constraints on their numbers and configurations.
Like threefold axes, sixfold axes always contain inversion centers, and since a sixfold axis is automatically a twofold axis, all twofold axes also contain inversion centers in the three subgroups.
According to the discussion in Sec. \ref{sec:twofold1}, any nodal ring in this case cannot be symmetric about a $C_2$-invariant line or a $C_6$-invariant line.
A minimal configuration, having the least number of rings, therefore has six nodal rings, each vertical and symmetric under the mirror symmetry $k_z\rightarrow-k_z$.
They are related to each other by sixfold rotation/screw, and may be centered either on the $k_z=0$ or the $k_z=\pi$-plane (depending on which space group).
In all three space groups, the mirror plane allows us to define the two additional $\mathbb{Z}_2$-indicators, as in the twofold axis case, $z^{(\pm)}_2$ for the positive and negative mirror sectors, respectively.

\subsubsection{$\mathbb{Z}_2$-indicators in space groups \#176, 192}

For \#176, \#192, $z'_2$ is the only indicator, and its nontrivial value corresponds to the six-ring configuration described above.
However, in \#176, the screw axis makes the entire $k_z=\pi$-plane double degenerate, inconsistent with having twelve discrete band touch points where the six rings cross this plane.
These rings are hence centered at the $k_z=0$-plane, each symmetric about the plane.
In \#192, the nontrivial indicator always corresponds to band structure having $z_2^{(\pm)}=1$, and this means that the six rings are located on the $k_z=\pi$-plane.

\subsubsection{$\mathbb{Z}_2\times\mathbb{Z}_2\times\mathbb{Z}_2$ in space group \#175}

The three indicators forming the indicator group of space group \#175 are $z_2^{(\pm)}$ and $z'_2$.
The topological information carried by their nonzero combinations can be derived in the same way as in Sec. \ref{sec:twofold3}.

\subsection{Inversion plus additional fourfold rotation/screw axis}

We now consider adding fourfold rotation/screw axes to space group \#2 and also to space groups having twofold axes: these are the tetragonal space groups.

Among them, the following ones have nontrivial indicator groups: \#83-88, 124, 128 and 130, nine in total.
Calculation shows that only \#86 and 88 allow band structures having $z_4=2$, while all the others have $z_4=0$.
Therefore, for the other seven space groups, three new $\mathbb{Z}_2$-indicators are found.

Before defining them, we need to introduce a new variation of the lemma presented in Eq. (\ref{eq:lemma1}).
Consider a 2D system having $C_4$, $P$ and $T$, and consider a loop shown in Fig. \ref{fig:BerryPhase1}(b), which encloses a quadrant of the BZ.
It is shown in Ref. [\onlinecite{Fang2012}] that the Berry phase associated with this loop is determined by the symmetry eigenvalues of the occupied bands in the following equation\\
\textit{Lemma: second variation}
\begin{equation}
\exp(i\Phi_B)=\prod_{n\in{occ.}}\xi_n(\Gamma)\xi_n(\mathrm{M})\zeta_n(\mathrm{X}),\label{eq:lemma2}
\end{equation}
where $\xi$ is the $C_4$-rotation eigenvalue and $\zeta$ the $C_2$-rotation eigenvalue.
While $\xi_n$ takes four possible values $\pm1$ and $\pm{i}$, time-reversal ensures that $\pm{i}$ appear in pairs, such that the RHS of Eq. (\ref{eq:lemma2}) can only be either $+1$ or $-1$.
If $-1$, we have $\Phi_B=\pi$, and since $PT$ makes Berry curvature vanish, this implies that in each quadrant of the BZ there must be 1 mod 2 Dirac points.

We can apply this lemma to 3D systems having the same symmetries.
For the $k_z=0$ and the $k_z=\pi$-slice in a simple tetragonal lattice and for $k_z=0$-slice in a body-centered tetragonal lattice, applying the lemma of Eq.(\ref{eq:lemma2}) simply tells us that if $\Psi_B=\pi$, the slice of question is crossed by nodal lines 1 mod 2 times in each quadrant, or 4 mod 8 times in total.
This observation enables the definition of a new $\mathbb{Z}_2$-indicator
\begin{equation}
\delta_2\equiv{}N_{\xi=-1}(\Gamma)+N_{\xi=-1}(\mathrm{M})+N_{\zeta=-1}(\mathrm{X})\;\mathrm{mod}\;2,
\end{equation}
where $N_{\xi=-1}$ and $N_{\zeta=-1}$ are the numbers of occupied bands having $\xi=-1$ and $\zeta=-1$, respectively.

If the $C_4$-axis contains inversion centers, then $M=C_4^2P$ is a mirror plane. 
In this case, the Hamiltonian on the $k_z=0$- and $k_z=\pi$-slices can be further divided into decoupled sectors having $M=+1$ and $M=-1$, so that the Berry phase for the two sectors can also be separately defined on the two planes.
Applying the lemma of Eq. (\ref{eq:lemma2}) to each sector, we have the definitions of the following $\mathbb{Z}_2$-numbers
{\small
\begin{align}
\delta_2^{(\pm),0}\equiv&{N}^{(\pm)}_{\xi=-1}(\Gamma)+N^{(\pm)}_{\xi=-1}(\mathrm{M})+N^{(\pm)}_{\zeta=-1}(\mathrm{X})\;\mathrm{mod}\;2,\\
\delta_2^{(\pm),\pi}\equiv&{N}^{(\pm)}_{\xi=-1}(\mathrm{Z})+N^{(\pm)}_{\xi=-1}(\mathrm{A})+N^{(\pm)}_{\zeta=-1}(\mathrm{R})\;\mathrm{mod}\;2.
\end{align}}
We emphasize that (i) if the lattice is body-centered tetragonal, then $\delta_2^{(\pm),\pi}$ is undefined, (ii) if $C_4$ is replaced by screw axis $C_{4_1}$, then again $\delta_2^{(\pm),\pi}$ is undefined and (iii) these four numbers satisfy
\begin{equation}
\delta_2^{(+),0}+\delta_2^{(-),0}=\delta_2^{(+),\pi}+\delta_2^{(-),\pi}=\delta_2\;\mathrm{mod}\;2.
\end{equation}
Depending on the specific groups, we choose from these four indicators one, two or three as generators of the indicator group.

\subsubsection{$\mathbb{Z}_2$-indicators in space groups \#85, 86, 88, 124, 128, 130}

Unlike the cases of twofold or sixfold axes, here we cannot find a common $\mathbb{Z}_2$-indicator for these space groups, although their indicator groups are the same.

We directly solve the compatibility relations for all the basis vectors of the symmetry data space, and check if $z'_2$ or any of the above new indicators, namely, $\delta_2$, $\delta^{(\pm),0/\pi}_2$ is well-defined in the given space group, and if yes trivial or nontrivial.
If any one of them is nontrivial for any one of the basis vectors of the band representation space, we know that this is the generator of the indicator group, because there is only one $\mathbb{Z}_2$-indicator.

The exhaustive search described above yields the indicator $\delta_2$ for \#85 and 86, and the indicator $\delta_2^{(+),\pi}$ for \#124, 128.
For \#130, the candidate indicators are either undefined or take zero value for all possible basis vectors, and we have to solve the some algebraic equations find the expression of its indicator
\begin{eqnarray}
\label{eq:130SI}
\theta_2&=&[N(\Gamma_{1}^+)+N(\Gamma_{1}^-)+N(\Gamma_{3}^+)+N(\Gamma_{3}^-)\\
\nonumber&+&N(\mathrm{M}_1)+N(\mathrm{M}_2)]/2\;\mathrm{mod}\;2,
\end{eqnarray}
where $N(irrep)$ stands for the number of appearance of a certain irreducible representation in the valence bands at a given high-symmetry point.
The labels of the irreducible representations follow the convention of Ref. \cite{Vergniory2017,Elcoro2017,Bradlyn2017}.

The configurations of nodal lines are easily found for \#85, 86.
Since $\delta_2=1$, the nodal lines pass the $k_z=0$-plane for 4 mod 8 times, and but since there is no band crossing along any high-symmetry line, the $C_4$-eigenvalues remain the same for any $k_z$-slice, such that all planes with fixed $k_z$ are crossed by nodal lines at four points.
Therefore, there must be four nodal lines running along the $k_z$-direction.

For \#124 and \#128, we find that if $\delta^{(\pm),\pi}_2=1$, then $\delta_2^{(\pm),0}=0$ is guaranteed by the compatibility relations.
Therefore, the nontrivial configuration has four nodal rings, each centered at and symmetric about the $k_z=\pi$-plane, related to each other by $C_4$.

For \#130, we notice that (i) it is a subgroup of \#124 and (ii) the band structure which corresponds to the nontrivial indicator $\delta_2^{(+),\pi}=1$ in \#124 gives $\theta_2=1$ when we reduce the symmetries of \#124 to those of \#130.
These two facts mean that the configuration of nodal lines in \#124 is also a possible configuration in \#130 having $\theta_2=1$.
The difference between the two space groups is that, while the four nodal rings must be centered on the $k_z=\pi$-plane in \#124, they can move to the positions shown in Table \ref{tab:1}: two of the four are centered on $\mathrm{XR}$ and two others on its $C_4$ equivalent line.
Through this configuration, the four rings may further move to the $k_z=0$-plane.
Based on this observation, we make a \textit{conjecture} that $e^{i\theta_2}$ is in fact the Berry phase of the loop illustrated in Fig. \ref{fig:BerryPhase1}(c).

The indicator for \#88 is, interestingly, the old inversion indicator $z'_2$.
If $z'_2=1$, there should be 2 mod 4 nodal rings.
Point P in the BZ is the only point whose multiplicity is two, therefore in the minimal configuration, there are two rings around the two P's.

\subsubsection{$\mathbb{Z}_2\times\mathbb{Z}_2$-indicators in space groups \#84, 87}

An exhaustive search for indicators in \#84 space groups reveal the two $\mathbb{Z}_2$-indicators to be $\delta^{(\pm),0}_2$.
If $\delta^{(+),0}_2=1$ and $\delta^{(-),0}_2=0$, or if $\delta^{(-),0}_2=1$ and $\delta^{(+),0}_2=0$, there are four lines running along the $k_z$-direction.
If $\delta^{(+),0}_2=\delta^{(-),0}_2=1$, there are four nodal rings centered and symmetric about the $k_z=0$-plane, and they are related to each other by the fourfold rotation/screw axis.

For space group \#87, one generator of the indicator group is found to be $\delta_2^{(+),0}$, and when $\delta_2^{(+),0}=1$, the compatibility relations require also $\delta_2^{(-),0}=1$.
Since the rotation axis contains inversion centers, in the minimal configuration, there are four nodal rings centered at and symmetric about the $k_z=0$-plane, related to each other by the fourfold rotation.
The other generator takes the form
\begin{equation}
\phi_2\equiv N_{\xi=-1}(\mathrm{M}) + N_{\chi=-1}(\mathrm{N}) + N_{\xi^\prime=-1}(\mathrm{P})\;\mathrm{mod}\;2, \label{eq:phi2}
\end{equation}
where $N_{\xi=-1}(\mathrm{M})$ is number of occupied states having $C_4$ eigenvalue $-1$ at $\mathrm{M}$, $N_{\chi=-1}(\mathrm{N})$ is number of occupied states having $P$ eigenvalue $-1$ at $\mathrm{N}$, and $N_{\xi^\prime=-1}(\mathrm{P})$ is the number of states having $S_4$ eigenvalue $-1$ at $\mathrm{P}$.
We find that when $\phi_2=1$ the Berry phase along the loop shown in Fig. \ref{fig:BerryPhase87} is $\pi$ (mod $2\pi$) and thus $\phi_2=1$ corresponds to a semimetal where an odd number of nodal lines pass through the curve enclosed by the loop.

\begin{figure}
\begin{centering}
\includegraphics[width=0.4\linewidth]{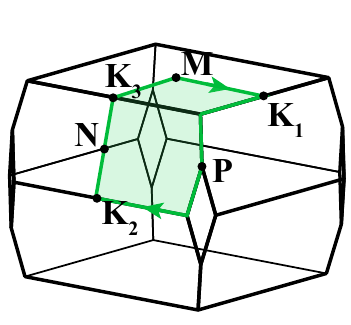}
\par\end{centering}
\protect\caption{\label{fig:BerryPhase87}A loop considered in body-centered tetragonal lattice for space group \#87.}
\end{figure}

Now let us prove the correspondence between the Berry phase and the symmetry eigenvalues.
Below we adopt the notation 
\begin{equation}
        W_{\mathbf{K}\to \mathbf{P}} \equiv \lim_{N\to \infty } U_{\mathbf{K}}^\dagger U_{\mathbf{k}_1} U_{\mathbf{k}_1}^\dagger U_{\mathbf{k}_2}\cdots U_{\mathbf{k}_{N}}^\dagger U_{\mathbf{P}},
\end{equation}
where $\mathbf{k}_{i=1\cdots N}$ give a path from $\mathbf{K}$ to $\mathbf{P}$, $U_\mathbf{k} = [ |u_{1,\mathbf{k}}\rangle, |u_{2,\mathbf{k}}\rangle\cdots ]$ where each column represents an occupied state. 
Then, since wave functions are all real due to the $PT$ symmetry, the Berry phase along the loop in Fig. \ref{fig:BerryPhase87}, $\Phi_B$, is quantized to $0,\pi$ (mod $2\pi$) and given by 
\begin{equation}
e^{i\Phi_B} = \det W_{\mathrm{K}_3\to\mathrm{M}} W_{\mathrm{M}\to\mathrm{K}_1} W_{\mathrm{K}_1\to\mathrm{P}} W_{\mathrm{P}\to\mathrm{K}_2} W_{\mathrm{K}_2\to\mathrm{K}_3}.
\end{equation}
Without loss of generality, we choose the gauge where wave functions in the path $\mathrm{K}_1\to\mathrm{M}$ (except for $C_4$ invariant point $\mathrm{M}$), are $C_4$ rotation counterparts of the wave functions in the path $\mathrm{M}\to\mathrm{K}_1$, such as $U_{\mathrm{K}_1}=C_4U_{\mathrm{K}_3}$. 
For the $\mathrm{M}$ point, we have $C_4 U_{\mathrm{M}} = U_{\mathrm{M}} D^\mathrm{M}(C_4)$, with $D^\mathrm{M}(C_4)$ the $C_4$ representation matrix.
Applying the $C_4$ operation on the wave functions in the path $W_{\mathrm{K}_3\to\mathrm{M}}$, we get
\begin{equation}
W_{\mathrm{K}_3\to\mathrm{M}} = U_{\mathrm{K}_3}^\dagger C_4^\dagger C_4 \cdots C_4^\dagger C_4 U_{\mathrm{M}} =   W_{\mathrm{M}\to\mathrm{K}_1}^\dagger D^\mathrm{M}(C_4).
\end{equation}
Taking similar gauges for other pathes, we get
\begin{equation}
W_{\mathrm{K}_1\to\mathrm{P}} = W_{\mathrm{P}\to\mathrm{K}_2}^\dagger D^\mathrm{P}(S_4),
\end{equation}
and
\begin{equation}
W_{\mathrm{K}_2\to\mathrm{N}} = W_{\mathrm{N}\to\mathrm{K}_3}^\dagger D^\mathrm{N}(P),
\end{equation}
where $D^\mathrm{P}(S_4)$ is the $S_4$ representation matrix at $\mathrm{P}$, and $D^\mathrm{N}(P)$ the $P$ representation matrix  at $\mathrm{N}$.
Therefore, we have
\begin{equation}
e^{i\Phi_B} = \det  D^\mathrm{M}(C_4) D^\mathrm{P}(S_4) D^\mathrm{N}(P).\label{eq:87-berry}
\end{equation}
Comparing Eq. (\ref{eq:phi2}) and (\ref{eq:87-berry}), we find that $\frac{\Phi_B}{\pi}=\phi_2\;\mathrm{mod}\;2$.

\subsubsection{$\mathbb{Z}_2\times\mathbb{Z}_2\times\mathbb{Z}_2$-indicators in space group \#83}

The three indicators for space group \#83 are chosen to be $\delta^{(\pm),\pi}_2$ and $\delta'_2\equiv\delta_2^{(+),0}-\delta_2^{(+),\pi}=\delta_2^{(-),0}-\delta_2^{(-),\pi}$, where the second equality is guaranteed by compatibility relations.
The configurations corresponding to each nontrivial set of indicators are the following.
If $\delta^{(+),\pi}_2=1$ and $\delta^{(-),\pi}_2=0$ or if $\delta^{(+),\pi}_2=0$ and $\delta^{(-),\pi}_2=1$, there are four nodal lines along the $k_z$-direction, related to each other by the fourfold rotation.
In this case the value of $\delta'_2$ is irrelevant, as one can redefine the origin as $\mathbf{a}_3/2$ such that the mirror eigenvalues at $k_z=\pi$ flip, leading to the interchange of $\delta^{(+)\pi}$ and $\delta^{(-)\pi}$ and thus the change of $\delta_2^\prime$. 
We can say the value of $\delta'_2$ is ``convention dependent'', in the same way that $z_4=0$ and $z_4=2$ are convention dependent if any one of $z_{2,i}\neq0$ (Sec. \ref{sub:SI-P-1} and \ref{sec:SPG2}).
Similar with $z_4=0$ and $z_4=2$ in space group \#2, in a fixed convention, $\delta_2^\prime=0$ and $\delta_2^\prime=1$ have a relative difference in the sense that we can tune one to another without closing gaps at high symmetry momenta.

We are left with three cases:
(i) $\delta^{(+),\pi}_2=\delta_2^{(-),\pi}=0,\delta'_2=1$, where there are four nodal rings centered at and symmetric about the $k_z=0$-plane, related to each other under $C_4$;
(ii) $\delta^{(+),\pi}_2=\delta_2^{(-),\pi}=1,\delta'_2=0$, where there are eight nodal lines along the $k_z$-direction, and
(iii) $\delta^{(+),\pi}_2=\delta_2^{(-),\pi}=1,\delta'_2=1$, where there are four nodal rings centered at and symmetric about the $k_z=\pi$-plane, related to each other under $C_4$.

\subsection{Cubic space groups \#201, 203}

There are two space groups in the cubic system that have nontrivial indicator groups ($\mathbb{Z}_2$).
The indicator is simply $z'_2=z_4/2$, and when $z'_2=1$, there are 2 mod 4 nodal rings.
However, due to the presence of threefold rotation axis, there would be in total 6 mod 12 nodal rings.
In these space groups, none of the twofold axes contains any inversion center, so that each nodal ring is symmetric about a $C_2$-invariant line on the boundary of BZ where $\{\hat{C}_2,\hat{P}\hat{T}\}=0$.

\section{Non-centrosymmetric space groups}\label{sec:noncentro}

In the absence of inversion, the nodal lines are unprotected, so that all band crossings at generic momenta are Weyl points.
There are only 12 noncentrosymmetric space groups that have nontrivial indicator groups. Except for \#81, the indicator group is $\mathbb{Z}_2$, while for \#81, the indicator group is $\mathbb{Z}_2\times\mathbb{Z}_2$.
In this section, we use rotation eigenvalues at high-symmetry points to express these indicators, and for each nonzero indicator, we give one possible configuration of the Weyl points in the Brillouin zone.
We have tried to provide the configuration with minimal number of Weyl points.

This section is further divided into subsections, treating three classes of space groups according to the generators: a single rotation axis, or a rotation axis plus vertical glide planes, or an $S_4$-symmetry defined by $S_4:(x,y,z)\rightarrow(-y,x,-z)$, respectively.
In Fig. \ref{fig:4}, we provide the minimal configurations of Weyl points for each nonzero set of indicators in each one of the 12 space groups.

\subsection{Single rotation/screw axis (\#3, 75, 77, 168, 171, 172)}

These space groups have only one rotation (\#3, 75, 168) or screw (\#77, 171, 172) axis.
We can simply apply the lemma of Eq. (\ref{eq:lemma}) to the rotation (or screw) invariant planes in \#3, 168, 171, 172, where the product of all $C_2$-eigenvalues at all TRIM implies if there are 2 mod 4 band crossing points (Weyl points) in the plane.
To be specific, we define for space group \#3, 168, 171, 172 the $\mathbb{Z}_2$-indicator
\begin{equation}
\alpha_2\equiv\sum_{K_i=0,\mathbf{K}\in\mathrm{TRIM}}N_{\zeta=-1}(\mathbf{K})\;\mathrm{mod}\;2, \label{eq:alpha2}
\end{equation}
where $i=y$ for \#3, and $i=z$ for \#168, 171, 172.
If $\alpha_2=1$, then the Berry phase associated with the loop enclosing half of the BZ on the $k_i=0$-slice is $\pi$.
Due to $C_2T$-symmetry, the $z$-component of the Berry curvature vanishes for the $k_i=0$-slice, and the $\pi$-Berry phase implies the existence of 1 mod 2 band crossing in each half of the $k_i=0$-slice.
There are hence 2 mod 4 Weyl points on the $k_i=0$-plane if $\alpha_2=1$.
For \#3, the minimal number is two, but for \#168, 171 and 172, the minimal number is six due to the sixfold rotation.
In the configuration having minimal number of Weyl points, the Weyl points on the $k_i=0$-plane are related to each other by twofold or sixfold rotations, thus having the same monopole charge.
Using the absence of band crossing along high-symmetry lines, the rotation eigenvalues are the same for the $k_i=0$ and the $k_i=\pi$ slices, so that $\alpha_2=1$ also implies that on the $k_i=\pi$-slice, there are 2 mod 4 Weyl points in \#3 and 6 mod 12 Weyl points in \#168, 171, 172.
In the configuration with minimal number of Weyl points, those Weyl points are related by rotation symmetries thus having the same monopole charge.
However, since the total charge must vanish in the entire BZ, the Weyl points on the $k_i=0$- and the $k_i=\pi$-slice have opposite charges.

Space groups \#75 and \#77 have a single fourfold rotation and screw axis, respectively.
According to the lemma in Eq. (\ref{eq:lemma2}), we can define on the $k_z=0$-slice the following $\mathbb{Z}_2$ indicator
\begin{equation}
\beta_2\equiv{N}_{\xi=-1}(\Gamma)+N_{\xi=-1}(\mathrm{M})+N_{\zeta=-1}(\mathrm{X})\;\mathrm{mod}\;2.
\end{equation}
The physical meaning of $\beta_2=1$ is that the Berry phase of the loop enclosing one quadrant of the BZ on the $k_z=0$-slice is $\pi$, and implies the presence of a Weyl point within the quadrant.
Following similar steps, we find that if $\beta_2=1$, there are, minimally, four Weyl points on the $k_z=0$-plane and four on the $k_z=\pi$-plane.
The four Weyl points having same $k_z$ have the same charge, while those on different $k_z$-slices have opposite monopole charges.

Here we take space group \#3 as an example to show how the Weyl points can be created, or, equivalently, annihilated.
The minimal configuration of Weyl points is shown in Fig. \ref{fig:4}(a).
In the following we present a process annihilating these Weyl points symmetrically. 
First, the two Weyl points in $k_y=0$-slice move toward $\Gamma$ and then meet each other at $\Gamma$. 
(For space group \#3 the $C_2$-invariant line, $\Gamma\mathrm{Z}$,  is along $k_y$-direction.)
After their meeting the $k_y=0$-slice become fully gapped.
According to lemma (\ref{eq:lemma}), this process changes the parity of the sum of $C_2$ eigenvalues in the occupied bands, leading to a change of $\alpha_2$.
The interchange of $C_2$ eigenvalues also break the compatibility relation along $\Gamma\mathrm{Z}$, leading to a band crossing protected by $C_2$.
Second, to remove the band crossing and recover the compatibility relation, we move the crossing point to $\mathrm{Z}$, causing a band inversion there, which changes $C_2$ eigenvalues at $\mathrm{Z}$.
Again, due to lemma (\ref{eq:lemma}), this band inversion creates two additional Weyl points in $k_y=\pi$-slice, possessing opposite charges of the two Weyl points that have been in $k_y=\pi$.
Third, we annihilate the four Weyl points pairwise at two $C_2$-related generic momenta.

\subsection{Rotation/screw axis plus vertical glide planes (\#27, 37, 103, 184)}

In this subsection, the space groups all have, aside from a rotation axis, vertical glide planes whose half-translation directions are along the rotation axis, denoted by $G_a:(x,y,z)\rightarrow(-x,y,z+1/2)$.
To find the expressions of the indicators, we will have to invoke new variations of the lemma in Eq. (\ref{eq:lemma}).
There are three variations for this type of space groups, in simple orthorhombic, base centered orthorhombic and simple tetragonal Bravais lattices, respectively.
(Here we consider the hexagonal lattice of \#184 a special case of based-centered orthorhombic.)
We start our discussion with the simple orthorhombic case, whose BZ is depicted in Fig. \ref{fig:3}(a).

On the $k_z=\pi$-slice, we have $G_a^2=e^{-ik_z}=-1$, so that its eigenvalues $g_a=\pm{i}$.
$C_2$ and $G_a$ automatically imply another glide plane $G_b\equiv{C}_2G_a$, which also satisfies $G_b^2=e^{-ik_z}=-1$ and $g_b=\pm{i}$.
The two lines $\mathrm{Z}\mathrm{T}$ and $\mathrm{U}\mathrm{R}$ are invariant under $G_a$, while $\mathrm{Z}\mathrm{U}$ and $\mathrm{T}\mathrm{R}$ are invariant under $G_b$. 
Therefore, along these lines, bands can be labeled by their respective eigenvalues, $|u_{a/b,\pm{i}}(\mathbf{k})\rangle$.
In fact, since $C_2T$ commutes with $G_a$ and $G_b$, all bands along these lines are doubly degenerate, as $C_2T$ maps an eigenstate having $g_{a,b}=\pm{i}$ to one having $g_{a,b}=\mp{i}$, leaving $\mathbf{k}$ unchanged on the $k_z=\pi$-slice.
Without additional symmetries, the bands along these lines appear in groups of two.
For simplicity we assume for now that there are only two bands (one group).

The four TRIM are not only both $G_{a,b}$-invariant but also $C_2$-invariant.
Since $C_2$-eigenvalues, $\zeta=\pm1$, are real, they are unchanged under $C_2T$, at each TRIM, the doublet states have the \textit{same} $\zeta$.
Another important fact is that, due to $G_b=C_2G_a$, at each TRIM we have
\begin{equation}
\label{eq:temp1}
G_b|u_{a,\pm{i}}(\mathbf{K})\rangle=\pm\zeta(\mathbf{K})|u_{a,\pm{i}}(\mathbf{K})\rangle,
\end{equation}
which implies
\begin{equation}
|u_{b,{\pm}i}(\mathbf{K})\rangle\propto|u_{a,\pm\zeta(\mathbf{K}){i}}(\mathbf{K})\rangle.
\end{equation}
Now we consider the Berry phase associated with the loop shown in Fig.\ref{fig:3}(a), enclosing a quarter of the BZ at $k_z=\pi$.
We pick a basis at $\mathrm{Z}$:
\begin{eqnarray}
|u_1(\mathrm{Z})\rangle&\equiv&|u_{a,+i}(\mathbb{Z})\rangle,\\
\nonumber
|u_2(\mathrm{Z})\rangle&=&(|u_1(\mathrm{Z})\rangle)^\ast.
\end{eqnarray}
The first segment of the path is $\mathrm{ZT}$, along this line, the Hamiltonian can be block-diagonalized by into $H_{a,+i}\oplus{H}_{a,-i}$.
$|u_1\rangle$ hence evolves under $H_{a,+i}$ and $|u_2\rangle$ evolves under $H_{a,-i}$, such that at $\mathrm{T}$, they are still eigenvectors of $G_a$, and since $H_{+i}=H^\ast_{-i}$ thanks to $C_2T$-symmetry, they still satisfy $|u_2\rangle=(|u_1\rangle)^\ast$.
Using Eq.(\ref{eq:temp1}), we know that $|u_{1,2}(\mathrm{T})\rangle$ are also eigenvectors of $G_b$ with eigenvalues $\pm\zeta(\mathrm{T})$.
Starting from $\mathrm{T}$ to $\mathrm{R}$, the two states evolve under $H_{b,+\zeta(\mathrm{T})i}$ and $H_{b,-\zeta(\mathrm{T})i}$ respectively.
Repeating this process, when the two states $|u_{1,2}\rangle$ go back to $\mathrm{Z}$, they become $|u'_1(\mathrm{Z})\rangle$ and $|u'_2(\mathrm{Z})\rangle$, which are eigenvectors of $G_b$, because the final segment of the path $\mathrm{UZ}$ is invariant under $G_b$.
Their eigenvalues are given by $g_b=\pm\zeta(\mathrm{T})\zeta(\mathrm{R})\zeta(\mathrm{U})i$ respectively.
Using Eq.(\ref{eq:temp1}) at $\mathbf{K}=\mathrm{Z}$, we find that $|u'_{1,2}\rangle$ are also eigenvectors of $G_a$
\begin{eqnarray}
\label{eq:temp2}
|u'_1(\mathrm{Z})\rangle&=&e^{i\theta}|u_{a,\zeta(\mathrm{T})\zeta(\mathrm{R})\zeta(\mathrm{U})\zeta(\mathrm{Z})i}\rangle,\\
\nonumber
|u'_2(\mathrm{Z})\rangle&=&(|u'_1(\mathrm{Z})\rangle)^\ast.
\end{eqnarray}
Eqs.(\ref{eq:temp2}) hence call for the definition of the new $\mathbb{Z}_2$-indicator
\begin{equation}
\gamma_2\equiv\sum_{\mathbf{K}\in\mathrm{Z,T,U,R}}{N}_{\zeta=-1}(\mathbf{K})/2\;\mathrm{mod}\;2,
\end{equation}
where the division by 2 is because each level is doubly degenerate.
When $\gamma_2=1$, we have
\begin{eqnarray}
\left(\begin{matrix}
|u'_1\rangle\\
|u'_2\rangle
\end{matrix}\right)=
\left(\begin{matrix}
0 & e^{i\theta}\\
e^{-i\theta} & 0
\end{matrix}\right)
\left(\begin{matrix}
|u_1\rangle\\
|u_2\rangle
\end{matrix}\right).
\end{eqnarray},
so that the Berry phase of the loop is
\begin{equation}
e^{i\Phi_B}=\det(\langle{u'_i}|u_j\rangle)=-1.
\end{equation}
Here we have established yet another variation of the lemma:\\
\textit{Lemma: third variation}
\begin{eqnarray}
e^{i\Phi_B}&=&\prod_{n\in{occ./2},\mathbf{K}\in{\mathrm{Z,T,U,R}}}\zeta_n(\mathbf{K})\\
\nonumber&=&e^{i\gamma_2\pi},
\end{eqnarray}
where $occ./2$ means that for each degenerate pair at these TRIM, we only take one band for the calculation.
\begin{figure}
\begin{centering}
\includegraphics[width=1\linewidth]{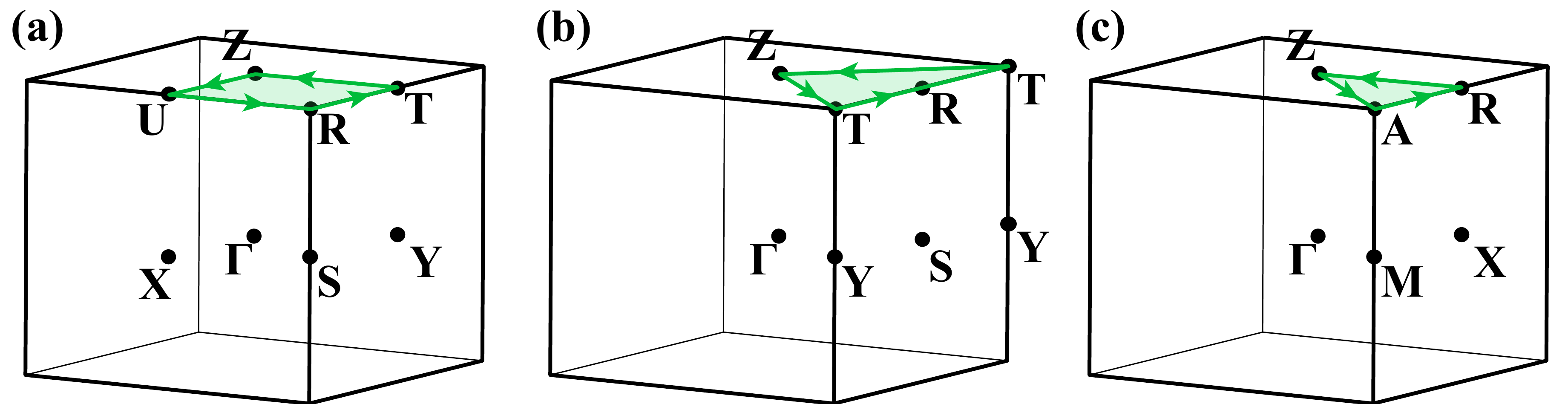}
\par\end{centering}
\protect\caption{\label{fig:3} Loops along which the Berry phases are used to define indicators in noncentrosymmetric space groups (a) \#27, (b) \#37, and (c) \#103.}
\end{figure}
Therefore, when $\gamma_2=1$, the Berry phase of the loop in Fig. \ref{fig:3}(a) is $\pi$, and since $C_2T$ ensures the vanishing of the $z$-component of the Berry curvature, there must be 1 mod 2 Weyl point in each quadrant.
In the configuration having least number (four) of Weyl points, the two Weyl points related by $G_a$ or $G_b$ have opposite monopole charges.

It may be a little counterintuitive why the two Weyl points of opposite charge cannot pairwise annihilate on the glide plane.
Consider two such Weyl points that are related to each other by $G_a$.
Then we remark that they are necessarily also related by $G_b*T$, and that on $k_z=\pi$-slice, there is
\begin{equation}
(\hat{G}_b\hat{T})^2=\hat{G}^2_b\hat{T}^2=-1.
\end{equation}
In Ref.[\onlinecite{Fang2016}] one of us shows that two Weyl points related by $G_b*T$ satisfying $(\hat{G}_b\hat{T})^2=-1$ cannot pairwise annihilate when they meet, but will form an accidental Dirac point.

The discussion of a base-centered orthorhombic case differs but a little from the simple orthorhombic case.
Again we consider the $k_z=\pi$-slice, where the TRIM are denoted $\mathrm{Z, R, T}$, whose multiplicities are one, two and one respectively.
While all TRIM are $C_2$-invariant, $\mathrm{Z}$ and $\mathrm{T}$ are also invariant under $G_a$ and $G_b$.
Consider the loop shown in Fig.\ref{fig:3}(b).
One should notice that while the loop encloses one quarter of the BZ, it is formed by two closed loop, labeled loop-1 from $\mathrm{T}$ to $\mathrm{R}$ to $\mathrm{T}$ and loop-2 from $\mathrm{T}$ to $\mathrm{Z}$ to $\mathrm{T}$.
The total Berry phase is the sum of the Berry phases of loop-1 and loop-2, each of which is quantized to 0 and $\pi$ due to $C_2T$.
Using the same steps, we can show that the second Berry phase is given by\\
\begin{equation}
\exp(i\Phi_2)=\prod_{n\in{occ.}/2}\zeta_n(\mathrm{Z})\zeta_n(\mathrm{T}),
\end{equation}
where $occ./2$ means that for each degenerate pair only one state is chosen.
Regarding loop-1, Ref.~\cite{Turner2010,Hughes2011} show that its Berry phase is given by
\begin{eqnarray}
\exp(i\Phi_1)&=&\prod_{n\in{occ.}}\zeta_n(\mathrm{T})\zeta_n(\mathrm{R})\\
\nonumber&=&\prod_{n\in{occ.}}\zeta_n(\mathrm{R}),
\end{eqnarray}
where the second equality comes from the fact that all $\zeta$ values at $\mathrm{T}$ appear in doubles.
These observations lead us to define a new $\mathbb{Z}_2$-indicator
\begin{equation}
\gamma'_2\equiv\frac{{N}_{\zeta=-1}(\mathrm{Z})+N_{\zeta=-1}(\mathrm{T})}{2}+N_{\zeta=-1}(\mathrm{R})\;\mathrm{mod}\;2,
\end{equation}
such that $e^{i(\Phi_1+\Phi_2)}=e^{i\gamma'_2\pi}$.
When $\gamma'_2=1$, there must be one Weyl point inside each quarter of the BZ at $k_z=\pi$-slice.
Therefore, there are four Weyl points in the minimal configuration, where any two related by either $G_a$ or $G_b$ have opposite monopole charges.

The indicator $\gamma'_2$ is well defined for both \#37 and \#183, since hexagonal lattice can be considered as a special case of base-centered orthogonal lattice.
However, in \#183, due to the sixfold rotation, the total number of Weyl points is 12 mod 24, and in the minimal configuration, 12 Weyl points are related to each other either by $G_a$ or by $C_6$, shown in Fig.\ref{fig:4}(h).
Two Weyl points related by $C_6$ have the same monopole charge, and two related by $G_{a,b}$ have opposite charges.

Similarly, we find the indicator for space group \#103:
\begin{equation}
\gamma''_2\equiv\frac{N_{\xi=-1}(\mathrm{Z})+N_{\xi=-1}(\mathrm{A})+N_{\zeta=-1}(\mathrm{R})}{2}\;\mathrm{mod}\;2.
\end{equation}
When $\gamma''_2=1$, the Berry phase of the loop in Fig.\ref{fig:3}(c) is $\pi$, so that there must be 1 mod 2 Weyl point in half of a quadrant, resulting in 8 mod 16 Weyl points in total.
In the minimal configuration, the eight Weyl points, four positive and four negative, are related to each other either by $C_4$ or $G_{a,b}$.
Two Weyl points related by $C_4$ have the same monopole charge, and the two related by $G_{a,b}$ have different ones.

\subsection{$S_4$-symmetry (\#81, \#82)}

$S_{2n}$-symmetries are the only type of (noncentrosymmetric) point group symmetries the invariant subspace of which consists of discrete points rather than lines or planes.
Aside from $S_2$ which is same as inversion, $S_4$ is also the only $S_{2n}$-symmetry that is consistent with 3D lattices.
Space group \#81 on a simple tetragonal lattice is singly generated by this symmetry.
In momentum space, we have $S_4:(k_x,k_y,k_z)\rightarrow(-k_y,k_z,-k_z)$, so that on $k_z=0$- and $k_z=\pi$-slice, they act the same way as $C_4$.
Therefore, we can use the same lemma in Eq.(\ref{eq:lemma2}) to define the following two $\mathbb{Z}_2$-indicators
\begin{equation}
\begin{aligned}
\omega_{2}^0=&N_{\xi=-1}(\Gamma)+N_{\xi=-1}(\mathrm{M})+N_{\zeta=-1}(\mathrm{X})\;\mathrm{mod}\;2\\
\omega_{2}^\pi=&N_{\xi=-1}(\mathrm{Z})+N_{\xi=-1}(\mathrm{A})+N_{\zeta=-1}(\mathrm{R})\;\mathrm{mod}\;2
\end{aligned},
\end{equation}
where $\xi$ is the $S_4$-eigenvalue, and we have observed that fact that $S_4^2=C_2$.

When $\omega_{2}^{0,\pi}=1$, there are 4 mod 8 points on the $k_z=0$ and the $k_z=\pi$-slice, respectively.
Since $S_4$ is an \textit{improper} rotation, two Weyl points related by $S_4$ have opposite monopole charges: this is a difference between $S_4$ and $C_4$.

For space group \#82, the only $\mathbb{Z}_2$ indicator is $\omega_{2}^0$.

\begin{figure*}
\begin{centering}
\includegraphics[width=1\linewidth]{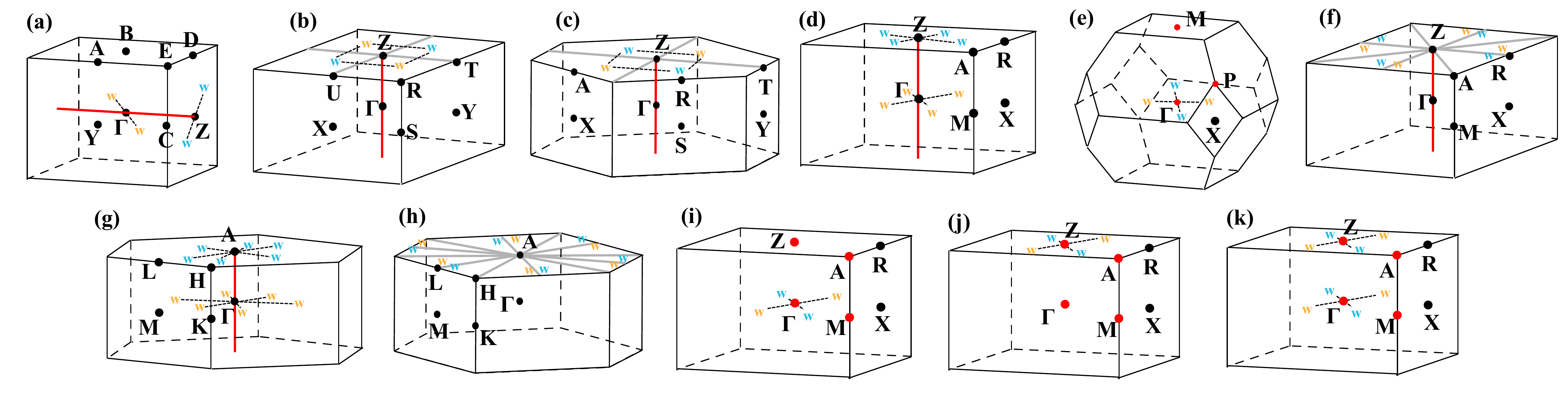}
\par\end{centering}
\protect\caption{\label{fig:4} Minimal configurations of the Weyl points for each nonzero set of indicators in various space groups. Red lines are rotation axes, grey lines the intersection between vertical glide planes and the $k_z=\pi$-plane, red dots $S_4$-centers, and ``w'' stands for Weyl, where different colors of ``w'' mean opposite monopole charge. (a-h) are for space groups with only one $\mathbb{Z}_2$-indicator: (a) \#3, (b) \#27, (c) \#37, (d) \#75, 77, (e) \#82, (f) \#103, (g) \#168, 171, 172 and (h) \#184. (i-k) are the minimal configurations for \#81 having indicator sets $(\omega_2^0,\omega_2^\pi)=(1,0), (0,1), (1,1)$, respectively.}
\end{figure*}

\section{Discussion and conclusion}

In this work, we find the explicit expressions for all symmetry-based indicators in terms of symmetry eigenvalues of valence bands at high-symmetry points in the momentum space for 3D systems with time-reversal symmetry and negligible spin-orbital coupling, and we characterize all topological states corresponding to every given set of nonzero indicators.
Somewhat to our surprise, all nonzero sets of indicators necessarily correspond to some topological semimetals, i.~e., none of them is compatible with a gapped band structure in three dimensions.
These semimetals are in general ``hidden'' from most first principles calculations, since in the latter only high-symmetry lines are routinely scanned, but in these semimetals all high-symmetry lines have gapped spectra.
Looking at these expressions, we realize that all the indicators except one are in fact \textit{topological invariants} for sub-manifolds of the 3D BZ.
(These expressions are checked via decomposing all linearly independent solutions of compatibility relations into elementary band representations, where both data can be found in Ref.[\onlinecite{Bradlyn2017}].)

Most of the indicators are equivalent to Berry phases: 
\begin{enumerate}[fullwidth,itemsep=0cm,label=(\roman*)]
\item $z_{2,i=1,2,3}\pi$ is the Berry phase of a loop enclosing half of the BZ at $k_i=\pi$ (all centrosymmetric space groups).
\item $z^{(\pm)}_2\pi$ is the Berry phase of a loop enclosing half of the BZ at $k_z=\pi$ in the $M=\pm1$-sector (\#10, 175, 176, 192).
\item $\delta_2\pi$ is the Berry phase of a loop enclosing quarter of the BZ at any fixed $k_z$ (\#83, 84, 85, 86).
\item $\delta^{(\pm),\pi}_2\pi$ is the Berry phase of a loop enclosing quarter of the BZ at $k_z=\pi$ in the $M=\pm1$-sector (\#83, 124, 128).
\item $\delta^{(\pm),0}_2\pi$ is the Berry phase of a loop enclosing quarter of the BZ at $k_z=0$ in the $M=\pm1$-sector (\#83, 84, 87). 
\item $\delta'_2\pi$ is the difference between the Berry phase of a loop enclosing a quarter of the BZ at $k_z=0$ in the $M=+1$-sector and that of a loop and that at $k_z=\pi$ (\#83).
\item $\phi_2\pi$ is the Berry phase of the loop indicated in Fig. \ref{fig:BerryPhase87} (\#87). 
\item $\theta_2\pi$ is \textit{conjectured} to be the Berry phase of the loop indicated in Fig.~\ref{fig:BerryPhase1}(c) (\#130).
\item $\alpha_2\pi$ is the Berry phase of a loop enclosing half of BZ at $k_z=0$ (\#3, 168, 171, 172).
\item $\beta_2\pi$ is the Berry phase of a loop enclosing quarter of BZ at $k_z=0$ (\#75, 77).
\item $\gamma_2\pi$ is the Berry phase of a loop enclosing quarter of BZ at $k_z=\pi$ (\#27).
\item $\gamma'_2\pi$ is the Berry phase of a loop enclosing quarter of BZ at $k_z=\pi$ (\#37,184).
\item $\gamma''_2\pi$ is the Berry phase of a loop enclosing one eighth of BZ at $k_z=\pi$ (\#103).
\item $\omega_{2}^0 \pi$ is the Berry phase of a loop enclosing quarter of BZ at $k_z=0$-plane (\#81,\#82).
\item $\omega_{2}^\pi \pi$ is the Berry phase of a loop enclosing quarter of BZ at $k_z=\pi$-plane (\#81).
\end{enumerate}

There is one special $\mathbb{Z}_4$-indicator, $z_4$, that is not a Berry phase but the number of nodal rings modulo four; 
$z'_2=z_4/2$ is a derivative of $z_4$ in space groups where $z_4$ is constrained by symmetries to be even.

Whenever any of the above Berry phases is $\pi$, we immediately know that there is 1 mod 2 robust band crossings in the area bounded by the loop.
Whenever $z_4$ or $z'_2$ is nonzero, it gives us the total number of rings or lines.
These observations help us determine all possible configurations of all the nodal loops and Weyl points given any set of symmetry-based indicators mentioned above.

Importantly, the definitions of these indicators in most cases only require several, but not all, symmetries in the space groups in which they are defined, and they only require that there is no band crossing along the loop on which the Berry phase is defined.
This means that even in band structures where compatibility relations are violated, i.~e., when there are band crossings along certain high-symmetry lines, as long as the above requirements hold, these indicators can still be applied to find band crossings at generic momenta, away from the ones on high-symmetry lines.

A very simple example helps illustrate this point.
In space group \#75, which has a simple tetragonal lattice and a single point group generator $C_4$, starting from a gapped state, we consider a band inversion at $\Gamma$ where the valence band with $\xi=+1$ becomes one having $\xi=-1$.
This band inversion is known to cause two band crossings along $\Gamma\mathrm{Z}$, since $C_4$ is a good quantum number along this line, so the band structure violates the compatibility relations of \#75.
However, one easily confirms that $\beta_2=1$ after the band inversion, so that away from $\Gamma\mathrm{Z}$, there are four additional Weyl points on the $k_z=0$-plane, having the same monopole charge.

Topological semimetals are known to be ``parent states'' for topologically gapped states.
In fact, two possible mechanisms have been established that give rise to nontrivial topology: the Kane-Mele mechanism \cite{Kane2005,Kane2005a} where topological band crossings open up a gap when spin-orbital coupling is turned on, and the Bernevig-Hughes-Zhang mechanism \cite{Bernevig2006} where band inversions occur due to strong spin-orbital coupling.
These two mechanisms (and their generalized versions) are used and so far considered as guiding principles for the search of topological materials that are systematically applied in Ref. [\onlinecite{Bradlyn2017}], which have successfully yielded many a new material candidate.
(Among them are new types of ``weak phases'' termed fragile topological phases, the understanding of which opens new directions \cite{Po2017a,Cano2017}.)
According to the Kane-Mele mechanism, some topological semimetals, like graphene, are just topological (crystalline) insulators disguised by the small SOC.
A question then naturally presents itself at this point: will the topological semimetals diagnosed by these indicators become topological gapped states, such as topological insulators or topological crystalline insulators \cite{Fu2011} when spin-orbital coupling is \textit{perturbatively} added?
For noncentrosymmetric space groups, the answer is very simple: each Weyl points splits into two having the same charge, so that the system remains a Weyl semimetal.
For centrosymmetric space groups, since neither Weyl point or nodal line at generic momenta is allowed, adding SOC will open full gaps for all topological semimetals discussed in this work.
But are all these gapped states topologically nontrivial?
A complete answer to this question justifies an independent work, but here we partly address this issue by connecting the indicators in the orthogonal (without SOC) and the symplectic (with SOC) classes.
The topological information carried by the nonzero indicators in the symplectic class is discussed in a parallel work \cite{Song2017a,Khalaf2017}.

To be specific, we inspect all space groups that have nontrivial indicator group both with and without SOC.
We have found the following quantitative relations between the indicators in the two classes of Hamiltonians.
First of all, the indicators that only depend on inversion eigenvalues $z_{2,i=1,2,3}$ and $z_4$ have exactly the same definition in both classes.
Fu-Kane formula shows that with SOC, $z_{2,i=1,2,3}$ and $z_4\;\mathrm{mod}\;2$ correspond to the three weak and one strong invariants of time-reversal topological insulators.
When $z_{2,i}=0$ and $z_4=2$, the corresponding gapped states with SOC belong to a new class of topological crystalline insulators, the surface states of which are discussed in Ref. \cite{Fang2017,Song2017a,Khalaf2017}.
Among centrosymmetric space groups having nontrivial indicator groups both with and without SOC, these four indicators have comprehended the indicator groups in both classes in all but the following five space groups: \#83, 87, 128, 175 and 192.

To find the relations between indicators without SOC and those with SOC, we first need to determine what an irreducible representation without SOC becomes when SOC is turned on.
To be specific, if a irreducible representation without SOC at $\mathbf{k}$ is given by $D(g)$, where $g$ belongs to the little group of $\mathbf{k}$, then the representation matrix after considering an infinite small SOC becomes $D(g)\otimes u(g)$, where $u(g)$ is the SU(2) rotation matrix of $g$.
In most cases, $D(g)\otimes u(g)$ becomes reducible and reduces to several irreducible representations with SOC.
With this information obtained, we are able to translate any symmetry data without SOC to a symmetry data with SOC.
%For each indicator without SOC, we can pick a representative symmetry data, translate it to the SOC side, and compute the indicator with SOC to determine the correspondence.
Following this method, we find:
\begin{enumerate}[fullwidth,itemsep=0cm,label=(\roman*)]
\item \#83. The indicators in absence of SOC are $\delta_2^{(+),\pi}$, $\delta_2^{(-),\pi}$, and $\delta_2^\prime$. The indicators in presence of SOC are $z_{2w,1}$, $z_{4m,\pi}$, and $z_8$, wherein $z_{2w,1}$ is the weak index in the $x$ direction, $z_{4m,\pi}$ is the Mirror Chern number (mod 4) at the $k_z=\pi$-plane, and $z_8\;\mathrm{mod}\;4$ is the sum of Mirror Chern numbers at the  $k_z=0$- and $k_z=\pi$-planes (mod 4). The remaining case, $z_8=4$, corresponds to either a $C_4$-protected topological crystalline insulator or a mirror Chern insulator with Mirror Chern numbers at the $k_z=0$- and $k_z=\pi$-planes differing from each other by 4 (mod 8) \cite{Song2017,Fang2017,Song2017a,Khalaf2017}. The mappings from indicators without SOC to indicators with SOC are $(100)\rightarrow(024)$, $(010)\rightarrow(024)$, $(001)\rightarrow(004)$.
\item \#87. The indicators in absence of SOC are $\phi_2$ and $\delta_2^{(+),0}$, and the indicators in presence of SOC are $z_{2w,1}$ and $z_8$, which have the same definitions in \#83 \cite{Song2017a,Khalaf2017}. The mappings are $(10)\rightarrow(04)$, $(01)\rightarrow(04)$.
%It is interesting that the two non SOC indicator sets are mapped to the same SOC indicator set $(04)$. 
%We conjecture that, after tuning on SOC, the one of them becomes the $C_4$-protected topological insulator whereas the other one becomes the mirror Chern insulator, both of which have the same indicator set $(04)$, as discussed above.
\item \#128. The indicator in absence of SOC is $\delta_2^{(+),\pi}$, and the indicator in presence of SOC is $z_8$, which has the same definition in \#83 \cite{Song2017a,Khalaf2017}. The mapping is $1\rightarrow4$.
\item \#175. The indicators in absence of SOC are $z_2^{(+)}$, $z_{2}^{(-)}$, and $z_2^\prime=z_4/2$. The indicators in presence of SOC are $z_{6m,\pi}$ and $z_{12}$, wherein $z_{6m,\pi}$ is the mirror Chern number (mod 6) at the $k_z=\pi$-plane, and $z_{12}\;\mathrm{mod}\;6$ is the sum of mirror Chern numbers at the $k_z=0$- and $k_z=\pi$-planes. The remaining case, $z_{12}=6$, corresponds to either a $C_6$-protected topological crystalline insulator or a mirror Chern insulator with mirror Chern numbers at the $k_z=0$- and $k_z=\pi$-planes differing from each other by 6 (mod 12) \cite{Song2017a,Khalaf2017}. The mappings are $(100)\rightarrow(30)$, $(010)\rightarrow(30)$, $(001)\rightarrow(06)$.
\item \#192. The indicator in absence of SOC is $z_2^\prime$, and the indicator in presence of SOC is $z_{12}$, which has the same definition in \#175. The mapping is  $1\rightarrow6$.
\end{enumerate}

\acknowledgements{The authors acknowledge support from Ministry of Science and Technology of China under grant numbers 2016YFA0302400 and 2016YFA0300600, National Science Foundation of China under grant number 11674370 and 11421092, and from Chinese Academy of Sciences under grant number XXH13506-202.}

\bibliography{Ref}

\end{document}